\documentclass[11pt]{article}
\usepackage{natbib}
\usepackage[toc,page,header]{appendix}
\usepackage{minitoc}

\usepackage[final]{acl}

\usepackage{times}
\usepackage{latexsym}
\usepackage{amsmath}
\usepackage{amssymb}

\usepackage{enumitem} %

\usepackage{amsmath,amsfonts,bm}

\def\eqref#1{equation~\ref{#1}}

\def\1{\bm{1}}

\DeclareMathAlphabet{\mathsfit}{\encodingdefault}{\sfdefault}{m}{sl}
\SetMathAlphabet{\mathsfit}{bold}{\encodingdefault}{\sfdefault}{bx}{n}

\usepackage[T1]{fontenc}

\usepackage[utf8]{inputenc}

\usepackage{microtype}

\usepackage{inconsolata}

\usepackage{graphicx}

\usepackage[dvipsnames]{xcolor}

\usepackage{subcaption}
\usepackage{tipa}
\usepackage{float}

\usepackage{booktabs} %
\usepackage{multirow} %
\usepackage{array}    %

\usepackage[table]{xcolor}

\title{AudioSAE: Towards Understanding of Audio-Processing Models with Sparse AutoEncoders}

\author{ Georgii Aparin\thanks{Equal contribution}, Tasnima Sadekova\footnotemark[1], Alexey Rukhovich, Assel Yermekova, \\ {\bf Laida Kushnareva, Vadim Popov, Kristian Kuznetsov, Irina Piontkovskaya} \\
        Huawei Noah's Ark Lab\\
        \texttt{aparingm@gmail.com},  \texttt{sadekova.t.r@gmail.com}}

\begin{document}
\doparttoc %
\faketableofcontents %
\maketitle
\begin{abstract}
Sparse Autoencoders (SAEs) are powerful tools for interpreting neural representations, yet their use in audio remains underexplored. We train SAEs across all encoder layers of Whisper and HuBERT, provide an extensive evaluation of their stability, interpretability, and show their practical utility. Over 50\% of the features remain consistent across random seeds, and reconstruction quality is preserved. 
SAE features capture general acoustic and semantic information as well as specific events, including environmental noises and paralinguistic sounds (e.g. laughter, whispering)
and disentangle them effectively, requiring removal of only 19--27\% of features to erase a concept. Feature steering reduces Whisper's false speech detections by 70\%
with negligible WER increase, demonstrating real-world applicability. Finally, we find SAE features correlated with human EEG activity during speech perception, indicating alignment with human neural processing.  The code
and checkpoints are available at \href{https://github.com/audiosae/audiosae_demo}{https://github.com/audiosae/audiosae\_demo}.
\end{abstract}

\section{Introduction}

Audio and speech representation learning has evolved significantly over the past decades. Early models relied on mel-spectrograms, while recent advances in deep learning, particularly Transformers, have enabled large-scale audio modeling for speech recognition \citep{whisper}, synthesis \citep{audiolm,valle}, and general audio understanding \citep{audioflamingo2}.  

A major milestone has been the development of self-supervised learning (SSL) models such as Wav2Vec~2.0, WavLM, and HuBERT~\citep{wav2vec, wav2vec2, wavlm, hubert}, which learn directly from raw audio by predicting masked segments. Their intermediate representations have proven useful for diverse downstream tasks including automatic speech recognition (ASR), speaker verification, translation, and audio generation~\citep{s2st_units1, s2st_units2, conversion_units, audiogen}.  

In parallel, neural audio codecs~\citep{soundstream, encodec, speechtokenizer} and encoder part of models like Whisper~\citep{whisper} have emerged as universal feature extractors for speech LLMs. Trained on multilingual and noisy data, these encoders produce robust, semantically rich embeddings applicable across a wide range of audio tasks.  

Since these models are optimized for different objectives, they encode information in distinct ways, motivating the need for systematic interpretability methods. Recently, \textit{SAEs} have gained attention as a means to decompose dense representations into sparse, interpretable components~\citep{cunningham2023sparseautoencodershighlyinterpretable, lieberum-etal-2024-gemma, patchsae}. While SAEs have been extensively studied in text and vision, their potential for analyzing audio models remains largely unexplored, with only isolated attempts in music modeling~\citep{sae_music} or informal Whisper analyses.  

Recent studies demonstrate that SAEs have been successfully applied across a wide range of models: from Gemma-2~\citep{lieberum-etal-2024-gemma} to GPT-4o~\citep{wang2025persona}, revealing interpretable internal mechanisms and enabling targeted behavior control. In NLP, SAEs help disentangle fine-grained semantic and stylistic features~\citep{cunningham2023sparseautoencodershighlyinterpretable}, while \citet{muhamed-etal-2025-decoding} used them to separate gender markers from professional content, mitigating spurious correlations. In computer vision, \citet{cywinski2025saeuroninterpretableconceptunlearning} achieved selective concept unlearning in diffusion models without additional fine-tuning. These results confirm that SAE features serve both as interpretable units of representation and as actionable levers for model steering.

In this work, we fill this gap by applying SAEs to large-scale audio representation models, Whisper and HuBERT, and conducting a systematic evaluation of their stability, interpretability, and practical usefulness.  

Our main contributions are as follows: \vspace{-0.15cm}
\begin{itemize}
    \item We train SAEs on activations of Whisper and HuBERT and release the models and code, providing the first large-scale analysis of audio representation interpretability via sparse autoencoders. \vspace{-0.15cm}
    \item We develop distributional and automatic validation methods showing that SAE features are stable across seeds and encode semantic, paralinguistic, and acoustic information.
    \vspace{-0.15cm}
    \item We demonstrate practical and neuroscientific relevance by steering Whisper to reduce hallucinations and revealing correlations between SAE features and human EEG activity.  
\end{itemize}

\section{Related Works}

\label{sec:related}

\subsection{Audio Representations}

In this work we consider two popular models widely used in downstream tasks related to speech processing: Whisper \cite{whisper} and HuBERT \cite{hubert}. The former is an encoder-decoder Transformer trained with multi-tasking on a large corpus of $680k$ hours of multi-lingual speech, while the latter is encoder-only Transformer trained on $60k$ hours of English speech iteratively to predict labels assigned by the model in the previous iteration given masked audio sequence. We also experimented with EnCodec \cite{encodec} representations but omit them from the paper because the obtained latent space was not sufficiently sparse.

\subsection{SAE in Various Domains and Applications}

\textbf{Natural Language Processing.} Since \cite{shakley2023superposition} introduced the idea, Sparse Autoencoders (SAEs) have been extensively studied in the analysis of NLP models, with pretrained variants now available for GPT-2, Pythia \citep{cunningham2023sparseautoencodershighlyinterpretable}, Gemma-2 \citep{lieberum-etal-2024-gemma}, LLaMA-3.1 \citep{he2024llamascopeextractingmillions}, and other LLMs.

\textbf{Spurious correlations detection in NLP.}
\citet{muhamed-etal-2025-decoding} trained SAEs that separated gender markers from professional content in career biographies, improving classifier accuracy by eliminating gender-based shortcuts. Bricken et al.\footnote{\url{https://transformer-circuits.pub/2024/features-as-classifiers/index.html}} trained a linear classifier on SAE features for bioweapon-related prompt classification and discovered an ``academic formatting'' feature that spuriously predicted harmlessness. Ablating it degraded classifier performance, and exploiting it allowed harmful prompts to be misclassified as harmless demonstrating SAE features' utility in identifying and eliminating spurious correlations.

\textbf{SAE utility in downstream NLP tasks.}
Results on SAE utility in downstream tasks are mixed. \citet{kantamneni2025sparseautoencodersusefulcase} report that SAE features provide no advantage over linear probes for classification in several tasks. In contrast, other works demonstrate benefits of SAE features: \citet{yang2025diversitydrivendataselectionlanguage} outperform heuristics in data selection, \citet{kuznetsov-etal-2025-feature} achieve strong results in AI-generated text detection, and \citet{oneill2024disentanglingdenseembeddingssparse} use SAE features for interpretable query steering in arXiv abstracts. Recently, \citet{wang2025persona} successfully applied SAE to explain an important ``emerging misalignment'' phenomenon in LLM fine-tuning.

\textbf{Computer Vision.}
SAEs enhance CV model interpretability \cite{kim2025textitreveliointerpretingleveragingsemantic} and image generation control \cite{stevens2025sparseautoencodersscientificallyrigorous}. In particular, \citet{journals-corr-abs-2503-09446} and \citet{cywinski2025saeuroninterpretableconceptunlearning} achieve selective concept unlearning in diffusion models without fine-tuning.

\textbf{Other domains.}
\citet{simon2024interplmdiscoveringinterpretablefeatures} and \citet{pnas.2506316122} apply SAEs to protein language models, generating interpretable features aligned with biological annotations. 

In the \textbf{audio domain},~\citet{sae_music} represents the only prior work, focusing on discovering concepts in music samples. We address this gap by training SAEs on more audio models and providing a substantially more comprehensive analysis of the learned features.

\subsection{Evaluating SAE Quality}

\citet{karvonen2025saebenchcomprehensivebenchmarksparse} propose a comprehensive evaluation framework for SAEs in NLP, spanning eight metrics across four capabilities: Concept Detection (Feature absorption \cite{chanin2025absorptionstudyingfeaturesplitting}, Sparse probing), Interpretability, Reconstruction Quality, and Feature Disentanglement (Unlearning, RAVEL \cite{huang-etal-2024-ravel}, Targeted Probe Perturbation, Spurious Correlations removal). Testing over 200 SAEs reveals that improvements on sparsity-fidelity trade-offs don't always translate to better performance on practical tasks like unlearning or bias mitigation. The framework highlights that assessing multiple aspects of SAE performance is improtant to determine which architecture is best suited for a given application, as SAEs show distinct task-specific strengths.

\subsection{Other Interpretability Methods in Audio Domain}
\citet{and} study neuron-level interpretability in acoustic models on environmental sound and music benchmarks. Their framework automatically generates natural-language explanations for acoustic neurons and leverages these explanations to cluster different sound types and to unlearn specific acoustic concepts. This work also sheds light on the kinds of features acoustic models rely on for sound classification and on how training strategies influence neuron interpretability.

Our work and~\citet{and} are complementary in terms of both the models considered and the types of features analyzed. First, rather than examining embedding spaces directly, we study SAE-disentangled representations of these embeddings. Second, we extend our analysis to speech-related models and datasets. As a result, we identify a number of new feature types and clusters that are not discussed in~\cite{and}.

\section{Background and Methodology} \label{methodology}

\subsection{SAE}
Recent studies have shown that in transformer models, particularly in LLMs, individual neurons are \textit{polysemantic} \citep{olah2020zoom}, meaning that these models learn more semantic features than there are available dimensions in a layer. This phenomenon is often explained by the \textit{superposition hypothesis}, which suggests that polysemantic features are linear combinations of monosemantic ones, allowing models to represent more features than they have dimensions \citep{elhage2022toy, hanni2024mathematical}. 

To recover these features, a Sparse Autoencoder (SAE) learns directions in the activation space such that each activation is a sparse linear combination of them. Given activations $\boldsymbol{x}$, the SAE encodes and reconstructs them as
\begin{equation}
\label{eq:sae}
\begin{aligned}
f(\boldsymbol{x}) &= \sigma(\boldsymbol{W}_{\text{enc}} \boldsymbol{x} + \boldsymbol{b}_{\text{enc}}), \\
\hat{\boldsymbol{x}}(f(\boldsymbol{x})) &= \boldsymbol{W}_{\text{dec}} f(\boldsymbol{x}) + \boldsymbol{b}_{\text{dec}},
\end{aligned}
\end{equation}
where $\sigma$ is an activation function such as ReLU, Top-$k$ or Batch-Top-$k$ \citep{cunningham2023sparse, gaoscaling, bussmann2024batchtopk}.

The SAE is typically optimized using the reconstruction loss
$
\mathcal{L}_{\text{rec}}(\boldsymbol{x}) = \Vert \boldsymbol{x} - \hat{\boldsymbol{x}}(f(\boldsymbol{x})) \Vert_{2}^{2},
$
with an additional sparsity loss $\mathcal{L}_{\text{sp}}(\boldsymbol{x})$ or auxiliary loss $\mathcal{L}_{\text{aux}}(\boldsymbol{x})$, depending on the architecture. 

In multiple previous works, listed in the preceding section, it was demonstrated that SAEs are indeed capable of extracting monosemantic features. A theoretical justification for this phenomenon was later provided in \cite{cui2025theoretical}. Most prior applications have focused on NLP and the visual domain, while audio SAEs remain comparatively less explored.

\subsection{Audio SAE Training Setup}

The SAE has a relatively straightforward architecture, with the main design choice being the form of the non-linearity $\sigma$, which influences the sparsity patterns learned by the model. Among the three explored options, Jump-ReLU, Top-$k$, and Batch-Top-$k$, we found the Batch-Top-$k$ variant to perform slightly better in terms of both reconstruction quality and sparsity. Therefore, it was selected for all subsequent experiments. All SAEs were trained using an $\mathcal{L}_{2}$ reconstruction loss without any auxiliary regularization.

\subsection{SAE Evaluation and Analysis}

Evaluating SAEs is challenging, as no single metric fully captures their quality. Following~\citet{karvonen2025saebenchcomprehensivebenchmarksparse}, we adopt a multi-faceted evaluation covering reconstruction, robustness, interpretability, and disentanglement. We report $\mathcal{L}_2$--$\mathcal{L}_0$ trade-offs for reconstruction quality, assess feature stability across random seeds, layers, and models, and apply a set of interpretability methods to analyze the latent space. %

We conduct both frame-level analysis and audio-level analysis using max-pooled feature aggregation.

\subsubsection{Feature Robustness via Distributional Semantics}\label{sec:distributional_semantics}
To measure feature stability, we introduce a distributional similarity metric inspired by distributional semantics~\citep{harris1954distributional, word2vec}. Two features $a_k, b_m$ are considered \textit{semantically similar} if their binary activation patterns over dataset $\mathcal{D}=\{d_i\}$ have high Intersection-over-Union (IoU):
\begin{equation*}\label{eq:iou_f2f}
\chi(a_k, b_m) =
\frac{|\{i \mid a_k(d_i)=1 \land b_m(d_i)=1\}|}
{|\{i \mid a_k(d_i)=1 \lor b_m(d_i)=1\}|}
\end{equation*}
A feature $a_k$ is \textit{covered} by set $B$ if $\exists b_m\in B: \chi(a_k, b_m) > \theta$, and \textit{coverage}
\begin{equation*}\label{eq:iou_sae}
c(A, B) = |\{k \mid \exists m: \chi(a_k, b_m) > \theta \}|
\end{equation*}
quantifies the proportion of transferable features. We apply this measure to compare sets of features of SAEs for different random seeds, layers, and model architectures (HuBERT, Whisper). The high coverage would indicate presence of robust and consistent features.
Also we define the \textit{duplicated features} of some SAE as those having high IoU with some other feature within the same SAE. The high amount of duplicates would indicate redundancy among SAE features.

\subsubsection{Domain Specialization}\label{sec:domain}

We analyze high-level organization by attributing features to one of three domains: \textit{speech}, \textit{music}, and \textit{environmental sounds}. A feature is considered \textit{domain-specialized} if its activation frequency is significantly higher within domain $D$ than in its complement $\overline{D}$. 
Activation frequency is assessed at two distinct levels for each domain $D$. At the frame level, it is defined as the proportion of frames within all samples in the domain that exhibit non-zero activation for a given feature. At the audio level, it is defined as the proportion of audios in the domain in which the feature is activated at least once.%

\subsubsection{Interpretability Analysis}\label{sec:classification}
To validate the general applicability of SAE features for audio-level analysis, we assess their performance across several \textbf{classification} tasks. These include gender identification, noise condition classification (clean vs. noisy speech), as well as accent and emotion recognition. All latents are ranked according to their influence on classification performance using the Fisher score \cite{fisher-measurments}. To examine the impact of the highest-ranked features, we employ two  techniques: \textit{top-k probing}, which masks all features except the top-k during activations' reconstruction, and \textit{unlearning}, which involves masking the top-k features \citep{karvonen2025saebenchcomprehensivebenchmarksparse}.

Additionally, fine-grained interpretability is assessed through the following methods:
\begin{itemize}
    \item Manual inspection of \textbf{top-activated samples} from \textit{reference set} of the most representative intervals. It consists of $1000$ samples from LibriTTS, Expresso, ESD, FSD50k and ESC50 datasets;
    \item \textbf{Semantic analysis}. To assess whether the SAE latent space captures speech semantics, we perform two experiments: vowel pronunciation classification and phoneme alignment. The first task addresses the complex phenomenon of how characters are pronounced, while the second task enables the mapping of some SAE features to specific linguistic units, namely phonemes;
    \item \textbf{Label-based search.} We identify features that demonstrate a strong individual correlation with specific label from the reference set (e.g., speech/non-speech, emotion, sound type) and lead to a high accuracy in separating data with target label from all others; 
    \item \textbf{Mel-interpretation} via averaging mel-spectrograms representation of activated frames to identify recurrent acoustic patterns;
    \item \textbf{Auto-interpretation  via captioning.} Active frames for a concrete feature from top-activated samples are concatenated to form 2-second chunks, which are processed by an audio captioning model. A large language model then aggregates all generated captions to produce a unified, high-level description of the feature;
\end{itemize}
Together, these analyses provide a comprehensive view of SAE robustness, interpretability, and functional relevance.

\subsection{Hallucination Reduction via SAE Steering}\label{sec:steering}
Steering refers to linear interventions in the latent space that guide model behavior toward desired outcomes. 
In the audio domain, we apply this approach to mitigate \textit{hallucinations} in the Whisper model~\citep{whisper_hallucinations}, where hallucination denotes false speech predictions in non-speech segments.
To reduce such errors, we add a directional vector during inference that biases activations away from hallucination-prone regions.

As a proxy metric for assessing the Whisper's tendency to hallucinate, we use an internal parameter no\_speech\_prob -- the probability that there is no speech in the audio. We aim to shift the distribution of this parameter toward 1 for non-speech samples and toward 0 for speech samples.

We search for SAE features that characterize hallucinations and modify them in the SAE space to minimize hallucinations. We define the Detection Rate (DR) metric as:
\[
\text{DR}(\mathcal{D}) = \frac{1}{|\mathcal{D}|} \sum_{i=1}^{|\mathcal{D}|} \mathbb{I}\{\text{no\_speech\_prob}_i < \tau\}.
\] 
This metric quantifies the fraction of samples classified as speech-containing. For non-speech datasets, DR estimates the False Positive Rate (FPR), capturing the proportion of hallucinations (non-speech samples incorrectly classified as speech). For speech datasets, DR estimates the True Positive Rate (TPR), the proportion of correctly identified speech samples.

To identify the top-$k$ SAE features related to hallucinations, we train a logistic regression classifier on SAE activations for non-speech dataset on target of predicting hallucinations ($\text{no\_speech\_prob} < \tau$). 
SAE feature $j$ is included in top-$k$ features if the absolute value of its regression coefficient $\beta_j$ is among largest $k$ such values.
Then the SAE steering vector is defined as:
\[
\vec{s}_{\text{SAE}}[j] = -\text{sign}(\beta_j), \quad j \in \text{top-}k,
\] and zero elsewhere. The negation operator ($-\text{sign}(\beta_j)$) ensures that the steering vector is opposing to features positively associated with hallucinations, thereby steering the model away from hallucination-prone representations.

During inference activations modified as 
\[
\text{act}_{\text{steered}} = \hat{\boldsymbol{x}}(f(\text{act}) + \alpha \vec{s}_{\text{SAE}}),
\]
where $\alpha$ controls steering intensity, \(f\) and \(\hat{\boldsymbol{x}}\) are SAE's Encoder and Decoder respectively, see Formula~\ref{eq:sae}.

\subsection{Correlation with EEG}\label{sec:eeg}
We closely follow the experimental setting from \cite{eeg-sem-diss} and use freely available data provided in this paper. Its authors tried to find correlation between semantic dissimilarity stimuli $s$ and EEG response $r$ by fitting a linear model representing response as a convolution of stimuli with some filter expressed in time domain by the so-called temporal-response function (TRF) $w$:
\begin{equation}
\label{eq:trf}
r(t) = \sum_{\tau}^{}{w(\tau)s(t-\tau) + \varepsilon(t)} ,
\end{equation}
where $\varepsilon$ is a residual signal whose energy is minimized during model fitting (i.e. finding TRF). If $w(\tau)$ differs from zero significantly for some $\tau$, then we can deduce that stimuli correlate with EEG response signal with time lag $\tau$. Note that, in contrast with our paper, \citet{eeg-sem-diss} used the term ``semantic'' in pure NLP sense corresponding to meanings of particular words -- they studied semantic dissimilarity stimuli $s$ measuring how unexpected a word is given its left context by computing word2vec embeddings \citep{word2vec}. In this paper we treat SAE features as stimuli without prior assumptions on what exactly they represent. We compute TRFs and do statistical tests to find out whether some of these features represent concepts aligned with EEG activity of people listening to speech.

\section{Experiments}

\label{sec:experiments}

\subsection{Training Details}

\textbf{Base models.}
We trained SAEs on the HuBERT and Whisper model families. For our downstream analysis, we focus on the HuBERT-base\footnote{\url{https://huggingface.co/facebook/hubert-base-ls960}} and Whisper-small\footnote{\url{https://huggingface.co/openai/whisper-small}} variants. Activations were extracted from every layer of the models' encoders.

\textbf{Data.}
The models were trained on a diverse corpus of approximately $2.8$k hours of audio, encompassing speech, music, and environmental sounds. A complete list of datasets included in the corpus is provided in Table~\ref{tab:audio_training_corpus} (Appendix~\ref{appendix:dataset}).

To improve robustness, we apply online augmentation by additively mixing noise and music at probabilities $p_{noise}=0.05$ and $p_{music}=0.025$, with signal-to-noise ratios uniformly sampled between 0 -- 20 dB. Activation vectors are stored in memory-mapped buffers and shuffled to enable efficient, randomized sampling.

\textbf{Hyperparameters.} All SAEs used a BatchTop-K architecture and were trained with the Adam optimizer for 200,000 steps using an $\mathcal{L}_{2}$ reconstruction loss. More details in Appendix~\ref{appendix:hyperparameters}.

\subsection{SAE Quality Evaluation}

\begin{figure}
    \centering
    \includegraphics[width=1.0\linewidth]{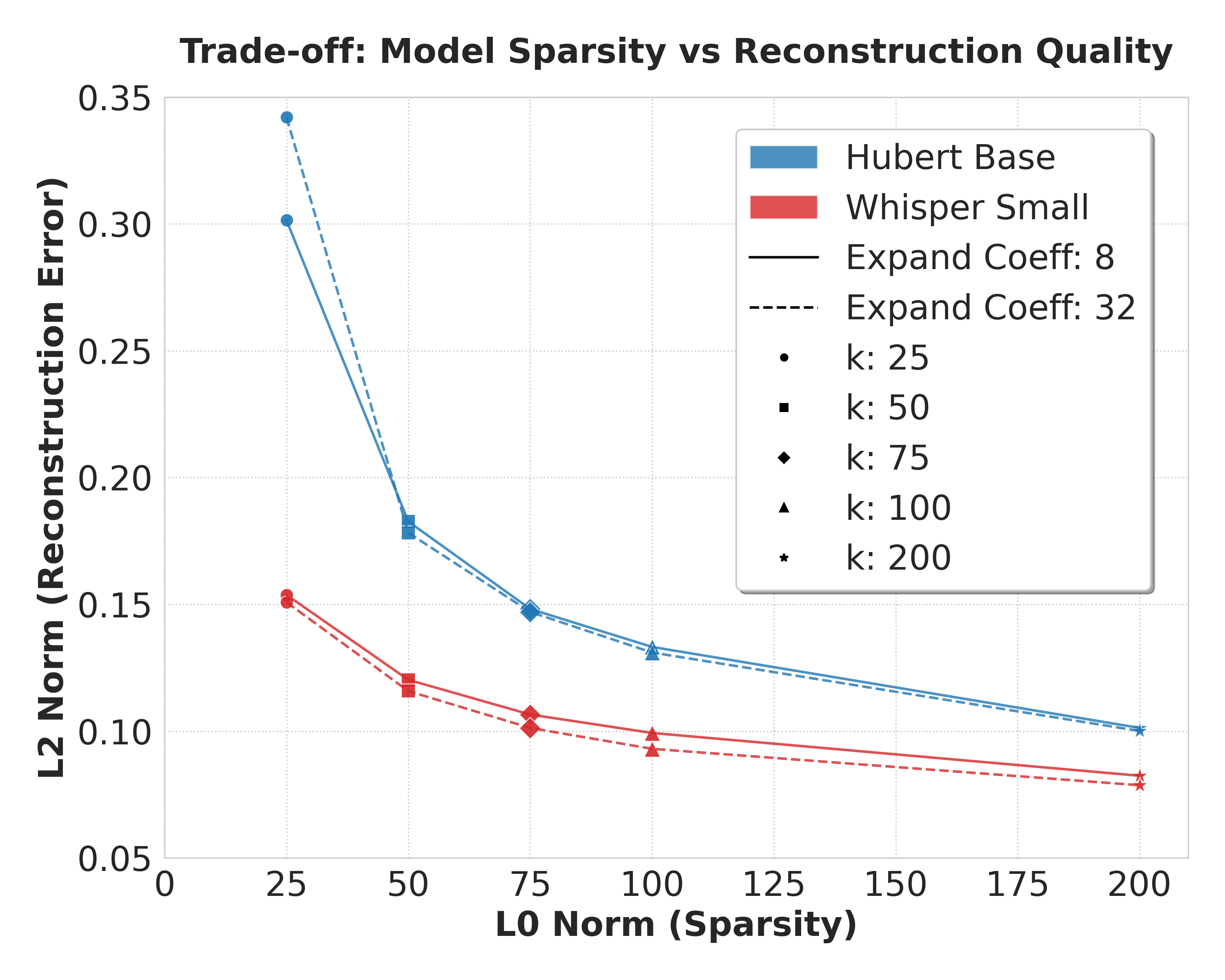}
    \caption{The trade-off between normalized reconstruction error (L2) and sparsity (L0) for models at layer 12.}
    \label{fig:trainnig_l0_l2}
\end{figure}

\textbf{Reconstruction–sparsity trade-off.}
Fig.~\ref{fig:trainnig_l0_l2} shows the trade-off between SAE sparsity and reconstruction quality under different training hyperparameters. Varying $k$ (the number of active neurons per token) yields Pareto frontiers at fixed expansion rates. Both expansion factors  demonstrate comparable results in term of reconstruction quality. We therefore adopt SAEs with $k=50$ and 8x expansion for all subsequent experiments. Additional experiments are provided in Appendix~\ref{appendix:sae-hyperparameters}.

\textbf{Feature Robustness Evaluation.} 
We evaluate feature stability using the distributional metrics described in Section~\ref{sec:distributional_semantics}.
Table~\ref{tab:duplicates_main} reports results for several layers of HuBERT and Whisper on LibriSpeech; additional layers and datasets are shown.

Amount of duplicates within a single checkpoint are low: below $5\%$ for the final layers of both models.
In contrast, feature coverage between checkpoints trained on the same layer but with different random seeds exceeds $50\%$, indicating strong stability.
However, features learned by HuBERT do not align with those from Whisper, likely due to differences in pre-training data and objectives: HuBERT  is trained via self-supervision on ASR data, whereas Whisper is trained through supervised learning for a broad range of audio tasks.
The inter-layer feature coverage is high only for latter layers.

For comparison, we measure the duplicate count and inter-layer coverage for GemmaScope SAEs~\cite{lieberum-etal-2024-gemma}, and show that our SAEs exhibit lower redundancy and comparable inter-layer coverage at last layers.
Overall, our SAEs extract a stable set of conceptual features from activations, although these concepts differ across models.
Additional results are provided in Appendix~\ref{appendix:feature_robustness}.

\begin{table}[h!]
\centering
\renewcommand{\arraystretch}{1.2} %
\scalebox{0.7}{
\begin{tabular}{lccccc}
\toprule
\textbf{Model} & \textbf{L1} & \textbf{L4} & \textbf{L7} & \textbf{L10} & \textbf{L12}\\
\midrule
Hub\_Hub${}^2$ & 419 & 1768 & 2930 & 4295 & 3164\\
Hub\_Hub${}_{L+n}$ & 92 & 258 & 453 & 1466 & \\
Wh\_Wh${}_{L+n}$ & 921 & 1296 & 634 & 2692 & \\
Hub\_Wh & 50 & 95 & 65 & 125 & 180\\
    \arrayrulecolor{black!30}\midrule
Hub (dup) & 27 & 47 & 102 & 352 & 219\\
Wh (dup) & 793 & 755 & 787 & 221 & 230\\
    \arrayrulecolor{black!30}\midrule
Gem\_Gem${}_{L+n}$ & ${}^\textbf{L2}$3365 & ${}^\textbf{L6}$3239 & ${}^\textbf{L12}$3210 & ${}^\textbf{L18}$2901 & \\
Gem (dup) & ${}^\textbf{L2}$4741 & ${}^\textbf{L6}$7825 & ${}^\textbf{L12}$3174 & ${}^\textbf{L18}$7115 & ${}^\textbf{L24}$2734\\
\arrayrulecolor{black}\bottomrule
\end{tabular}
}
\caption{SAE feature set coverage between models and layers. Suffix ${}^2$ is for SAE trained on the same activations but initialized with different random seeds; 
suffix ${}_{L+n}$ is for coverage between different layers of the model (each layer is compared with the layer from the next column, i.e. in L4 column we show the coverage between features from layer 4 with features from layer 7); (dup) means the amount of duplicates. For Gemma model we took SAEs from layers 2, 6, 12, 18 and 24. Note that the number of features for our SAEs in 6144, and for Gemma-Scope is 16384.}
\label{tab:duplicates_main}
\end{table}

\subsection{Domain Specialization Analysis}

\begin{figure}[h!]
\centering
\includegraphics[width=1.0\linewidth]{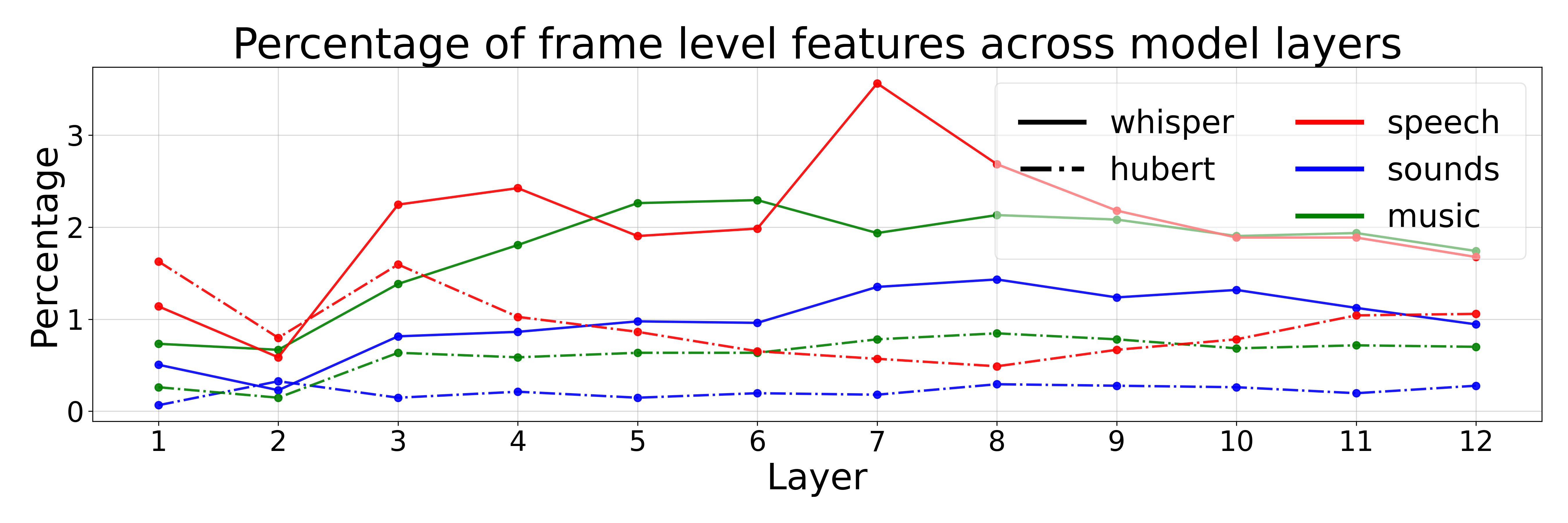}
\includegraphics[width=1.0\linewidth]{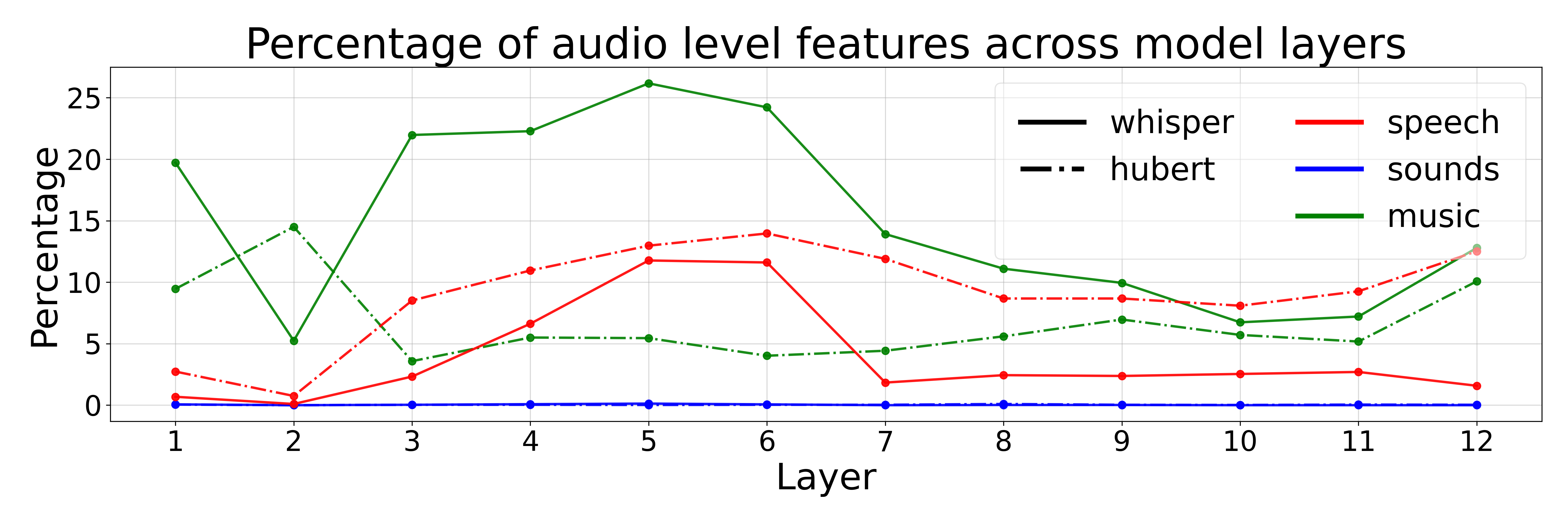}
\caption{Layer-wise feature specialization ratio by \textcolor{red}{speech}, \textcolor{blue}{sounds}, and \textcolor{ForestGreen}{music} domains for Whisper (solid line) and HuBERT (dashed) at frame (top) and audio (bottom) levels.}
\label{fig:clustering_4}
\end{figure}

Our domain analysis of feature type distributions across all 12 layers of both Whisper and HuBERT revealed distinct layer specialization trends (Fig.~\ref{fig:clustering_4}).

Whisper exhibits pronounced audio-level specialization for music, with music features comprising roughly 20--28\% of the detected audio-level activations and peaking around layer 5 (Fig.~\ref{fig:clustering_4}, bottom). Speech-related audio-level features are also most prominent in mid layers (peaking at roughly 13\%), but decrease sharply after layer 6, reaching only a few percent by layer 7. 

At the frame level, Whisper’s speech specialization peaks later: the speech-features proportion rises from about 2\% at layer 6 to approximately 3.5\% at layer 7 (Fig.~\ref{fig:clustering_4}, top), with the concurrent drop in audio-level speech activations. This divergence suggests that some layers encode speech information more locally (frame-level) even when global (audio-level) features are less frequently activated. Further analysis and additional experiments are described in Appendix~\ref{appendix:domain-layer-wise} and Appendix~\ref{appendix:domain-freq} respectively.

\subsection{Classification-based Analysis}

First, we analyze learned features via classification (Section~\ref{sec:classification}) on four audio tasks: gender (2 classes), clean vs.\ noisy speech (2), accents (5), and emotions (5). Logistic regression is chosen as the classifier, with a one-vs-all strategy applied for the $5$-class tasks. First, a classifier is trained on SAE features to rank their importance for a specific task. To assess the influence of these features on model activations, classifiers are subsequently trained on the reconstructed activations. Results on accents are shown in Fig.~\ref{fig:classification}; additional details are in Appendix~\ref{app:classification}.

The full SAE does not degrade performance and may even improve it (e.g., on emotion tasks). The top-$k$ curves rise sharply and saturate quickly, indicating that a small number of features ($k \approx 10$--$150$ out of $6144$ for binary tasks and $500$--$3000$ for more complex multi-class objective) captures most task-relevant information.

However, removing this information completely requires suppressing many more features ($\sim$2000), showing redundancy and distributed encoding: complex traits such as accent depend on multiple cues (phonemes, prosody, intonation). Compared to random selection, both top-$k$ and unlearning curves converge faster, confirming that learned features encode meaningful structure.

\begin{figure}[htb]
\includegraphics[width=1\linewidth]{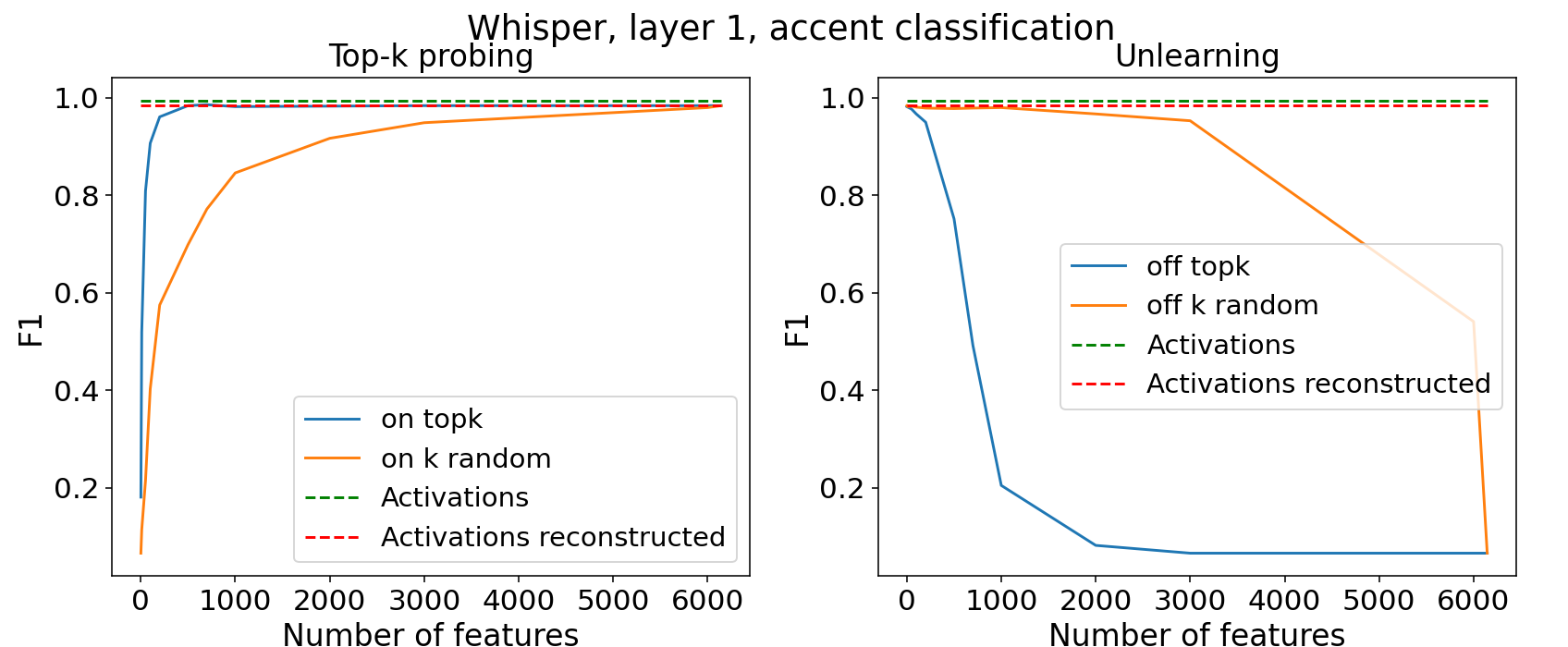}%
\caption{Top-$k$ probing and unlearning for accent classification. More results in Appendix~\ref{app:classification}.}
\label{fig:classification}
\end{figure}

\begin{figure}[htb]
\includegraphics[width=0.95\linewidth]{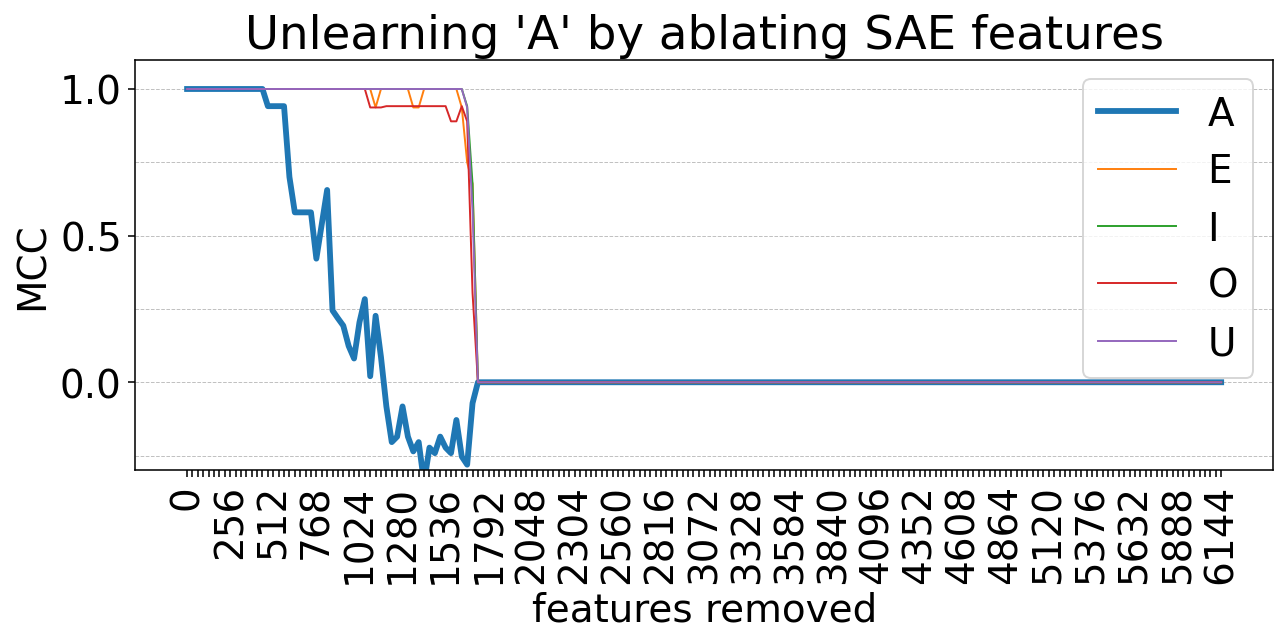}
\caption{Selective unlearning of letter 'A' via iterative feature removal. 
Feature indices on x-axis ordered by discriminative importance for target vowel.}
\label{fig:unlearning_vowels_main}
\end{figure}

\subsection{Semantic Analysis}
\textbf{Letter pronunciation classification.}
To demonstrate \emph{disentanglement}, we employ a vowel unlearning experiment: if disentangled features exist for different phonemes, we should be able to selectively unlearn one phoneme class while preserving recognition of others. Using AVLetters2~\cite{avletters2} -- recordings of five speakers pronouncing English letters -- we focus on vowels due to their simpler, atomic articulations.
Fig.~\ref{fig:unlearning_vowels_main} shows sequential removal of features most discriminative for letter ``A''. Deleting the first 1152 features ($\approx$19\%) almost entirely erases ``A'', while recognition of other vowels (MCC (Matthews Correlation Coefficient)~$>$~0.75) remains stable until over 27\% of features are removed -- showing some \textit{disentangled} features. See Appendix~\ref{sec:unlearning_app} for more results.

Overall, erasing speech concepts requires far more features (hundreds to thousands) than text-based SAEs, where abstract notions like gender or occupation can be removed with only tens~\cite{farrell2024applyingsparseautoencodersunlearn}. This reflects both (a) higher redundancy and (b) the inherently distributed nature of phonetic and paralinguistic information.

\textbf{Phonemes encoding.} 
To additionally verify which SAE features encode semantic information we work with text-audio alignments\footnote{Pre-trained aligner from \url{https://montreal-forced-aligner.readthedocs.io/}} extracted from $1000$ audio samples from LibriTTS. A phoneme label is assigned to a latent feature when this phoneme appears in a majority ($>50\%$) of its aligned, activated frames.

The final 12th layers of the Whisper and HuBERT models are analyzed on a test set consisting of $1000$ audio samples from LibriTTS. A frame is considered correctly classified if at least one activated SAE feature (determined by a threshold) has a label matching the ground-truth phoneme. The final accuracy scores achieved for these models are $0.92$ and $0.89$, respectively.

\subsection{Frame-level Features Interpretation}

\begin{figure}[htb]
\subfloat[Whisper, layer 6, "laughter"]{\includegraphics[width=0.9\linewidth]{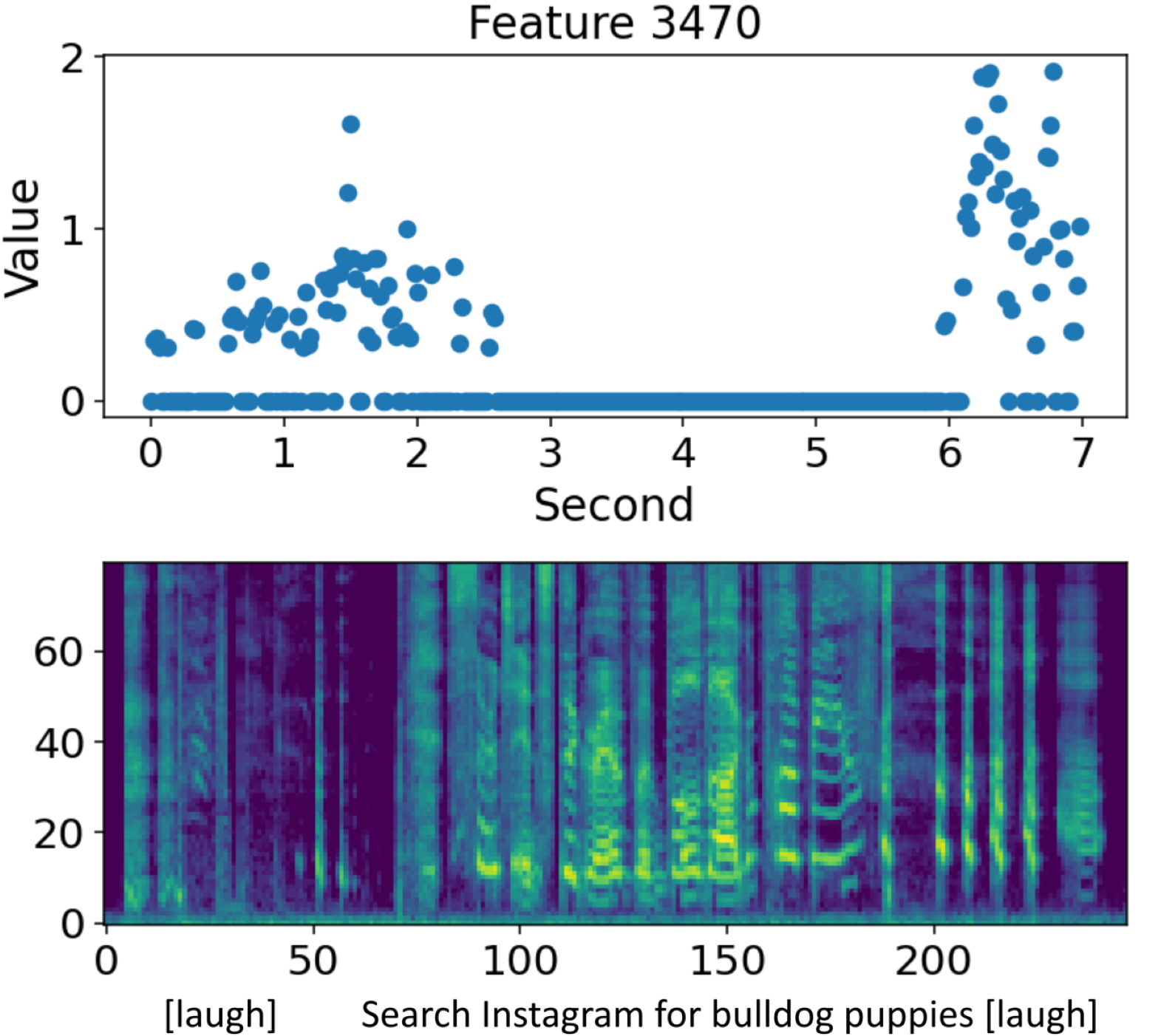}}\hfill
\subfloat[Whisper, layer 6, "sneezing"]{\includegraphics[width=0.9\linewidth]{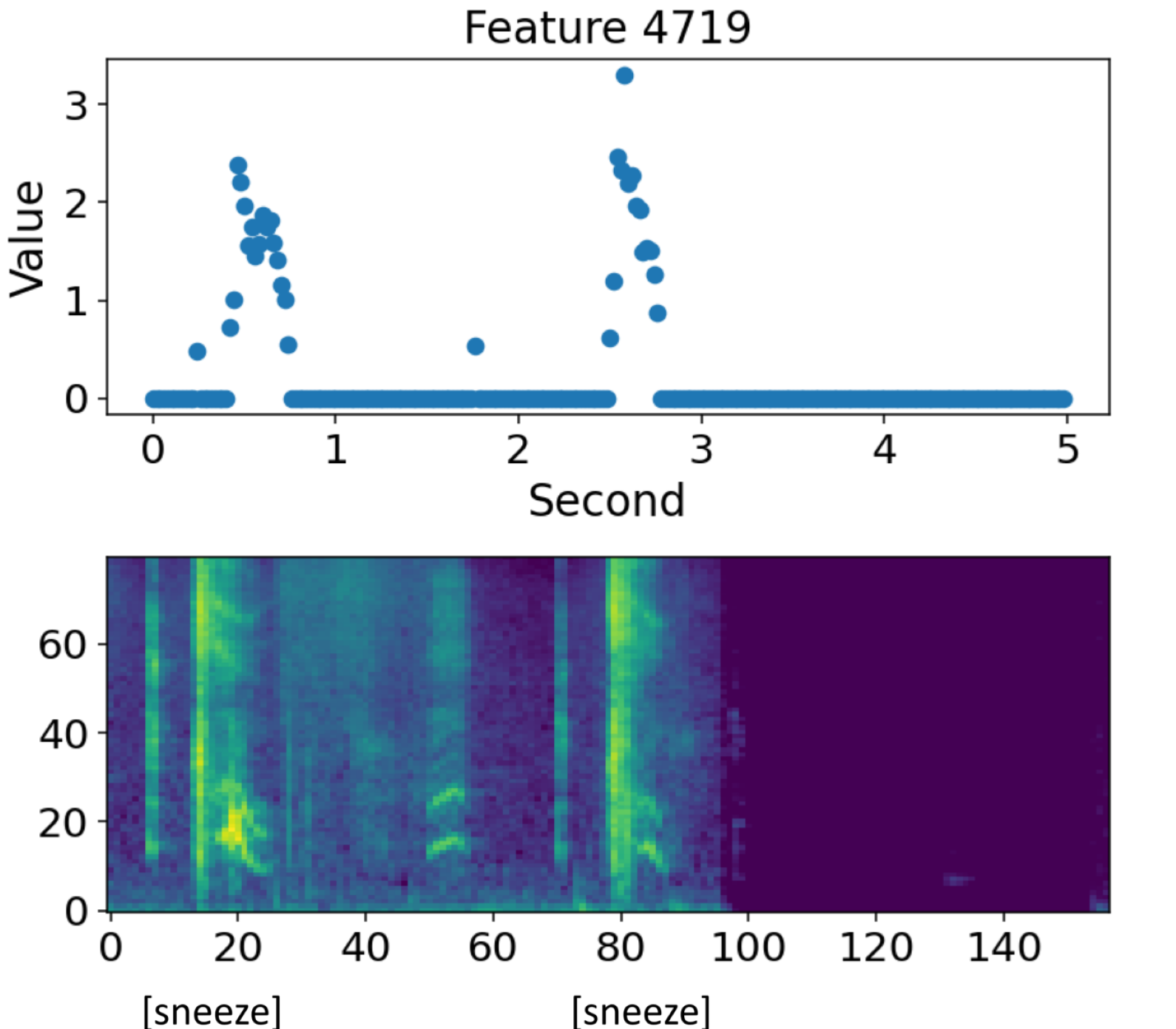}}
\caption{Features identified via the label-based classification experiment. 
Each sub-figure displays a feature's values above the corresponding audio mel-spectrogram.}
\label{fig:clf_by_label_main}
\end{figure}

By \textbf{label-based classification}, we aim to identify individual SAE features that independently encode concepts present in the dataset and separate label-specific samples from all others by some threshold. The label \textit{whisper} was successfully detected by HuBERT SAEs on layers~1-5, with feature 6106 (layer~4) achieving an F1-score of~0.6. Similarly, \textit{laughter} is captured by both models -- on layers~1–6 in HuBERT and layers~1,~6,~and~9 in Whisper. The phenomena \textit{sigh} and \textit{sneezing} are also represented, with relevant features appearing across layers~1-7 in both models. In contrast, \textit{animal sounds} and \textit{breathing} were not identified reliably, suggesting that these concepts are encoded in a distributed manner.
Frame-level examples are shown in Fig.~\ref{fig:clf_by_label_main} and Appendix~\ref{appendix:clf_by_label}, with an extended version available on the demo page.

Using the \textbf{mel-interpretation methodology} we further analyze salient features identified through domain specialization. Specifically, we extract $1$-second log-mel spectrogram windows centered at activated frame from the audio samples with highest activation magnitude. The element-wise average of these windows reveals the core acoustic pattern. Thus we found that HuBERT layer~11 features~3249 and~3081 exhibit specialized speech-boundary detection. Their temporal alignment with speech segments is illustrated in Appendix~\ref{appendix:mel-int}.

\begin{figure}
    \centering
    \includegraphics[width=1.0\linewidth]{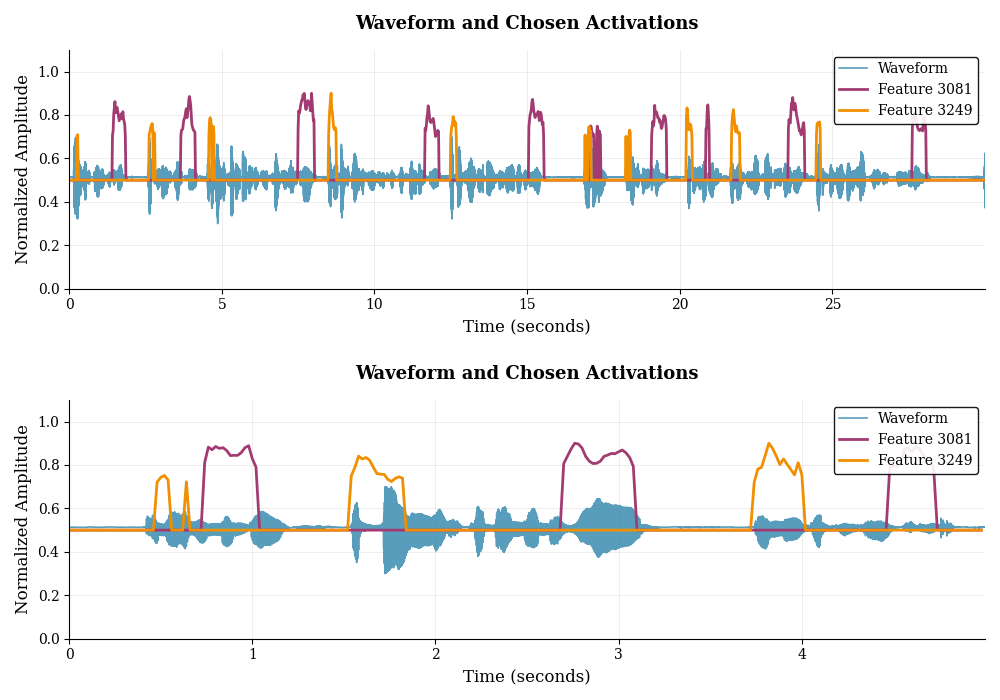}
    \caption{Activation of features responsible for the beginning (3249) and the end (3081) of speech, aligned with corresponding waveform. HuBERT, layer 11.}
    \label{fig:MM1_main}
\end{figure}

For the \textbf{automatic interpretation} of features, we first extract all time frames in the audio where a feature’s value exceeds the threshold of 0.1, marking them as activated frames. These frames are concatenated and divided into 2-second segments, each processed by captioning model \cite{xu2024efficientaudiocaptioningencoderlevel}. The resulting captions are then unified and aggregated using model GPT-4o mini \cite{openai2024gpt4ocard} to produce a final interpretative label for the feature. The entire pipeline is visualized in Appendix \ref{appendix:auto-interpretation}, Fig. \ref{fig:autointerpretation_pipeline}.

Auto interpretation helped identify unique features not present in the dataset annotations, such as \textbf{"ringing alarms"}, \textbf{"high-pitched beeping"}, \textbf{"birds chirping"}, and \textbf{"guitar playing"}. However, due to the limitations of the caption model, which was trained mainly on music and sound data, specific speech features, particularly, phonetic details were missed. As an example, the feature, which activates on the \textbf{ "ba"} sound, resulted in a more general interpretation \textbf{"a man is speaking."} More detailed observation of the results can be seen in the Appendix \ref{appendix:auto-interpretation}.

\begin{figure*}[htb]
\setkeys{Gin}{width=0.32\linewidth}
\subfloat[only negative correlation]{\includegraphics{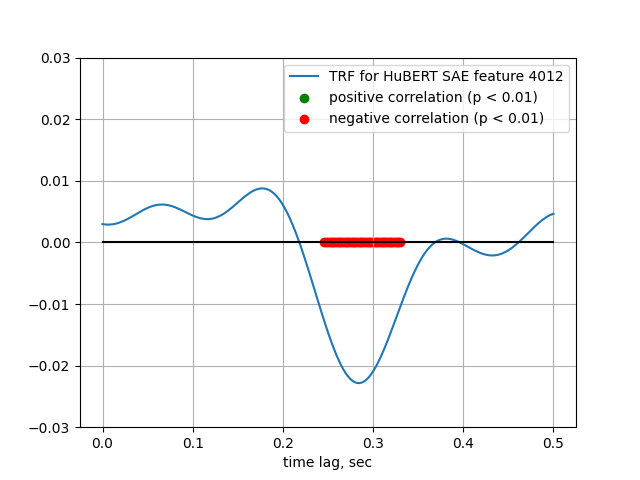}}\hfill
\subfloat[only positive correlation]{\includegraphics{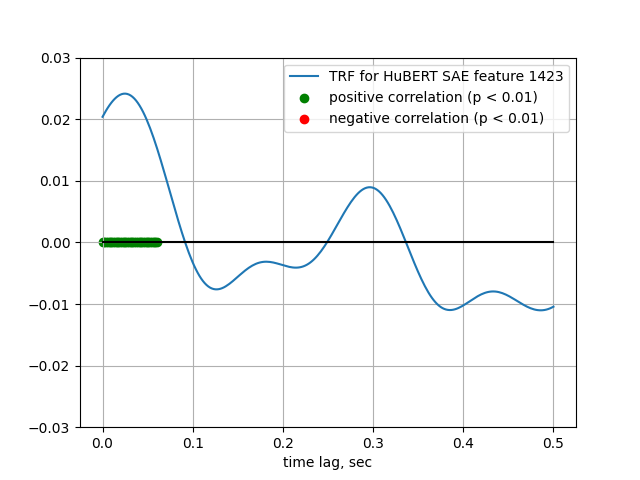}}\hfill
\subfloat[both positive and negative correlation]{\includegraphics{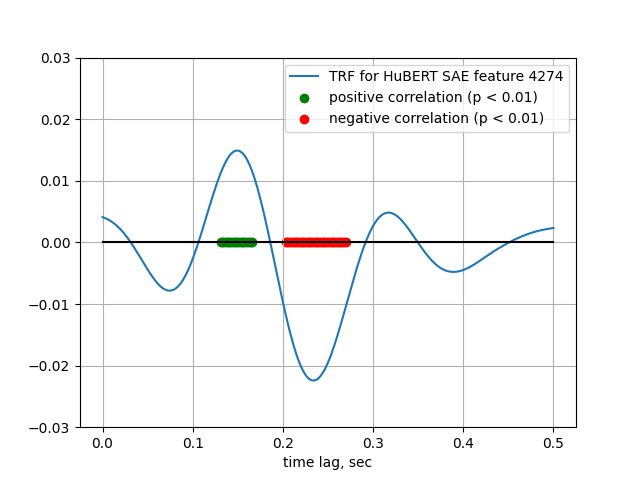}}

\caption{Temporal response functions for different SAE features for HuBERT model.}
\label{fig:trf}
\end{figure*}

\subsection{Steering for Hallucination Reduction}
 
To evaluate the effectiveness of SAE in reducing hallucinations (Sec.~\ref{sec:steering}), we measure the False Positive Rate (FPR) with \(\tau = 0.5\) on non-speech datasets (FSD50k, Musan and WHAM) and control Word Error Rate (WER) on LibriSpeech test-clean to ensure that speech recognition performance remains unaffected.

Optimal performance is achieved by steering the top-$100$ SAE features. We report SAE steering with different strengths ($\alpha = 1$ and $\alpha = 3$) and compare it with the baseline $S$-vector steering approach (described in Appendix~\ref{appendix:steering}) with $\alpha = 3$. Both methods substantially reduce hallucinations. SAE-based steering with moderate $\alpha$ provides a good balance between FPR and WER, achieving a threefold reduction in FPR by 70\% (0.37~$\rightarrow$~0.11) in average across datasets with only a negligible increase in WER (5.1\%~$\rightarrow$~5.5\%). However, aggressive steering with large $\alpha$ severely impairs speech comprehension, revealing a clear trade-off between efficacy and safety. Table~\ref{tab:her_results} summarizes the results. More details in Appendix~\ref{appendix:steering}.

\begin{table}[h]
\centering
\small
\setlength{\tabcolsep}{3pt}
\begin{tabular}{lccccc}
\toprule
\textbf{Dataset} & \textbf{No SAE} & \textbf{No Steer.} & \textbf{S-Vec.} & \textbf{SAE} & \textbf{SAE} \\
& & & ($\alpha$=3) & ($\alpha$=1) & ($\alpha$=3) \\
\midrule
Musan    &-     & 0.33 & 0.20 & 0.12 & 0.00 \\
FSD50K   &-     & 0.26 & 0.15 & 0.09 & 0.01 \\
WHAM    &-     & 0.51 & 0.32 & 0.14 & 0.03 \\
\arrayrulecolor{black!30}\midrule
LS (WER)      & 5.1 & 5.2  &  5.3  &  5.5  &  98.4 \\
\arrayrulecolor{black}\bottomrule
\end{tabular}
\caption{FPR (\(\tau=0.5\)) for steering configurations. No SAE: Whisper inference without modification. No Steer.: Whisper with injected SAE on the last layer. S-Vec.: S-Vector, calculated on Musan. Optimal and Best SAE: SAE S-Vector, top-100 features from FSD50k dataset with $\alpha$ equals to 1 and 3 respectively. LibriSpeech (LS) line represents WER. Lower is better.}
\label{tab:her_results}
\end{table}

\subsection{Correlation with EEG}

Here we present EEG signal correlation experiments (Sec.~\ref{sec:eeg}). Open-source EEG data was collected from $19$ participants listening to audiobooks. We studied both HuBERT and Whisper SAE features and found that some of them have statistically significant correlation with midline parietal electrode Pz (chosen as one of the most indicative in \cite{eeg-sem-diss}) at certain time lags $\tau$ as verified by one-tailed t-tests ($p$-value less than $0.05$) with Holm-Bonferroni correction for multiple comparisons. As illustrated in Fig.~\ref{fig:trf}, this correlation can be both positive and negative and occur at quite different time lags between $0$ and $500$ms. We further analyzed these features and found that many of them activate mostly on particular vowels (like IPA phonemes ``\textipa{O}'' or ``\textipa{A}''), but not on all of such vowels. This analysis, more details of which can be found in Appendix~\ref{appendix:eeg}, shows that at least some of features learned by SAE are generic features well-aligned with brain activity. Interpreting them and analyzing their TRF patterns, as well as studying EEG channels other than Pz and applying more sophisticated non-linear models rather than (\ref{eq:trf}) can be a future research direction.
\section{Conclusion}

This work presents a comprehensive investigation into the application of SAEs for interpreting the HuBERT and Whisper audio models. We introduce a novel metric for cross-layer and cross-model SAE evaluation and a bunch of methods for resulting latent analysis. Our proposed metric confirms that the resulting SAE features are robust and encode meaningful information. We found high-level features related to broad categories like speech and music, as well as more fine-grained features corresponding to semantic content (phonemes), paralinguistic phenomena (e.g., laughter, sigh, sneezing), and acoustic properties (discovered via auto-interpretation). Furthermore, steering on Whisper model SAE features reduced the false positive rate on hallucinations by $70$\%. An additional experiment demonstrated correlation between EEG signals and specific SAE features.

\section*{Limitations}

\begin{itemize}
\item Our downstream evaluation covers a limited set of classification tasks and applications. Future work should explore SAE features across broader audio processing tasks including speaker verification, speech enhancement, and audio generation.
\item Detailed analysis focuses on base/small model variants. Larger architectures and additional models (Wav2Vec 2.0, WavLM) were not comprehensively studied due to computational constraints.
\item The auto-interpretation method inherits limitations from its underlying audio captioning model, which was trained primarily on music and sound data. As a result, it tends to generate generic captions for speech-related features, losing fine-grained phoneme-level information.
\item EEG correlation analysis is limited to a single electrode (Pz) with linear temporal response models. More comprehensive brain imaging and non-linear modeling could reveal additional relationships.
\end{itemize}

\bibliography{custom}

\clearpage

\appendix

\part{Appendix} %
\parttoc %

\section{Extended SAE training details}\label{appendix:training}
This section provides additional details on model selection, dataset construction, and architectural choices referenced in the main text. During architecture search and hyperparameter sweeps, we evaluated trade-offs among the following metrics: reconstruction quality ($L_2$ loss, lower is better), sparsity ($L_0$ loss, lower is better), and the proportion of features activated at least once during $N$ steps (“alive”; higher is better).

\textbf{Base model Selection.} Our study included SAE training on four model variants: HuBERT-base, HuBERT-large, Whisper-small, and Whisper-large-v3-turbo. While SAEs were trained for all variants to ensure a comprehensive foundation, the downstream analysis in the main paper is conducted on HuBERT-base and Whisper-small for a focused comparison. We also initially considered the EnCodec model. However, a SAE trained on the final layer of its encoder yielded a number of active ("alive") features comparable to the source embedding dimension, suggesting it was not learning a sufficiently sparse representation for our purposes, and it was excluded from further analysis.

The HuBERT, Whisper, and Montreal Forced Aligner (MFA) software packages are distributed under the MIT license.

\textbf{Dataset.}\label{appendix:dataset}
Our training corpus is designed to comprehensively represent diverse acoustic environments. The dataset is composed of multiple publicly available sources, with each assigned a sampling weight to control its prevalence during activation extraction and training. All datasets used are described in the Table~\ref{tab:audio_training_corpus}.

The high weights for datasets like MUSAN, FSD50K, and Nonspeech7k were chosen to strongly bias the SAEs towards learning features for non-speech audio, music, and environmental sounds, complementing the speech-dominant datasets. After analyzing the audio types in the datasets, we decided to divide them into three types: \textit{speech} (LibriSpeech \citep{librispeech}, LibriHeavy \citep{libriheavy}, ESD \citep{esd}, Expresso \citep{expresso}, CREMA \citep{crema}, MELD \citep{poria2019meld}, IEMOCAP \citep{iemocap}), \textit{music} (MTG-Jamendo \citep{mtg}), and \textit{sounds} included noise, sound events and non-speech sounds (MUSAN \citep{musan}, WHAM \citep{wham}, FSD50K \citep{fsd50k}, Nonspeech7k \citep{nonspeech7k}, DEMAND \citep{demand}, VGGSound \citep{vggsound}, VocalSound \citep{vocalsound}, ESC-50 \citep{esc50k}). Taking into account the weights, each batch averaged approximately 40\% activations from speech data, 45\% from music, and 15\% from sounds. With a batch size of 2500, a step count of 200,000, and a model frame rate of 50 frames per second, the total data amount for each SAE was just under 2800 hours.

\begin{table}[t]
\centering
\small
\caption{Composition of the Training Corpus}
\label{tab:audio_training_corpus}
\begin{tabular}{l l c c}
\toprule
Dataset & Description & Weight & Hours \\
\midrule
LibriSpeech  & Read Speech & 1.0 & 960 \\
LibriHeavy  & Noisy Speech & 0.1 & 50000 \\
MUSAN & Music/Speech/Noise & 5 & 112 \\
WHAM!  & Domestic Noise & 2 & 58 \\
FSD50K  & Sound Events & 5 & 108 \\
Nonspeech7k  & Non-Speech Sounds & 5 & 125 \\
ESD  & Emotional Speech & 6 & 11 \\
Expresso  & Audiobooks & 5 & 41 \\
CREMA-D  & Emotional Speech & 4 & 5 \\
DEMAND & Env. Noise & 3 & 6 \\
MELD & Emotional Dialogues & 3 & 12 \\
VGGSound  & Audio Events & 1 & 550 \\
VocalSound  & Vocalizations & 3 & 18 \\
MTG-Jamendo  & Music & 2 & 3777 \\
IEMOCAP  & Emotional Speech & Test only & 12 \\
ESC-50  & Sound Classification & Test only & 3 \\
\bottomrule
\end{tabular}
\smallskip
\end{table}

\textbf{Batching.} 
Our training employs a dynamic batching strategy: audio samples are drawn from our weighted dataset mixture using a probability-proportional-to-size sampling scheme, where each dataset's selection probability is determined by its configured weight multiplied by the number of samples in each dataset. We randomly select a dataset determined by its weight and size, randomly select audio samples from a chosen dataset, then fulfill the buffer by model’s activations on this audio sample and sample fixed-size batches by randomly selecting unread indices from the buffer, ensuring diverse training examples. This approach helps prevent overfitting to the sequence order of any single audio file.

\textbf{Infrastructure.}
Training was conducted on a multi-GPU server. The inference of the base audio models (Whisper and HuBERT) was distributed across several GPUs to parallelize the computationally intensive forward passes required for activation generation. The subsequent SAE training was also sharded across available devices. SAEs were trained in parallel across all model layers within a single run using multi-threaded execution on identical data. We implemented an asynchronous data loading and buffering pipeline which pre-computes and stores activations in a memory buffer (holding 100 batches of 2500 activation vectors each), which is then sampled randomly to feed the SAE trainers. All experiments were performed on 8 NVIDIA V$100$ GPUs.

\textbf{Training hyperparameters.}\label{appendix:hyperparameters}
We employ the Adam optimizer \( (\beta_1, \beta_2) = (0.9, 0.999) \) with a fixed learning rate of \( 2 \times 10^{-4} \) and a linear warmup of the sparsity coefficient over the first 10,000 steps. Training proceeds for 200,000 update steps, with a linear learning rate decay schedule initialized for the terminal 20\%\ of training, progressively reducing the rate from its initial value to zero. Each training batch comprises 2,500 activation vectors, corresponding to 50 seconds of audio.

\textbf{SAE architecture Selection.} We tested three different SAE variants, depending of the form of non-linearity function: the Jump-ReLU, Top-K, and Batch-Top-K. Our preliminary analysis indicated that the Batch-Top-K SAEs demonstrated a slightly better performance in terms of reconstruction quality and sparsity control. Consequently, the Batch-Top-K was selected as the primary architecture for our investigation. All SAEs were optimized using an $\mathcal{L}_{2}$ reconstruction objective, without any auxiliary regularization.

\textbf{SAE-specific hyperparameters.}\label{appendix:sae-hyperparameters}
Key SAE hyperparameters include the \emph{expansion factor} (ratio of SAE to source embedding dimensions) and the \emph{sparsity level} $k$ (number of active features per training sample). We perform a structured sweep over expansion factors (8x, 32x) and sparsity levels $k \in {25, 50, 75, 100, 200}$ across all layers. Input activations are normalized to unit norm for stable training and metric comparability across layers.

\begin{figure}
    \centering
    \includegraphics[width=1.0\linewidth]{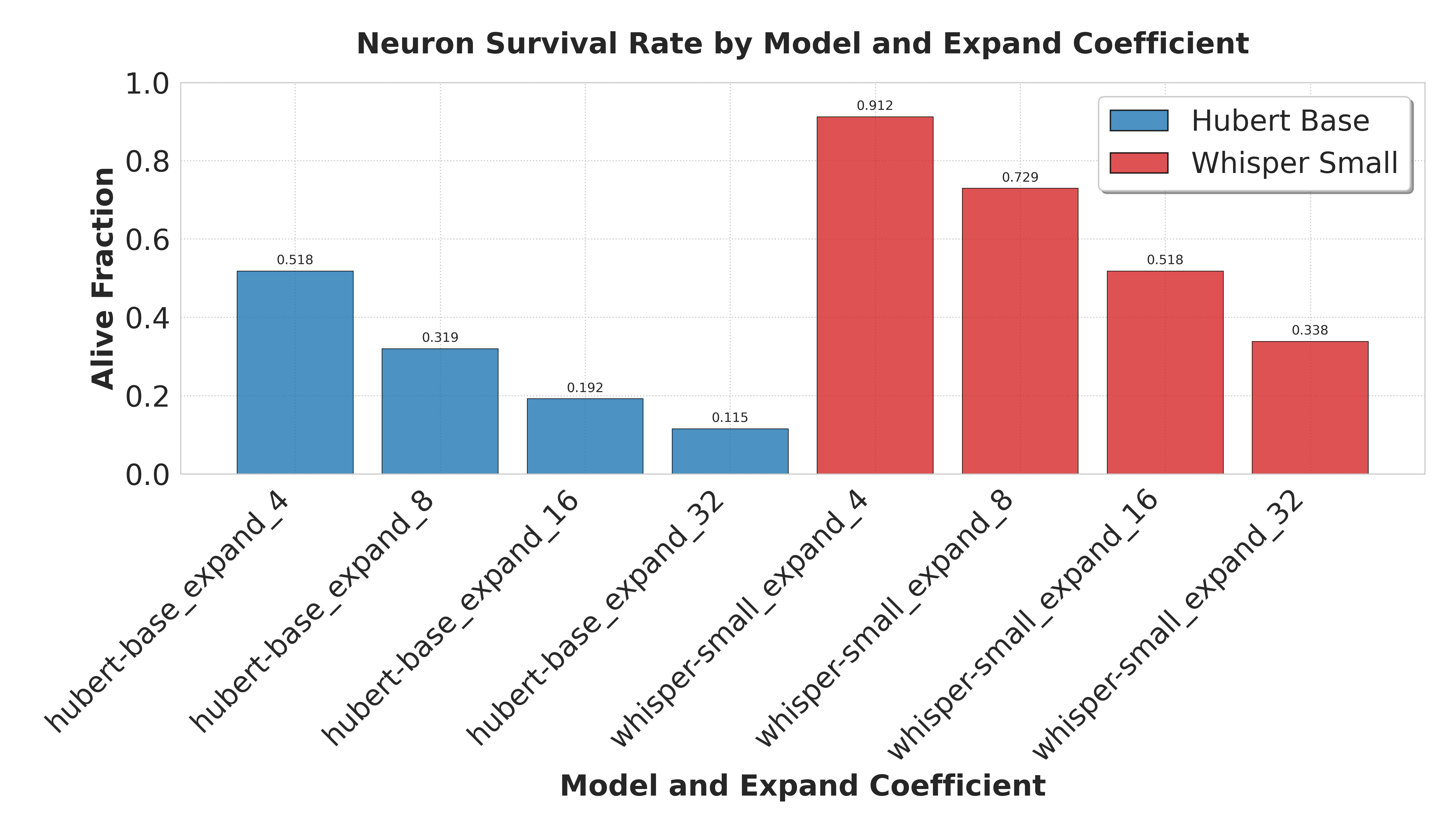}
    \caption{Influence of the expansion rate to the number of "alive" features}
    \label{fig:expansion_vs_alive}
\end{figure}

\begin{figure}
    \centering
    \includegraphics[width=1.0\linewidth]{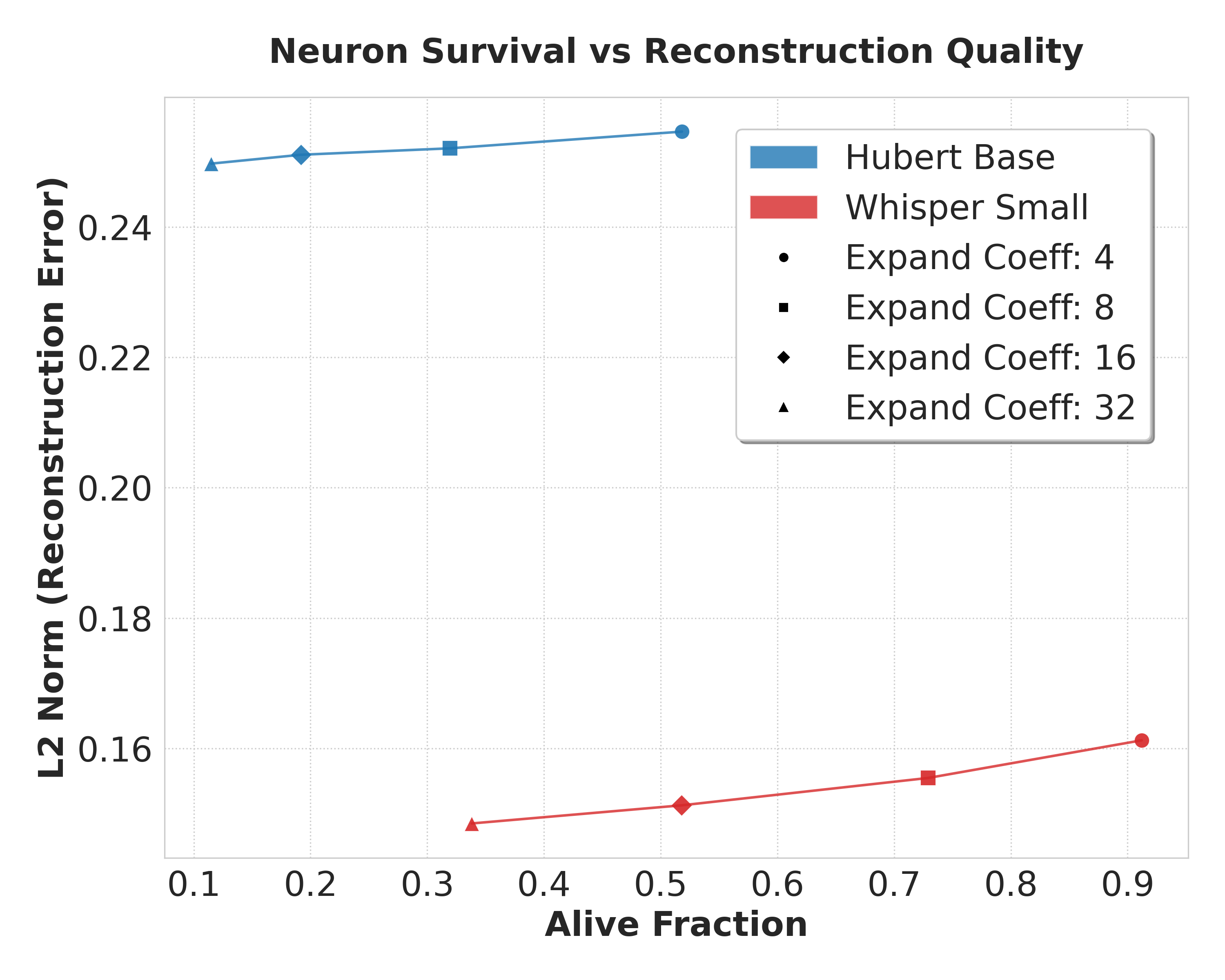}
    \caption{Connection between reconstruction quality and neurons survival rate}
    \label{fig:reconstruction_vs_alive}
\end{figure}

\begin{figure}
    \centering
    \includegraphics[width=1.0\linewidth]{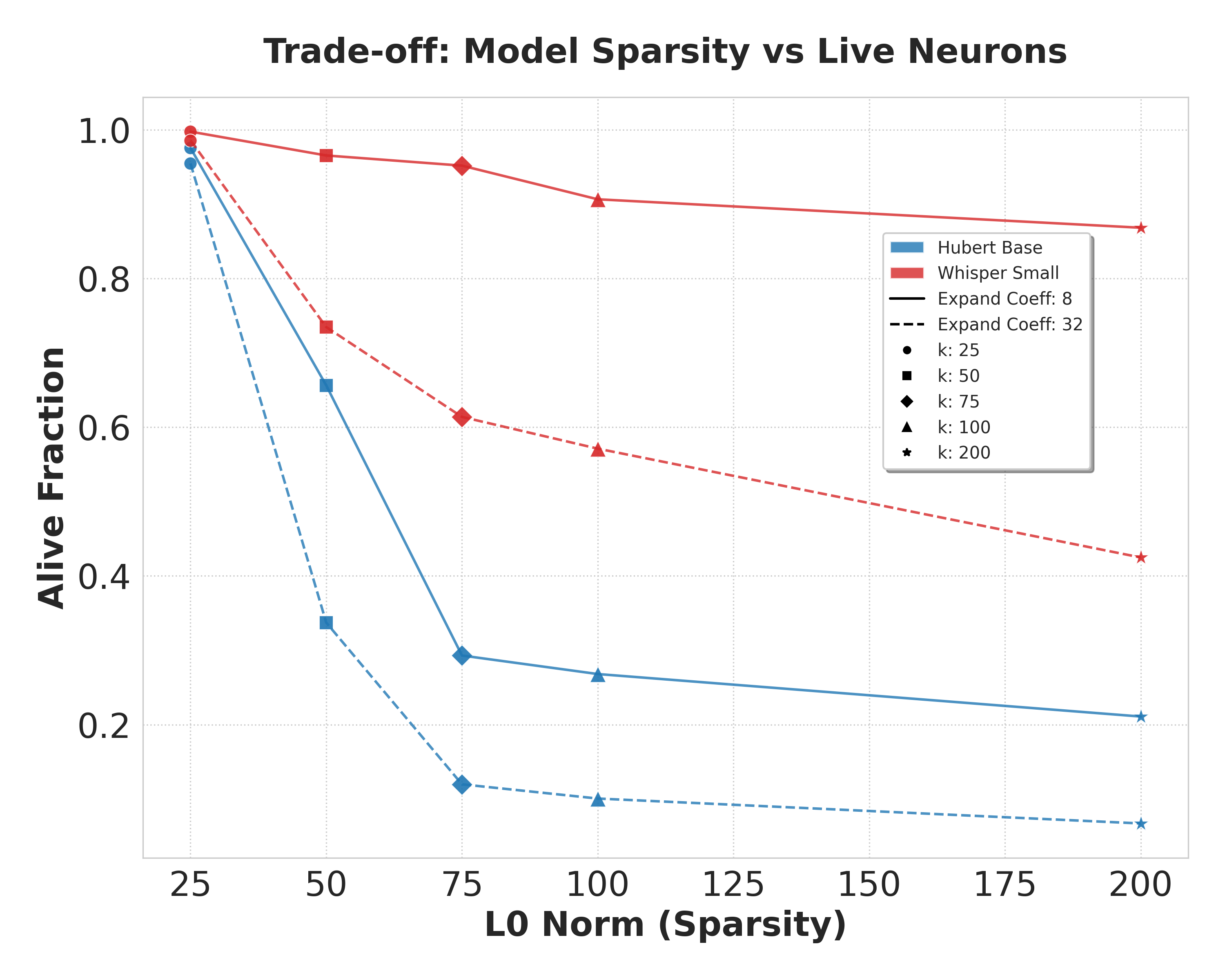}
    \caption{Connection between sparsity and neurons survival rate}
    \label{fig:sparsity_vs_alive}
\end{figure}

Figures~\ref{fig:expansion_vs_alive}–\ref{fig:sparsity_vs_alive} summarize the results. We observe a clear trade-off between sparsity and reconstruction quality (Fig.~\ref{fig:trainnig_l0_l2}), with little difference between 8x and 32x expansions, with 8x performing even better for HuBERT under high sparsity (low $L_0$). While neuron survival decreases with higher expansion, the total number of active neurons still grows, staying at least twice the base model’s size (Fig.~\ref{fig:expansion_vs_alive}). Notably, this ratio shows only weak correlation with $L_2$ quality, indicating that smaller values correspond to a suboptimal size–quality balance. Finally, Figs.~\ref{fig:sparsity_vs_alive} and~\ref{fig:trainnig_l0_l2} demonstrate that $k=50$ with 8x expansion provides the best compromise between reconstruction fidelity, sparsity, and compression efficiency, minimizing both memory cost and inactive features.

\label{appendix:training_details}

\section{Feature robustness}\label{appendix:feature_robustness}
For this set of experiments, we use datasets LibriSpeech, FSD and MTG for analysis of feature similarity (formulas \ref{eq:iou_f2f} and \ref{eq:iou_sae}).
From these datasets we sample $500$ audios with total of $749450, 74700 \text{ and } 86220$ frames, respectively.
We use BatchTopK with $k=50$ on inference, meaning that for a single input audio of $n$ frames, top $50\times n$ of all $6144 \times n$ features will be denoted as active.
For feature coverage we take Intersection-over-Union threshold $\theta=0.5$, meaning that features with $\chi(a, b)>0.5$ (see formula \ref{eq:iou_f2f}) are considered similar. 

See results in tables \ref{tab:coverage} and \ref{tab:coverage_self}.

\begin{table}[h!]
\centering
\renewcommand{\arraystretch}{1.2} %
\scalebox{0.7}{
\begin{tabular}{lcccccc}
\toprule
\textbf{Model} & \textbf{DS} & \textbf{L1} & \textbf{L4} & \textbf{L7} & \textbf{L10} & \textbf{L12}\\
\midrule
\multirow{3}{*}{Hub\_Hub2} & LS & 419 & 1768 & 2930 & 4295 & 3164\\
    & FSD & 447 & 1395 & 2292 & 2725 & 1826\\
    & MTG & 363 & 1517 & 2537 & 3390 & 2267\\
    \arrayrulecolor{black!30}\midrule
\multirow{3}{*}{Hub\_Hub100K} & LS & 1532 & 2495 & 3637 & 5049 & 4340\\
    & FSD & 1544 & 2220 & 2956 & 3314 & 2919\\
    & MTG & 1672 & 2467 & 3464 & 4396 & 4009\\
    \arrayrulecolor{black!30}\midrule
\multirow{3}{*}{Hub\_Hub${}_{L+n}$} & LS & 92 & 258 & 453 & 1466 & \\
    & FSD & 131 & 171 & 146 & 263 & \\
    & MTG & 97 & 184 & 122 & 242 & \\
    \arrayrulecolor{black!30}\midrule
\multirow{3}{*}{Wh\_Wh100K} & LS & 1277 & 1987 & 1610 & 3746 & 4650\\
    & FSD & 1354 & 1931 & 1550 & 2816 & 2606\\
    & MTG & 1888 & 2598 & 1923 & 3653 & 3783\\
    \arrayrulecolor{black!30}\midrule
\multirow{3}{*}{Wh\_Wh${}_{L+n}$} & LS & 921 & 1296 & 634 & 2692 & \\
    & FSD & 960 & 1267 & 629 & 1032 & \\
    & MTG & 1420 & 1748 & 583 & 1849 & \\
    \arrayrulecolor{black!30}\midrule
\multirow{3}{*}{Hub\_Wh} & LS & 50 & 95 & 65 & 125 & 180\\
    & FSD & 25 & 25 & 12 & 11 & 13\\
    & MTG & 44 & 29 & 15 & 10 & 6\\
\arrayrulecolor{black}\bottomrule
\end{tabular}
}
\caption{SAE feature set coverage between models and layers. 
LS (LibriSpeech), FSD (FSD50K) and MTG mean datasets for coverage score calculation.
Suffix ${}^2$ is for SAE trained on the same activations but initialized with different random seeds; 
suffix 100K is for early stage of SAE training (100K iterations);
suffix ${}_{L+n}$ is for coverage between different layers of the model (each layer is compared with the layer from the next column, i.e. in L4 column we show the coverage between features from layer 4 with features from layer 7).}
\label{tab:coverage}
\end{table}

\begin{table}[h!]
\centering
\renewcommand{\arraystretch}{1.2} %
\scalebox{0.7}{
\begin{tabular}{lcccccc}
\toprule
\textbf{Model} & \textbf{DS} & \textbf{L1} & \textbf{L4} & \textbf{L7} & \textbf{L10} & \textbf{L12}\\
\midrule
\multirow{3}{*}{Hub} & LS & 27 & 47 & 102 & 352 & 219\\
    & FSD & 122 & 82 & 66 & 101 & 171\\
    & MTG & 33 & 30 & 22 & 42 & 56\\
    \arrayrulecolor{black!30}\midrule
\multirow{3}{*}{Hub100K} & LS & 23 & 49 & 88 & 367 & 232\\
    & FSD & 66 & 71 & 47 & 113 & 132\\
    & MTG & 19 & 25 & 41 & 45 & 60\\
    \arrayrulecolor{black!30}\midrule
\multirow{3}{*}{Hub2} & LS & 40 & 45 & 88 & 355 & 227\\
    & FSD & 112 & 72 & 57 & 94 & 135\\
    & MTG & 38 & 34 & 23 & 53 & 68\\
    \arrayrulecolor{black!30}\midrule
\multirow{3}{*}{Wh} & LS & 793 & 755 & 787 & 221 & 230\\
    & FSD & 878 & 947 & 733 & 306 & 286\\
    & MTG & 1368 & 1328 & 736 & 101 & 65\\
    \arrayrulecolor{black!30}\midrule
\multirow{3}{*}{Wh100K} & LS & 844 & 718 & 736 & 144 & 166\\
    & FSD & 962 & 868 & 608 & 191 & 135\\
    & MTG & 1470 & 1375 & 754 & 91 & 69\\
\arrayrulecolor{black}\bottomrule
\end{tabular}
}
\caption{Numbers of features having duplicates within same SAE. Dataset and model names are as in Table~\ref{tab:coverage}}   
\label{tab:coverage_self}
\end{table}

\section{Domain-level feature specialization}\label{appendix:domain}

We characterize each feature by two activation metrics: \textit{activation frequency} and \textit{average non-zero activation value}, computed at both frame (per token) and audio (per sample) levels for datasets representing three predefined domains: \textit{speech}, \textit{sounds}, and \textit{music}.

For each domain combination (e.g., [speech, sounds, music], [speech, sounds], [music, sounds]), features are assigned to domains through threshold-based comparison. Specifically, for each feature $j$ and domain combination $c$, we compute the activation frequency $f_{i,j}$ for each domain $i$ in that combination. A feature is assigned to domain $i^*$ if:
\[
i^* = \arg\max_i f_{i,j} \quad \text{and} \quad \forall k \neq i^*: f_{i^*,j} - f_{k,j} \geq \tau
\]
where $\tau$ is a threshold from the progressive threshold set. Assignment occurs when the maximum activation frequency exceeds all others by at least $\tau$, providing graded confidence levels based on the threshold used. Features that fail to meet any threshold are marked as \textit{unassigned}, while inactive ones (with $f_{i,j} = 0$ for all domains) are labeled \textit{dead}. This procedure yields categorical labels and frequency-weighted color codes for visualization, with color intensity modulated by the threshold index $k$:
\[
\text{color}_j = \text{RGB}_{\text{base}} \times (1 - c_{\text{coeff}} \cdot k), \quad c_{\text{coeff}} = 0.2
\]

Final labels are aggregated across all domain combinations (three-way and pairwise), ensuring consistent categorization across contexts. The resulting assignments are visualized using t-SNE projections of SAE encoder weights, with colors corresponding to final domain labels. We additionally construct Venn diagrams to quantify overlap and exclusivity and to track the distribution of specialized features across model layers.

\subsection{Experimental setup}

\textbf{Datasets.} Seven datasets were used, grouped into three primary categories:
\begin{itemize}
\item \textbf{Speech:} LS-test-clean, IEMOCAP , ESD, Expresso, MELD, Demand
\item \textbf{Sounds:} WHAM!, FSD50k, VocalSound, Nonspeech7k, ESC-50, VGGSound
\item \textbf{Music:} MTG-Jamendo
\end{itemize}

\textbf{Thresholds.} Progressive thresholds were applied to ensure robust, confidence-graded specialization:
\begin{itemize}
\item Frame-level: $\tau \in \{0.2, 0.1, 0.04\}$
\item Audio-level: $\tau \in \{0.5, 0.3\}$
\end{itemize}
Frame-level thresholds identify fine-grained feature specialization across individual tokens, while audio-level thresholds target coarser patterns observable at the sample level.

\textbf{Audio-level analysis.} Audio-level domain specialization captures features responsive to global acoustic properties and long-range dependencies. Thresholds of 0.5 and 0.3 were selected to emphasize features with substantial full-sample activation while maintaining discrimination between domains. This level of analysis complements frame-level detection by revealing features whose specialization is consistent across entire audio samples rather than transient in individual tokens.

Detection across multiple domain combinations (\textbf{[speech, sounds, music]}, \textbf{[speech, sounds]}, \textbf{[speech, music]}, \textbf{[sounds, music]}) allows disambiguation of overlapping feature roles that remain hidden in single-domain analysis. Formally, this multi-combination strategy leverages pairwise comparisons to identify features salient for two domains but not the third, enabling detection of subtle cross-modal relationships. For instance, features activated by both speech and environmental sounds but not music are properly identified in the [speech, sounds] experiment while remaining unassigned in the comprehensive three-domains analysis.

\subsection{Frequency analysis}\label{appendix:domain-freq}
\begin{figure*}[t]
\centering

\begin{subfigure}[t]{0.48\linewidth}
  \centering
  \includegraphics[width=\linewidth]{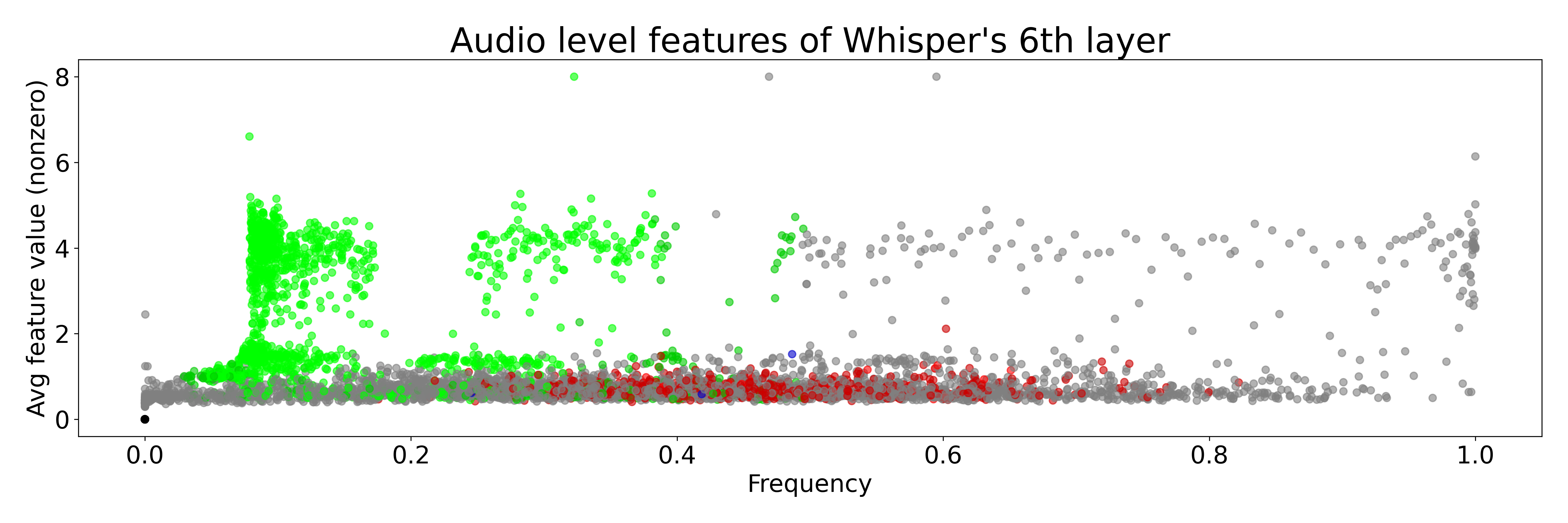}
\end{subfigure}\hfill
\begin{subfigure}[t]{0.48\linewidth}
  \centering
  \includegraphics[width=\linewidth]{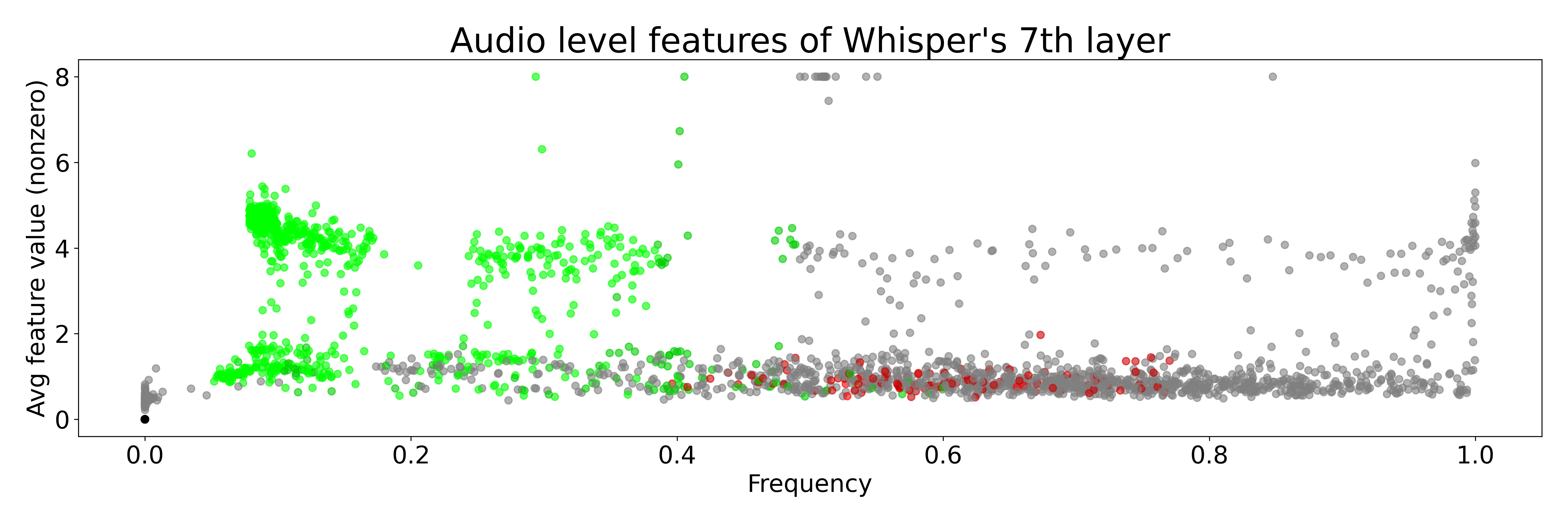}
\end{subfigure}

\begin{subfigure}[t]{0.48\linewidth}
  \centering
  \includegraphics[width=\linewidth]{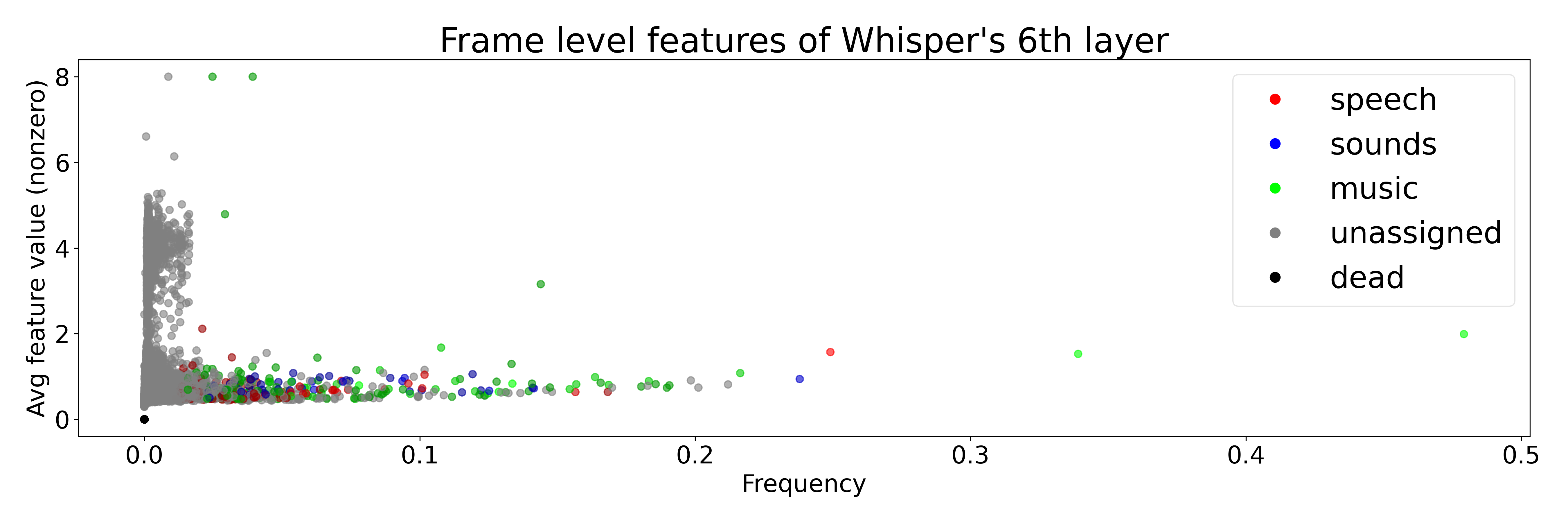}
\end{subfigure}\hfill
\begin{subfigure}[t]{0.48\linewidth}
  \centering
  \includegraphics[width=\linewidth]{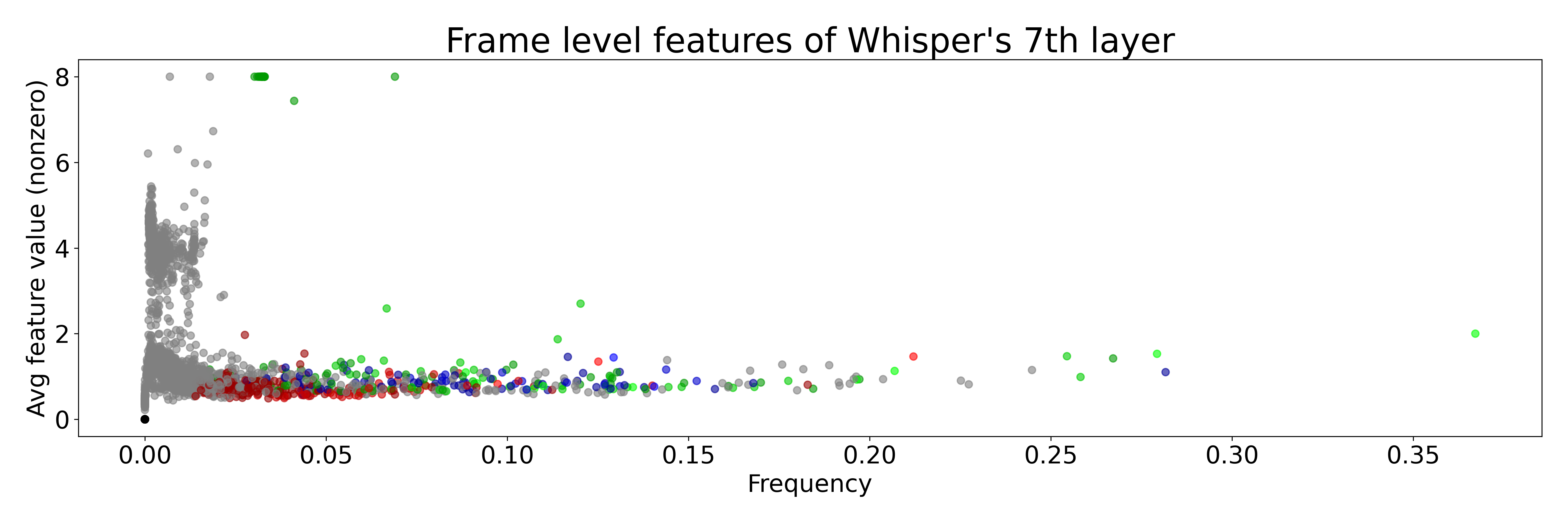}
\end{subfigure}

\begin{subfigure}[t]{0.48\linewidth}
  \centering
  \includegraphics[width=\linewidth]{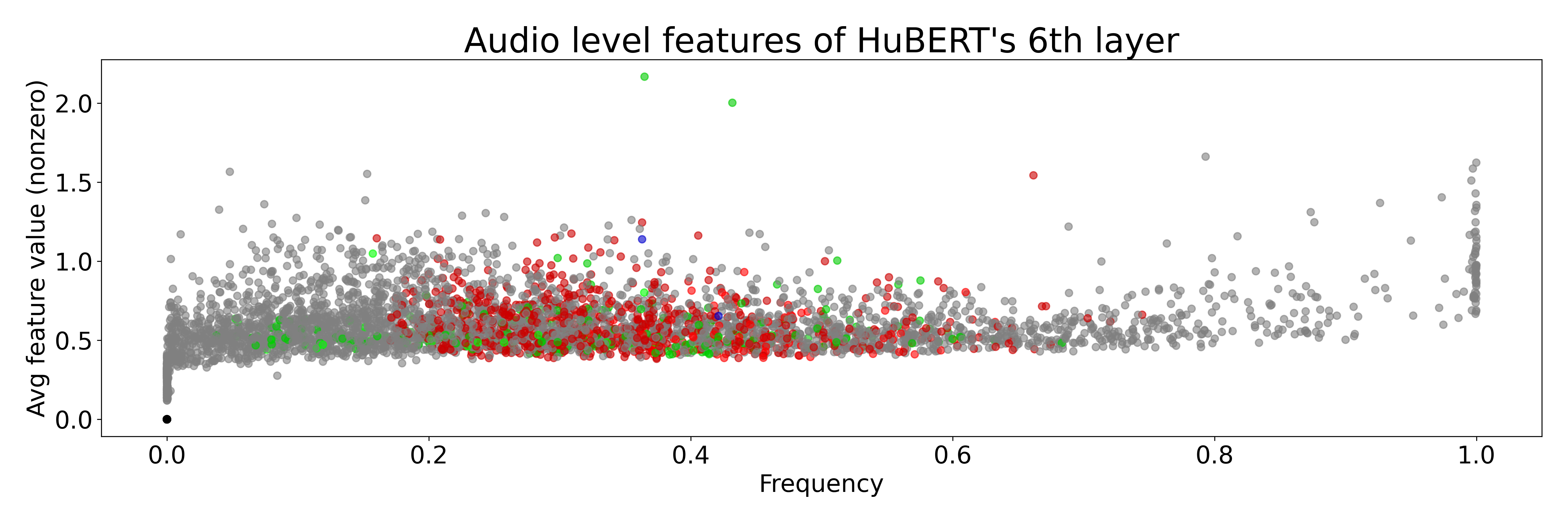}
\end{subfigure}\hfill
\begin{subfigure}[t]{0.48\linewidth}
  \centering
  \includegraphics[width=\linewidth]{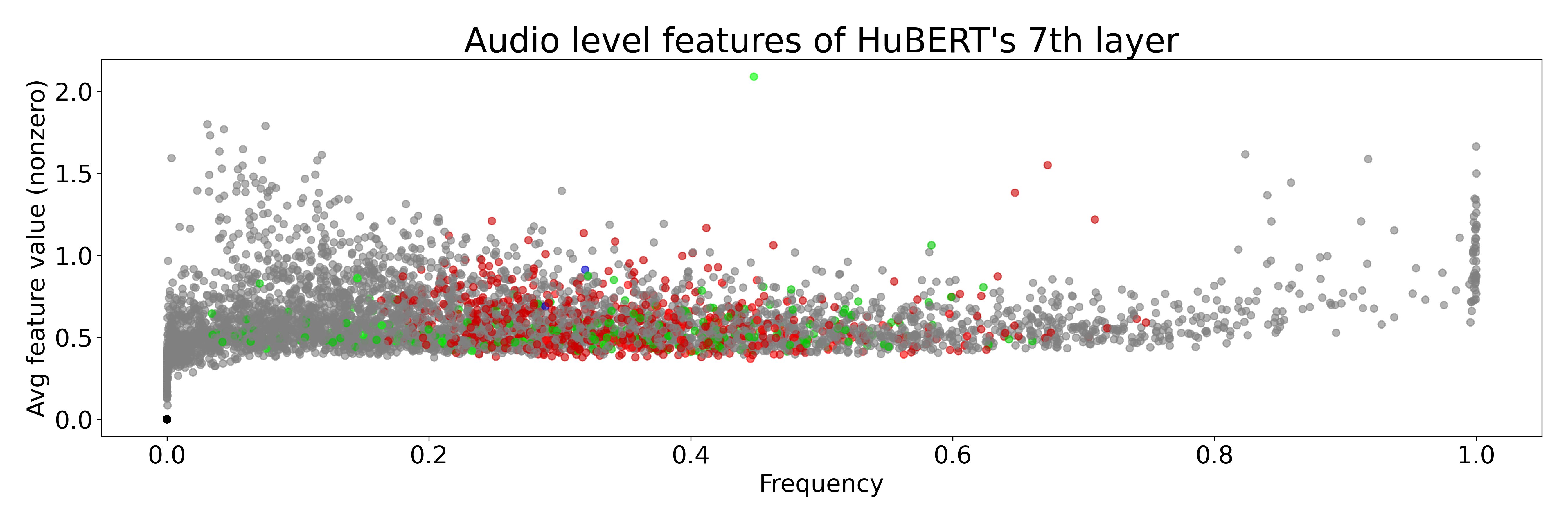}
\end{subfigure}

\begin{subfigure}[t]{0.48\linewidth}
  \centering
  \includegraphics[width=\linewidth]{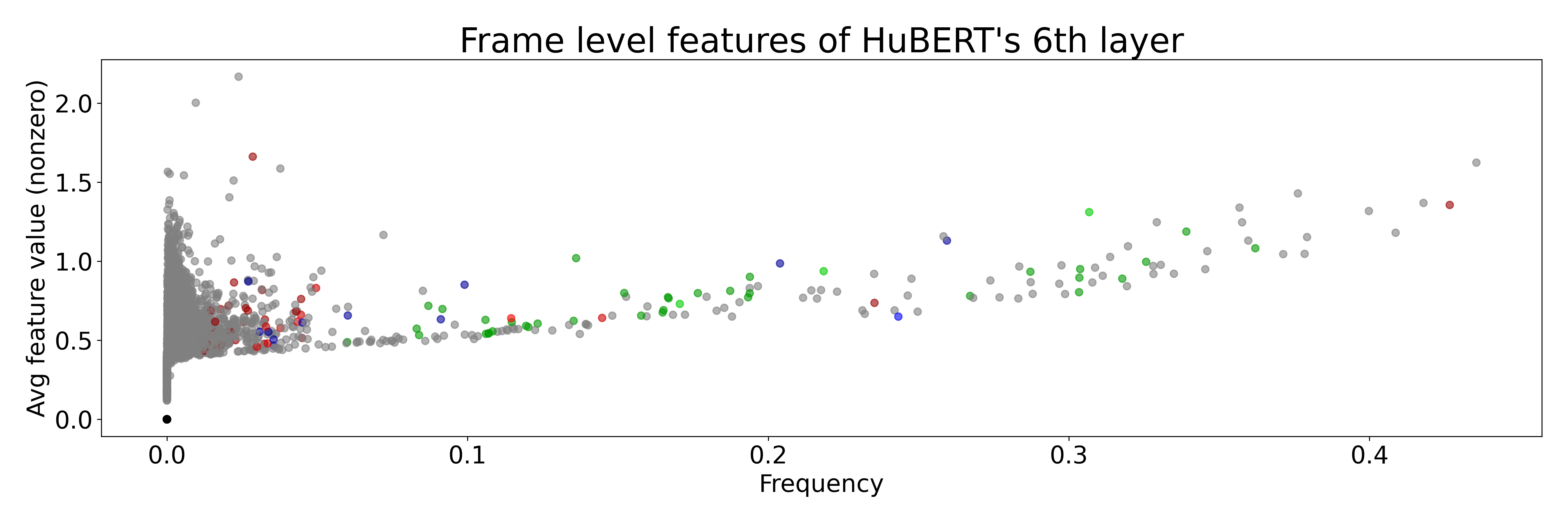}
\end{subfigure}\hfill
\begin{subfigure}[t]{0.48\linewidth}
  \centering
  \includegraphics[width=\linewidth]{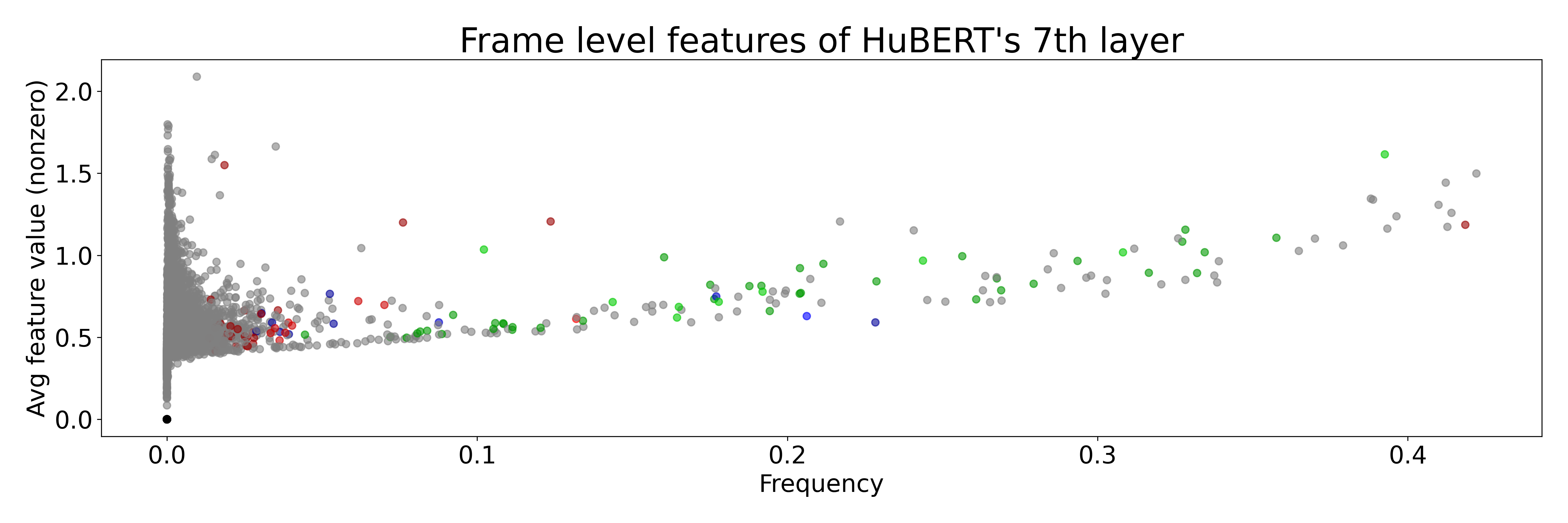}
\end{subfigure}

\caption{\textbf{Frequency-based domain specialization.} Audio-level (1st and 3rd rows from the top) and frame-level (2nd and 4th rows from the top) activation frequency versus activation magnitude for Whisper (1st and 2nd rows from the top) and HuBERT (3rd and 4th rows from the top) layers 6–7 (left and right columns respectively). Colors: red (speech), blue (sounds), green (music), gray (unassigned), black (dead).}
\label{fig:clustering_freq}
\end{figure*}

We present frequency-based domain specialization analysis across Whisper and HuBERT layers 6 and 7 (Fig.~\ref{fig:clustering_freq}), two depths where acoustic feature learning remains interpretable yet incorporates substantial linguistic context. Scatter plots of activation frequency (x-axis) versus average non-zero activation value (y-axis) reveal distinct frequency distributions, domain-specific patterns, and model-dependent differences in activation magnitude.

Audio-level frequency exhibits a markedly different distribution than frame-level frequency. Features in both Whisper and HuBERT span the full frequency range from $f^{\text{audio}}_{i,j} = 0$ (never activated in any samples from any domains) to $f^{\text{audio}}_{i,j} = 1.0$ (activated in all samples from every domain). This coverage reflects the aggregative nature of audio-level frequency: features that activate sparsely at the frame level may still reach high sample-level frequency if their activations are distributed across many different samples.

Whisper exhibits pronounced clustering of music features (Fig.~\ref{fig:clustering_freq}, top) at high average activation values ($\approx 3$–6) with frequencies spanning $0.1$ to $0.2$ and $0.25$ to $0.4$. Speech (red) and sounds (blue) features are at lower activation values ($\approx 0.5$–2) with frequencies distributed across the full range. This pattern suggests that Whisper has learned a dedicated set of music-responsive features with high activation magnitude.

HuBERT exhibits a different audio-level profile (Fig.~\ref{fig:clustering_freq}, bottom): the feature space shows dense, undifferentiated specialization at low average activation values ($\approx 0.4$–1.0) across all frequencies up to $1.0$. Domain specialization is visible through domain colors (red (speech), blue (sounds), green (music)) but without any separation in activation magnitude observed in Whisper. This suggests HuBERT distributes specialization across more features with comparable activation strengths.

Frame-level frequency exhibits differences from audio-level analysis, with maximum frequencies typically not exceeding $f^{\text{frame}}_{i,j} \approx 0.5$, and substantial feature specialization at frequencies near zero. This sparsity pattern reflects the discrete, context-dependent nature of feature activation in sparse models: a feature may activate in only a small fraction of frames within a domain, even if it appears in most samples at the audio level. Whisper's and HuBERT's frame-level distribution are compressed, with most features concentrated in $f^{\text{frame}} < 0.1$.

\subsection{Encoder matrix decomposition analysis}

To directly probe how domain-specialized features are organized in representation space, encoder matrix decomposition is applied to the SAE encoder weights corresponding to layers 6 and 7 of Whisper and HuBERT. This analysis operates on a filtered subset of features, selected using the same activation-frequency statistics as in Section~\ref{appendix:domain-freq} to ensure that only active and interpretable units are retained. Only those features are used that have been applied at least once to any domain in any domain combination.

The filtered encoder weight matrix $W_{\text{enc}} \in \mathbb{R}^{M \times d}$ (with $M = 8 \times 768$ latent units for Whisper and HuBERT) is then projected to two dimensions using t-SNE. 

Each point in the resulting 2D embedding corresponds to a single SAE feature and is colored according to its domain assignment from the threshold-based procedure. Then for every domain combination, for example, for the combination [speech, sound, music] the features active for this combination are colored, while unassigned gray features will be active for one of the combinations, e.g. for [speech, sound]. Also, brighter dots reflect features with greater frequency differences.

\begin{figure*}[t]
\centering

\begin{subfigure}[t]{0.48\linewidth}
  \centering
  \includegraphics[width=\linewidth]{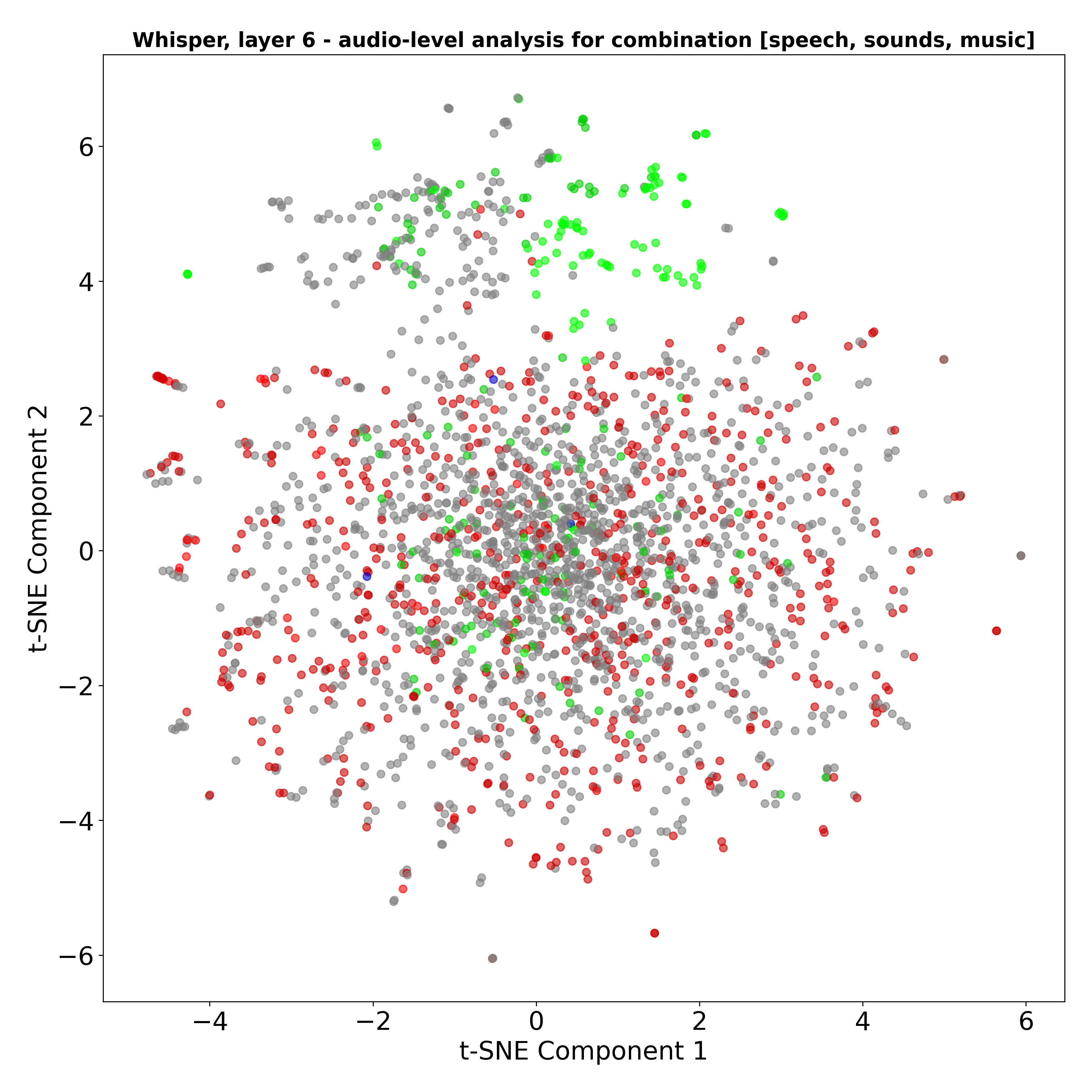}
\end{subfigure}\hfill
\begin{subfigure}[t]{0.48\linewidth}
  \centering
  \includegraphics[width=\linewidth]{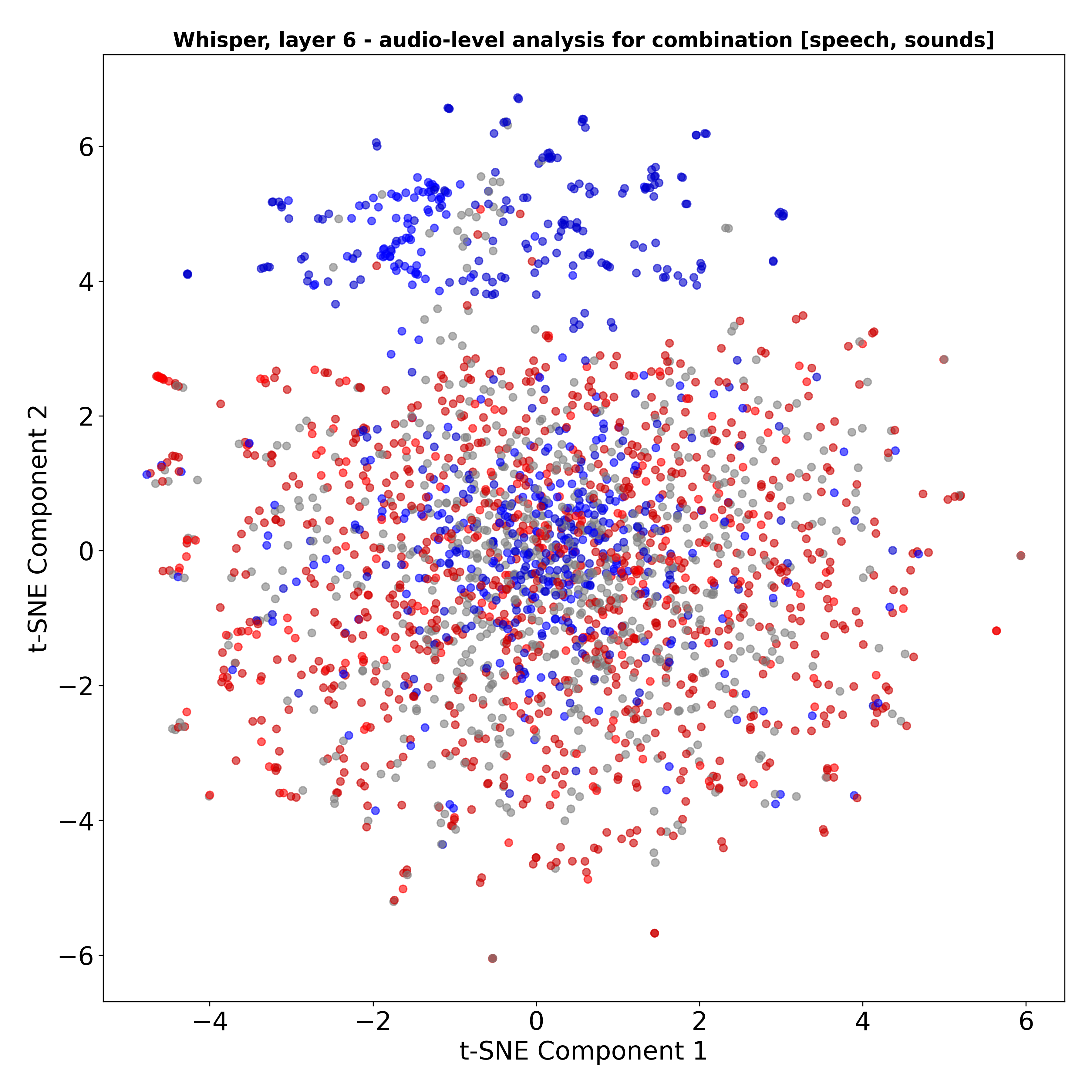}
\end{subfigure}

\begin{subfigure}[t]{0.48\linewidth}
  \centering
  \includegraphics[width=\linewidth]{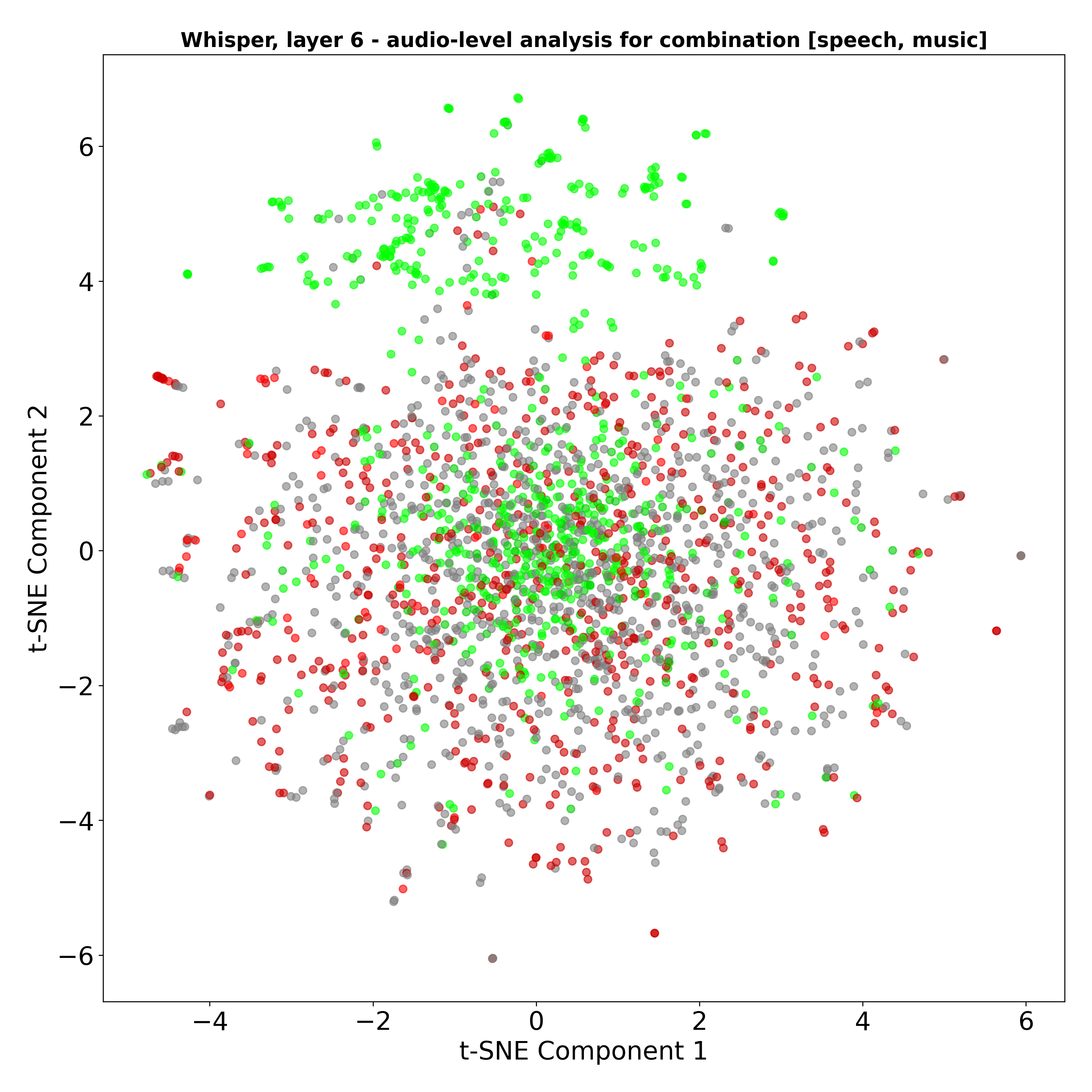}
\end{subfigure}\hfill
\begin{subfigure}[t]{0.48\linewidth}
  \centering
  \includegraphics[width=\linewidth]{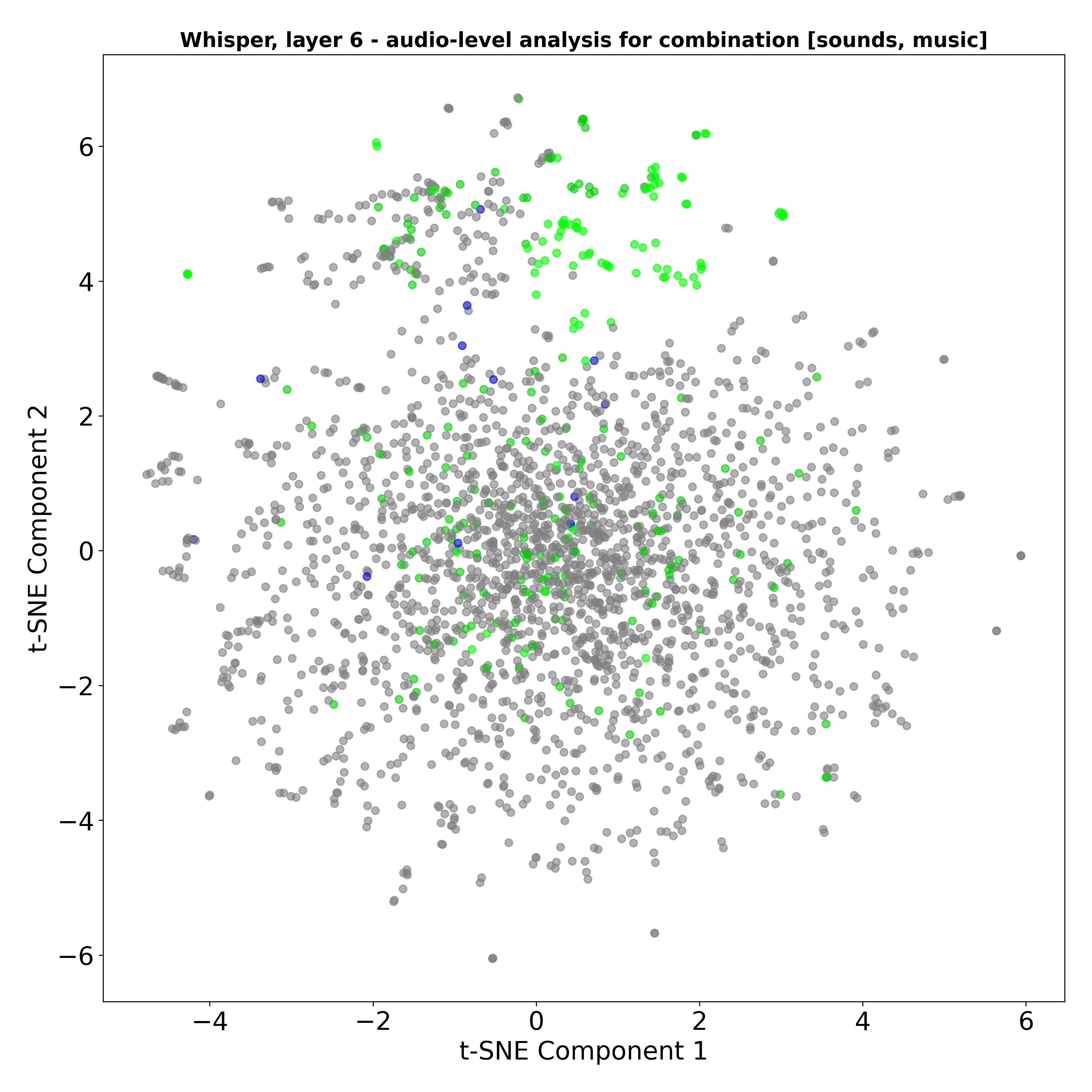}
\end{subfigure}

\caption{t-SNE decomposition of SAE encoder weights for Whisper layer 6 in the audio-level setup. Each point corresponds to a single latent feature, colored by its domain assignment (speech, sounds, music, or unassigned) obtained from activation-frequency–based specialization. Brighter dots indicate features with larger activation frequency differences between domains, highlighting the most strongly specialized units in the representation space.}
\label{fig:tsne-w}
\end{figure*}

\begin{figure*}[t]
\centering

\begin{subfigure}[t]{0.48\linewidth}
  \centering
  \includegraphics[width=\linewidth]{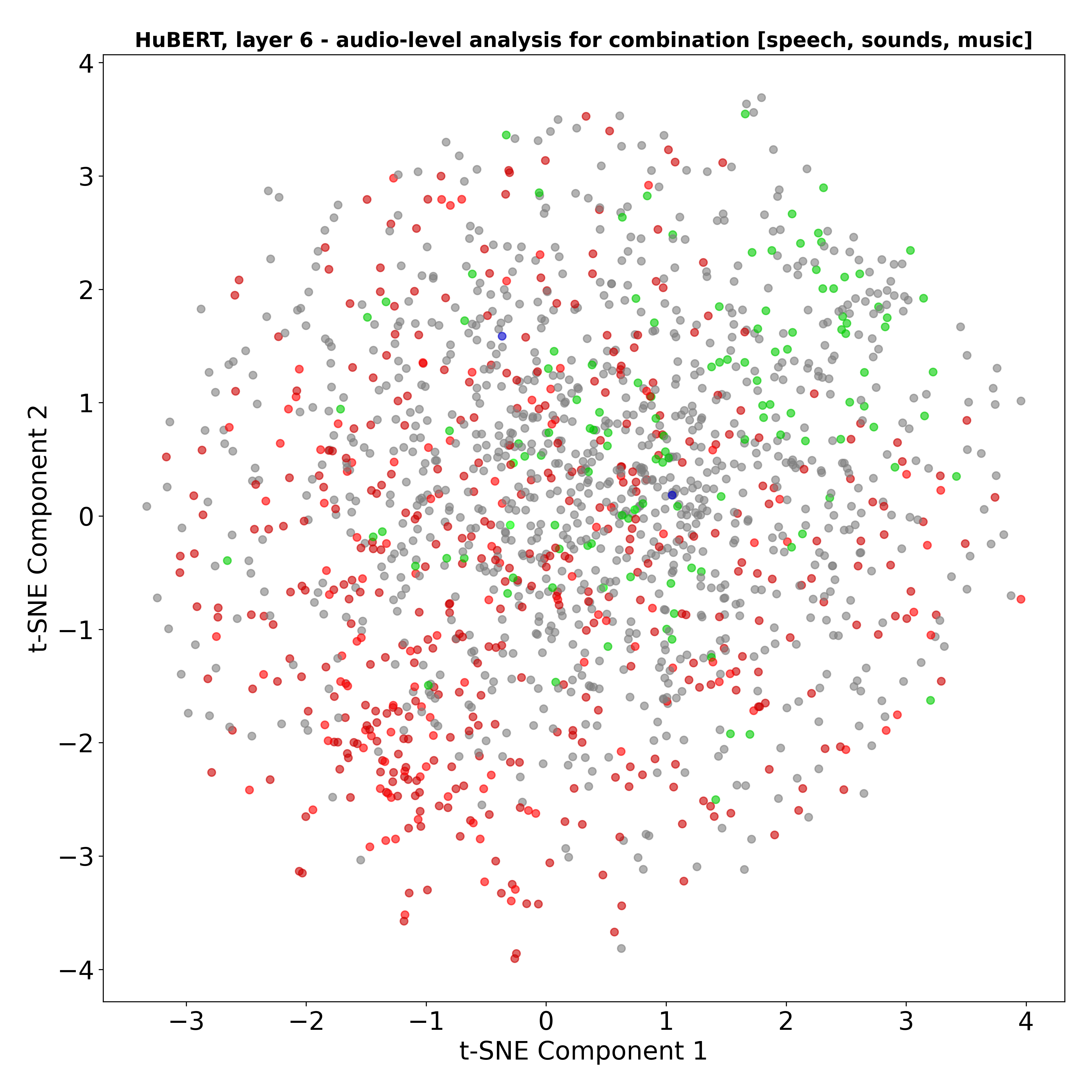}
\end{subfigure}\hfill
\begin{subfigure}[t]{0.48\linewidth}
  \centering
  \includegraphics[width=\linewidth]{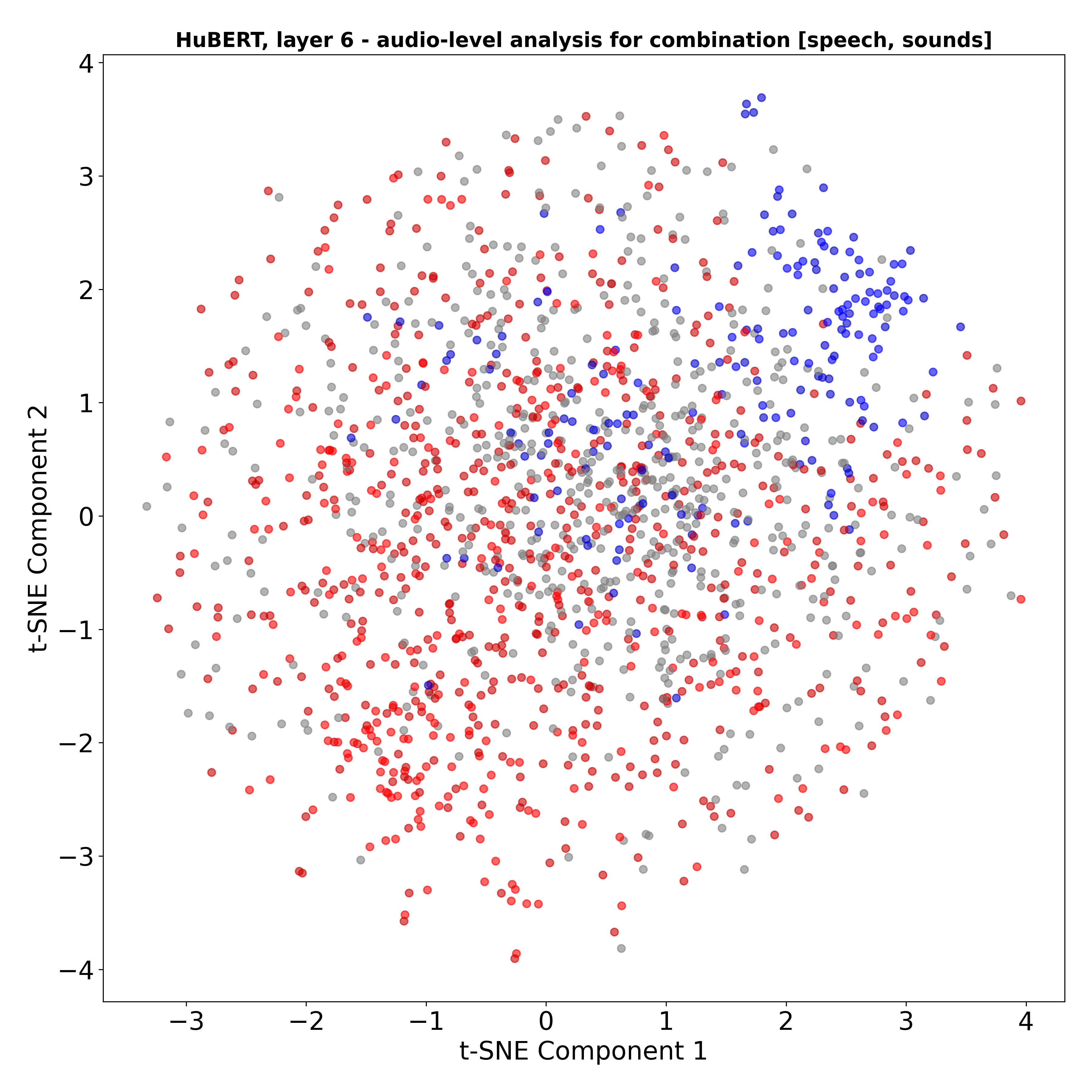}
\end{subfigure}

\begin{subfigure}[t]{0.48\linewidth}
  \centering
  \includegraphics[width=\linewidth]{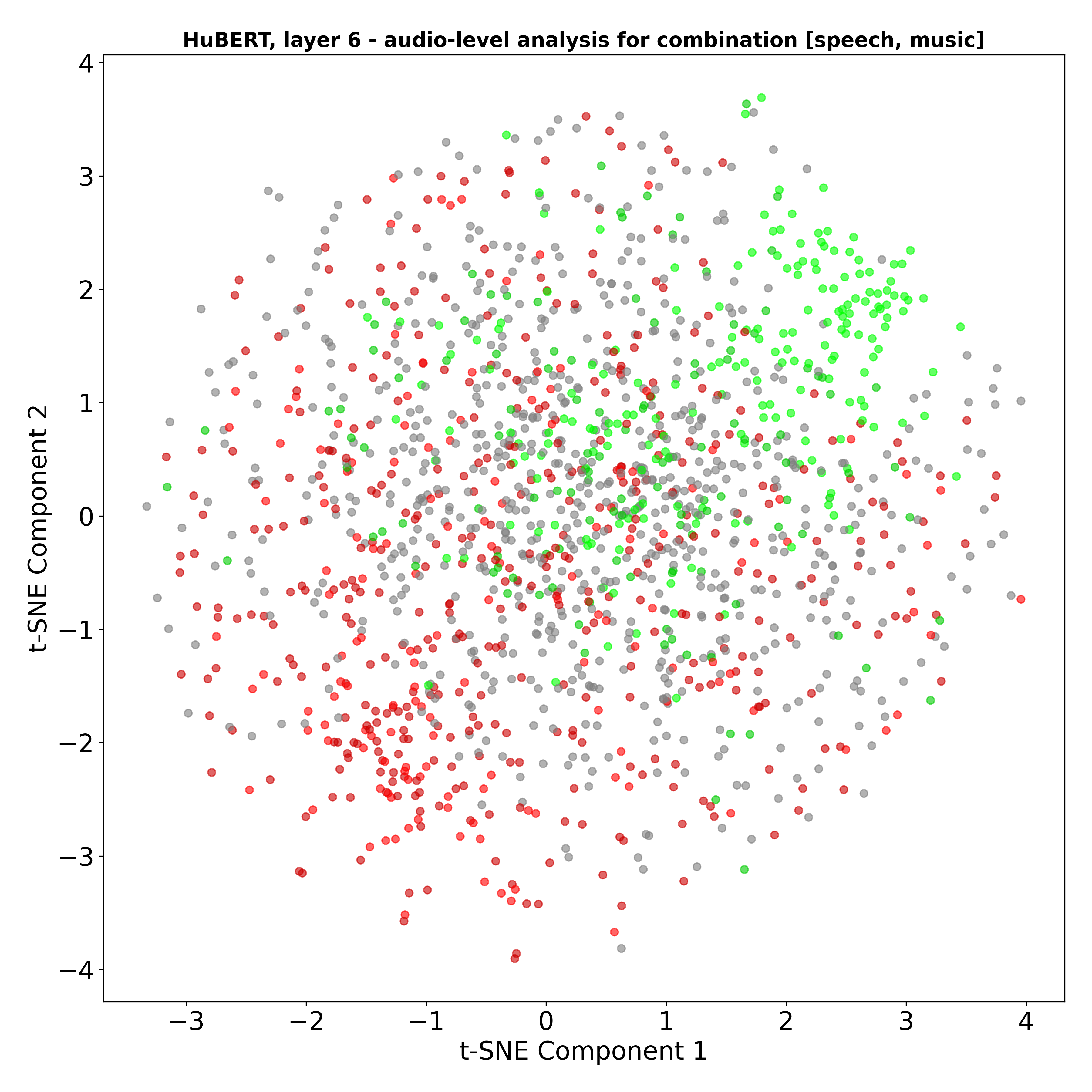}
\end{subfigure}\hfill
\begin{subfigure}[t]{0.48\linewidth}
  \centering
  \includegraphics[width=\linewidth]{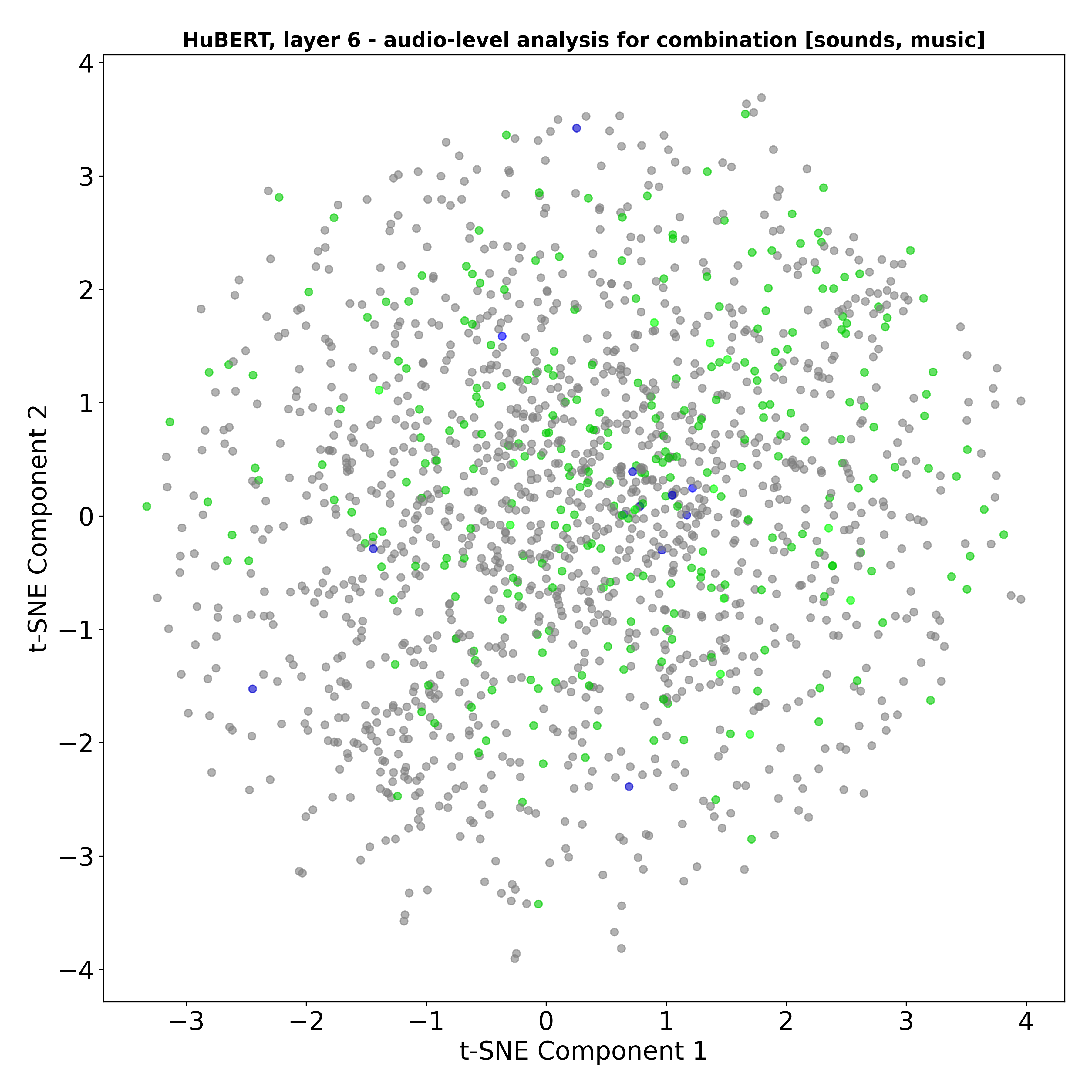}
\end{subfigure}

\caption{t-SNE decomposition of SAE encoder weights for HuBERT layer 6 in the audio-level setup. Points represent individual latent features colored by their domain assignments, with unassigned gray features active only in alternative domain combinations.}
\label{fig:tsne-h}
\end{figure*}

The t-SNE decomposition of the SAE encoder matrix is presented for audio-level setup for Whisper layer 6 (see Fig.~\ref{fig:tsne-w}) and HuBERT layer 6 (see Fig.~\ref{fig:tsne-h}).

\subsection{Multi domain features analysis}
Feature specialization overlap is quantified through set-theoretic analysis. Define $\mathcal{S}_D = \{j : \text{feature } j \text{ assigned to domain } D\}$ for each domain $D \in \{\text{speech}, \text{sounds}, \text{music}\}$. Venn diagrams visualize sets $|\mathcal{S}_D|$ and all pairwise intersections, revealing the structure of cross-modal feature dependencies. The main observation is that for most layers for both models the speech set is separated from the sound and music sets, and the sound set is almost completely absorbed by the music set at both the audio and frame levels. See Fig.~\ref{fig:clustering_2}.

Those features that are strictly related to speech, sounds, or music (lie outside the intersections) form a Fig.~\ref{fig:clustering_4} for layer-wise analysis.

\subsection{Layer-wise analysis}\label{appendix:domain-layer-wise}

In addition to the main text (Section~\ref{sec:domain}).

Lower proportion of Whisper's speech features (compared to music) may reflect either stronger compression of speech information or a more efficient internal representation for speech relative to music.

HuBERT shows more speech features at the audio level and lower at the frame level than Whisper. This pattern is consistent with the interpretation that HuBERT is more sensitive to global audio attributes, whereas Whisper contains a richer set of frame level features, related to local semantic information, what is caused by the difference in pre-training objectives and data composition.

\begin{figure*}[t]
\centering

\begin{subfigure}[t]{0.48\linewidth}
  \centering
  \includegraphics[width=\linewidth]{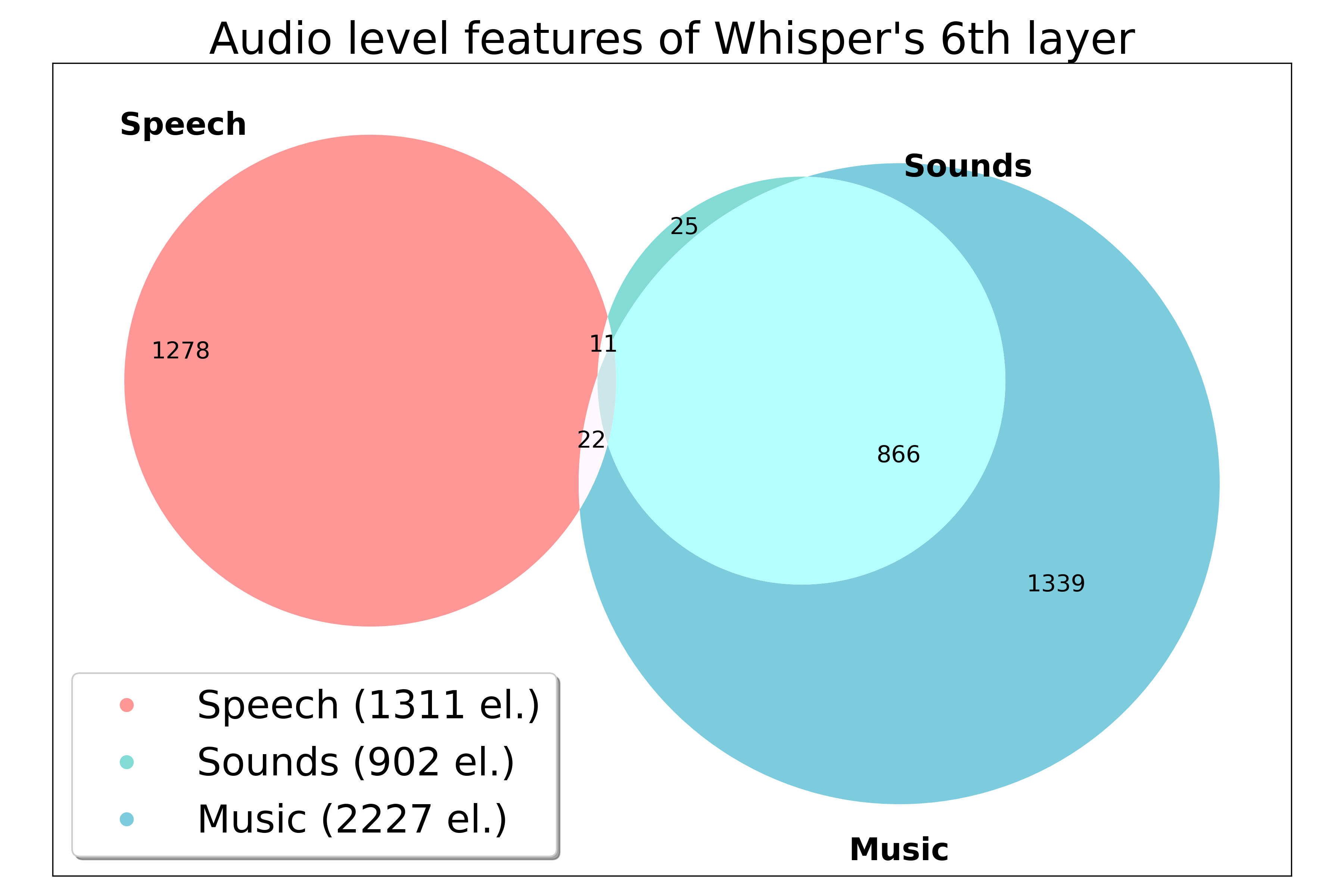}
  \label{fig:clustering_2a}
\end{subfigure}\hfill
\begin{subfigure}[t]{0.48\linewidth}
  \centering
  \includegraphics[width=\linewidth]{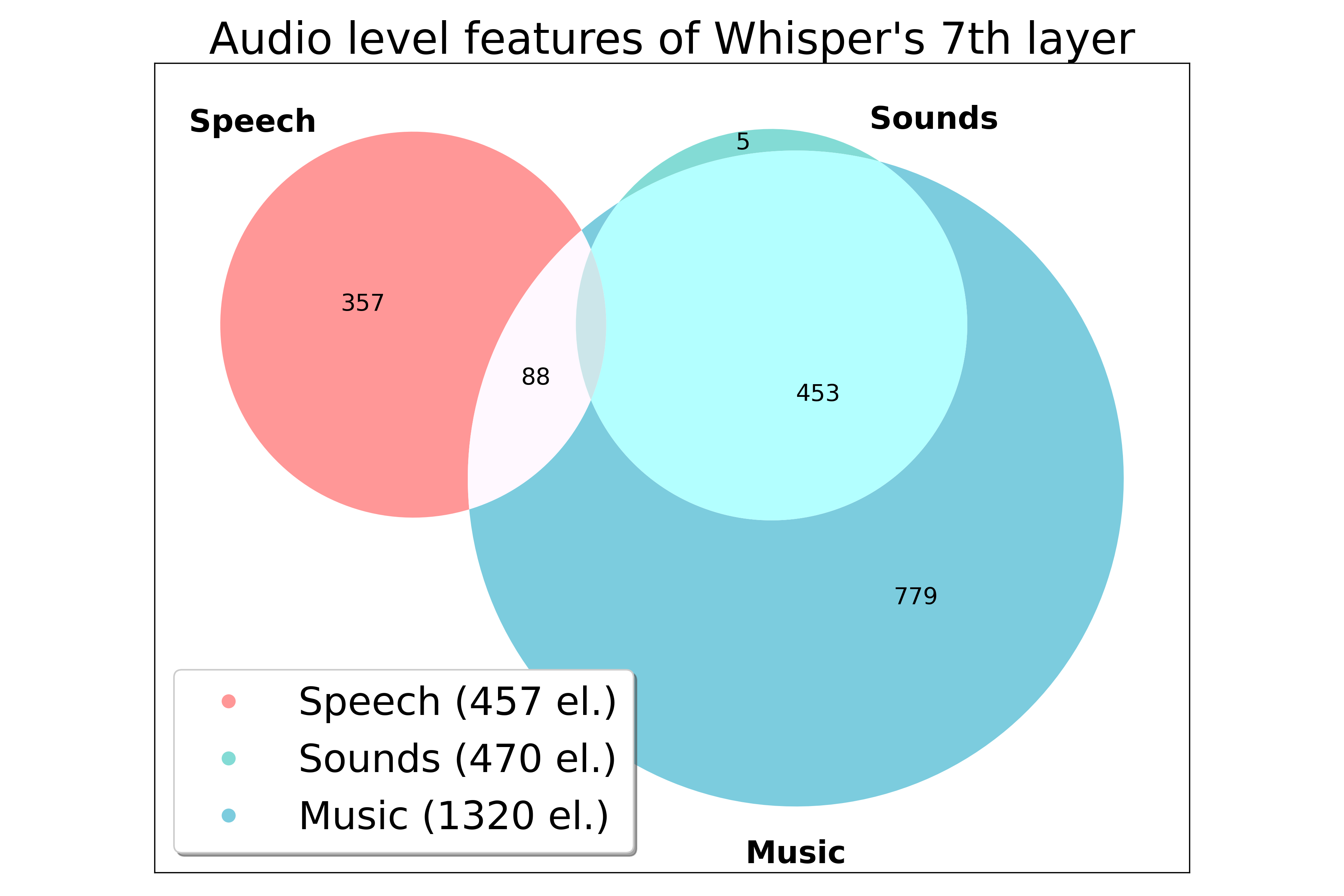}
  \label{fig:clustering_2b}
\end{subfigure}

\begin{subfigure}[t]{0.48\linewidth}
  \centering
  \includegraphics[width=\linewidth]{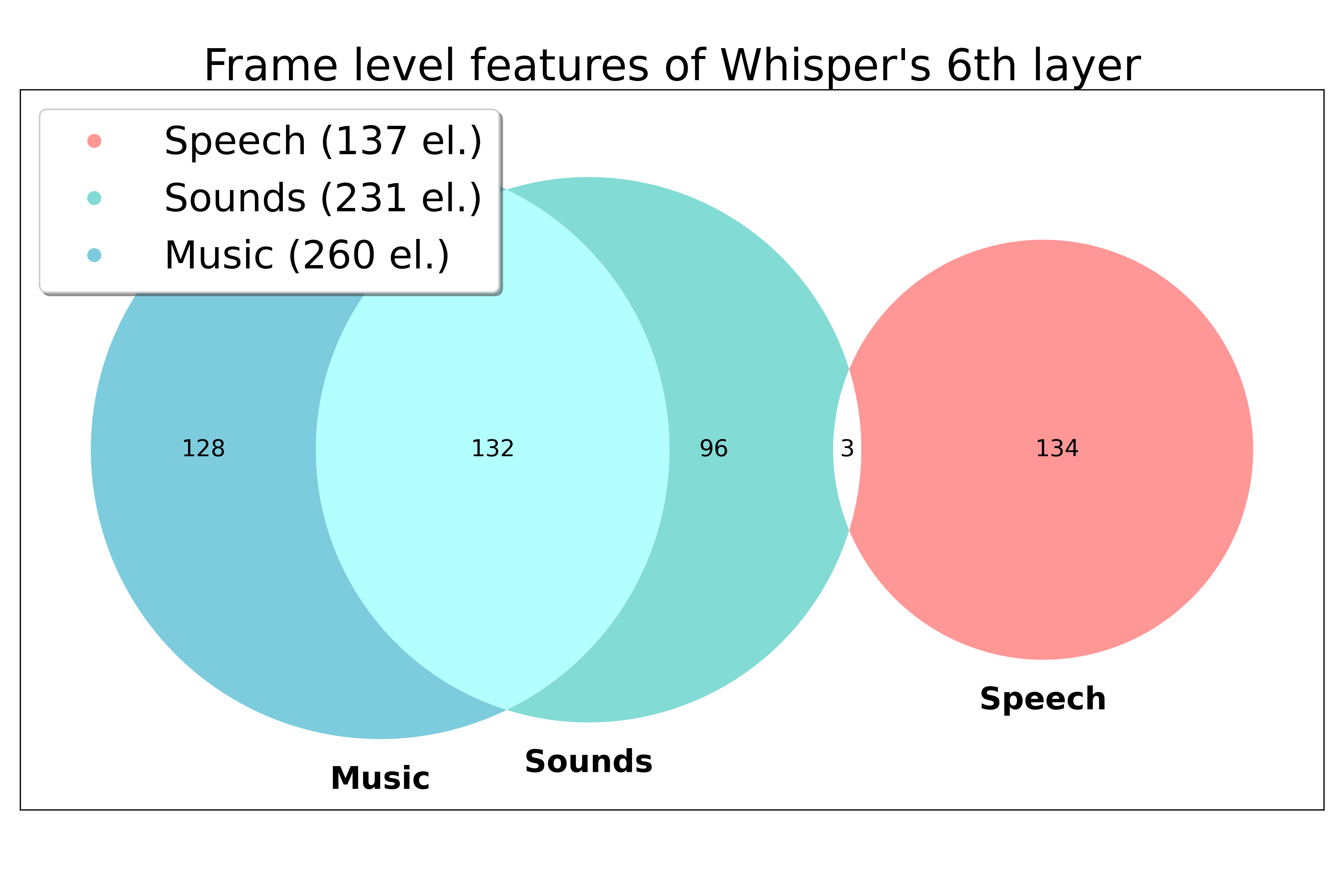}
  \label{fig:clustering_2c}
\end{subfigure}\hfill
\begin{subfigure}[t]{0.48\linewidth}
  \centering
  \includegraphics[width=\linewidth]{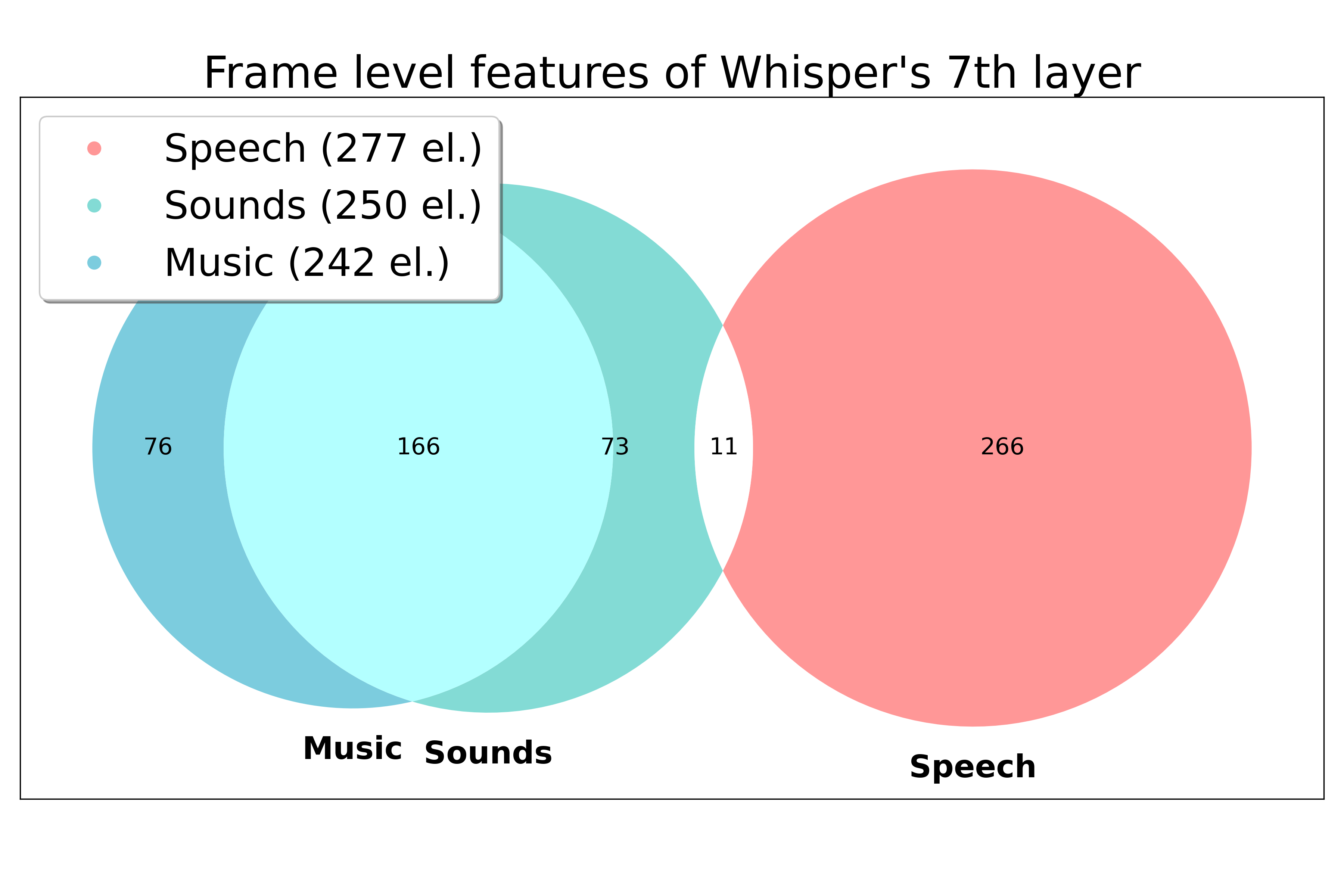}
  \label{fig:clustering_2d}
\end{subfigure}

\caption{\textbf{Feature overlap} for Whisper (layers 6 and 7): Venn diagrams for audio and frame levels.}
\label{fig:clustering_2}
\end{figure*}

Sound features (blue) are consistently underrepresented in both models: they are nearly absent in HuBERT and appear only sparsely in Whisper, primarily at the frame level. An explanation is that many sound features co-occur more frequently with music than with the sound domain, causing the sound feature set to be effectively subsumed by music-associated activations. This conclusion is supported by an additional Fig.~\ref{fig:clustering_2}.

\section{Classification}\label{app:classification}

The following datasets are selected for the classifiers training: $5000$ audios from LibriTTS \cite{libritts} dataset for gender classification; $2500$ clean and $2500$ speech samples from Demand dataset; $1500$ for each of five accents -- American, British, Indian, Irish, Scottish -- from VCTK for accent classification; and English part of ESD dataset for a $5$ class emotion classification, encompassing the emotions angry, happy, neutral, sad, and surprise.

We used \texttt{LogisticRegression} class from \texttt{scikit-learn} with parameters \texttt{max\_iter=10000, penalty='none', solver='newton-cg'}.

\begin{figure*}[htb]

\includegraphics[width=0.5\linewidth]{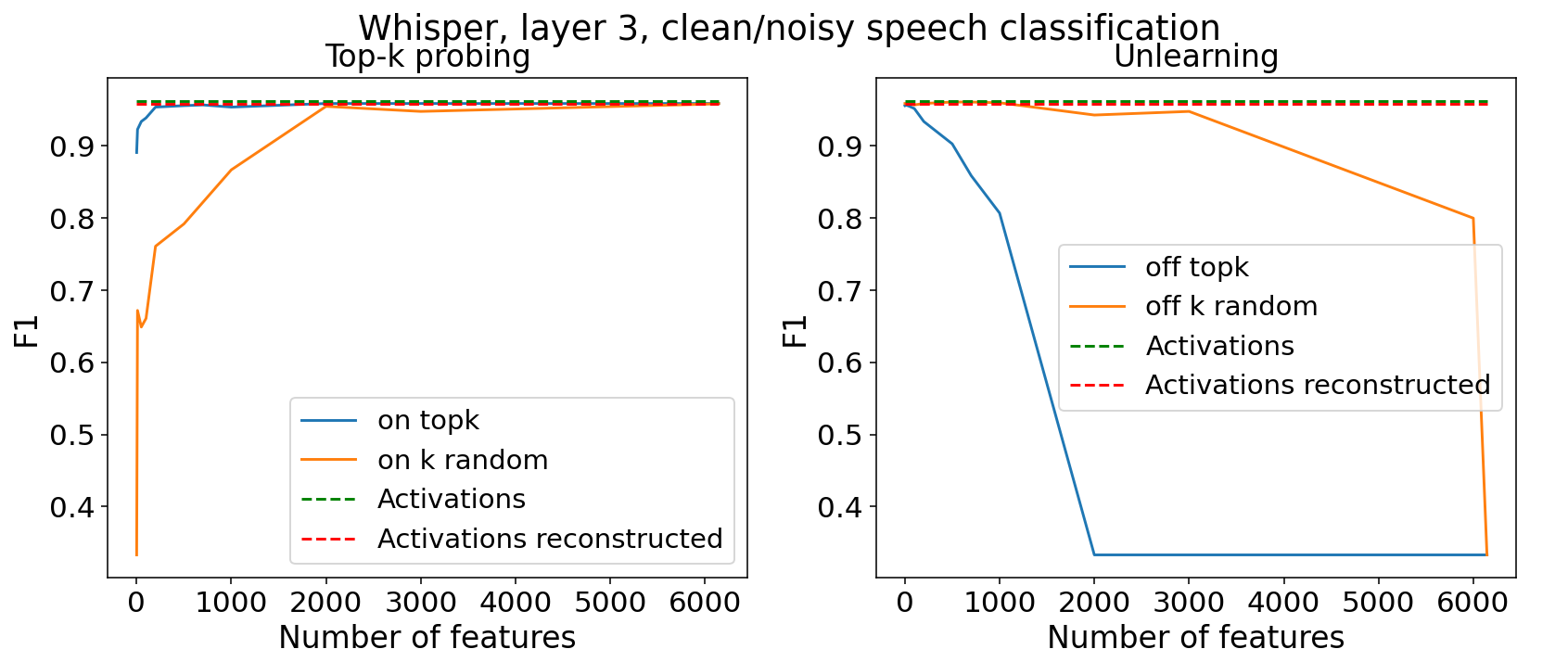}%
\includegraphics[width=0.5\linewidth]{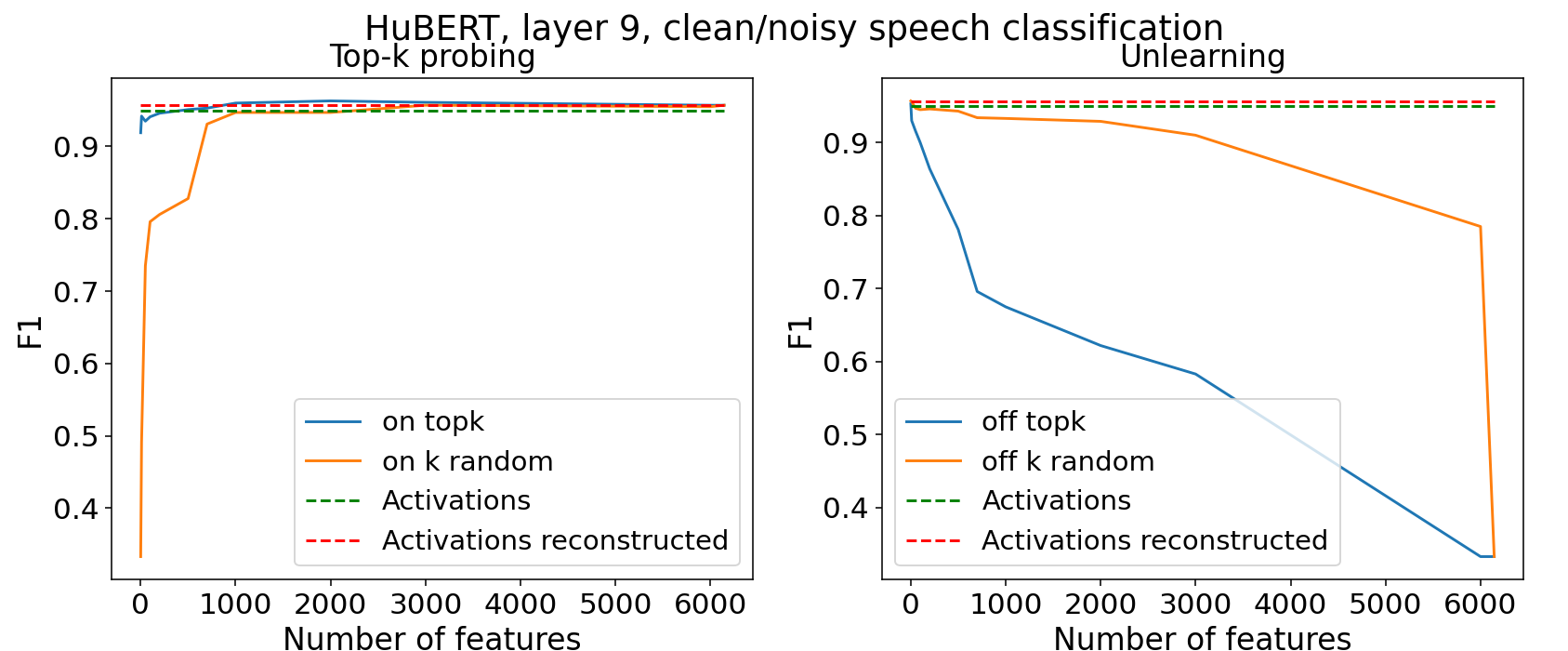}
\includegraphics[width=0.5\linewidth]{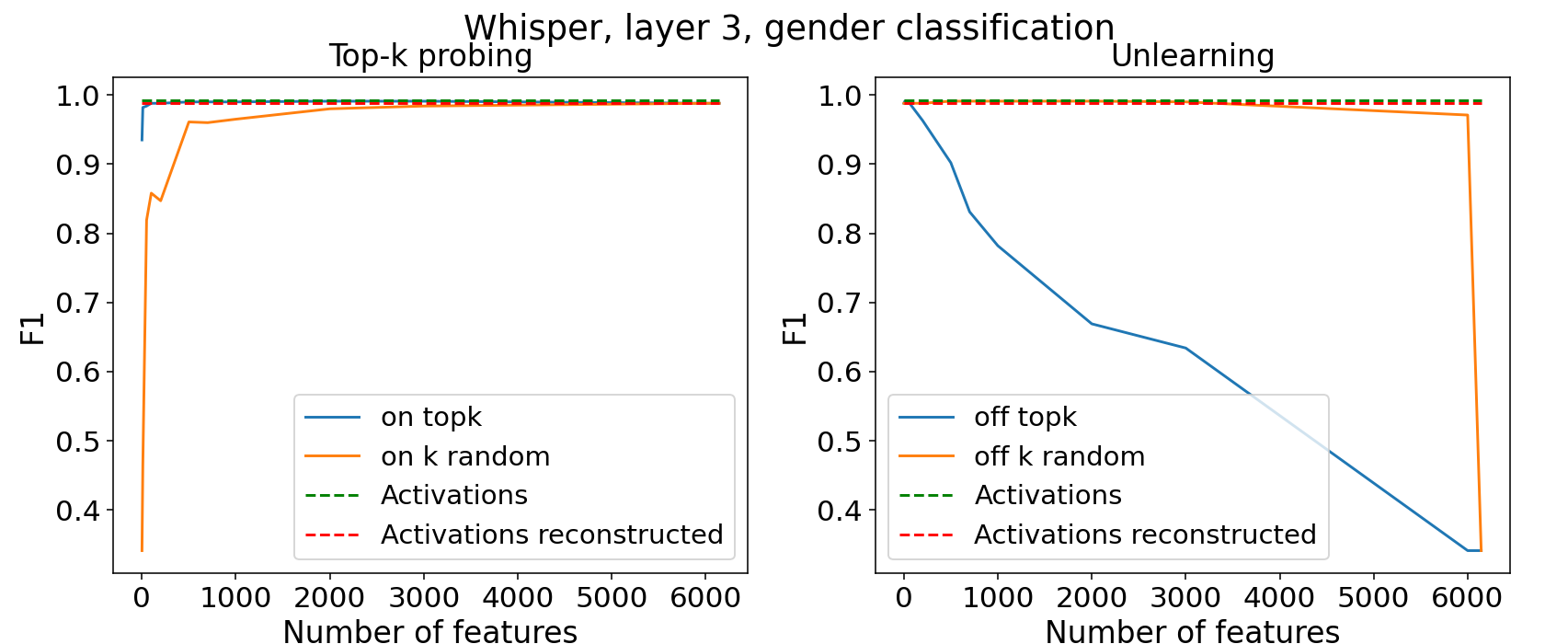}%
\includegraphics[width=0.5\linewidth]{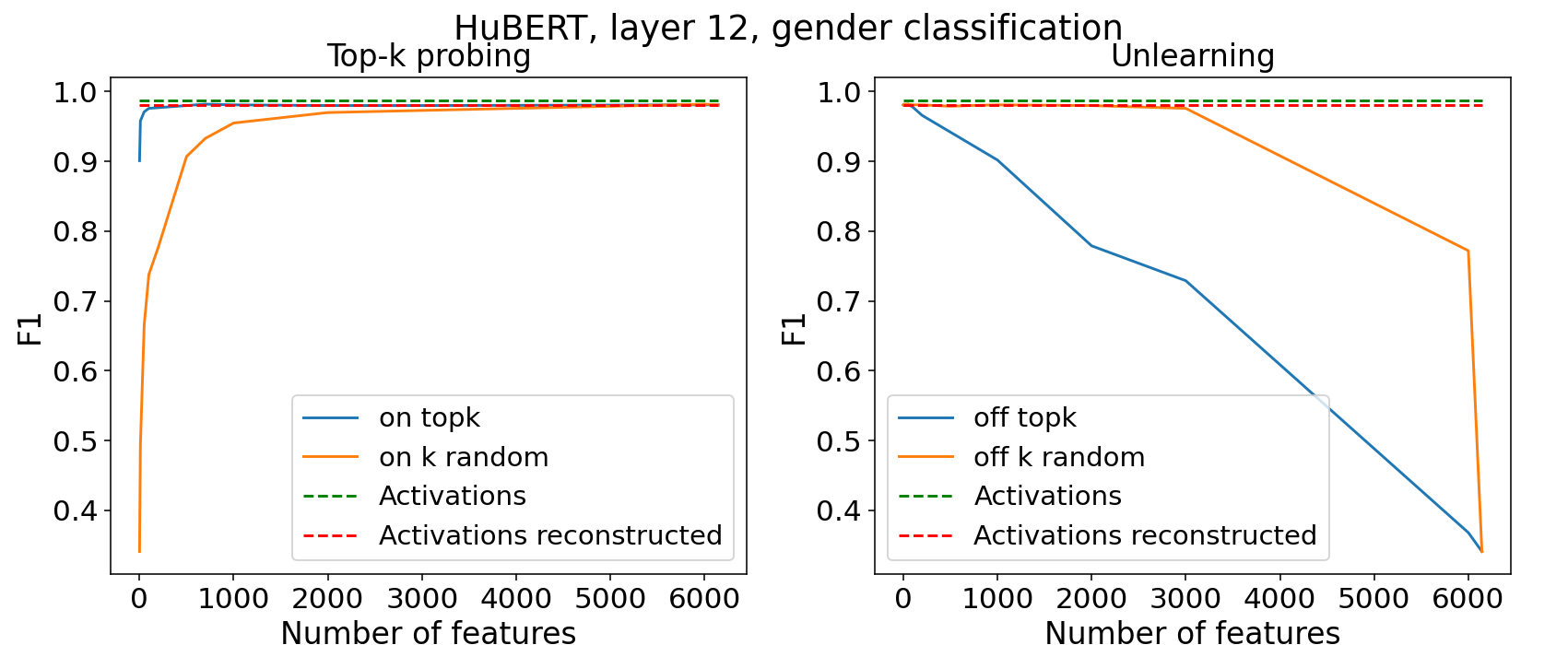}
\includegraphics[width=0.5\linewidth]{images/whisper_layer0_modif1.png}%
\includegraphics[width=0.5\linewidth]{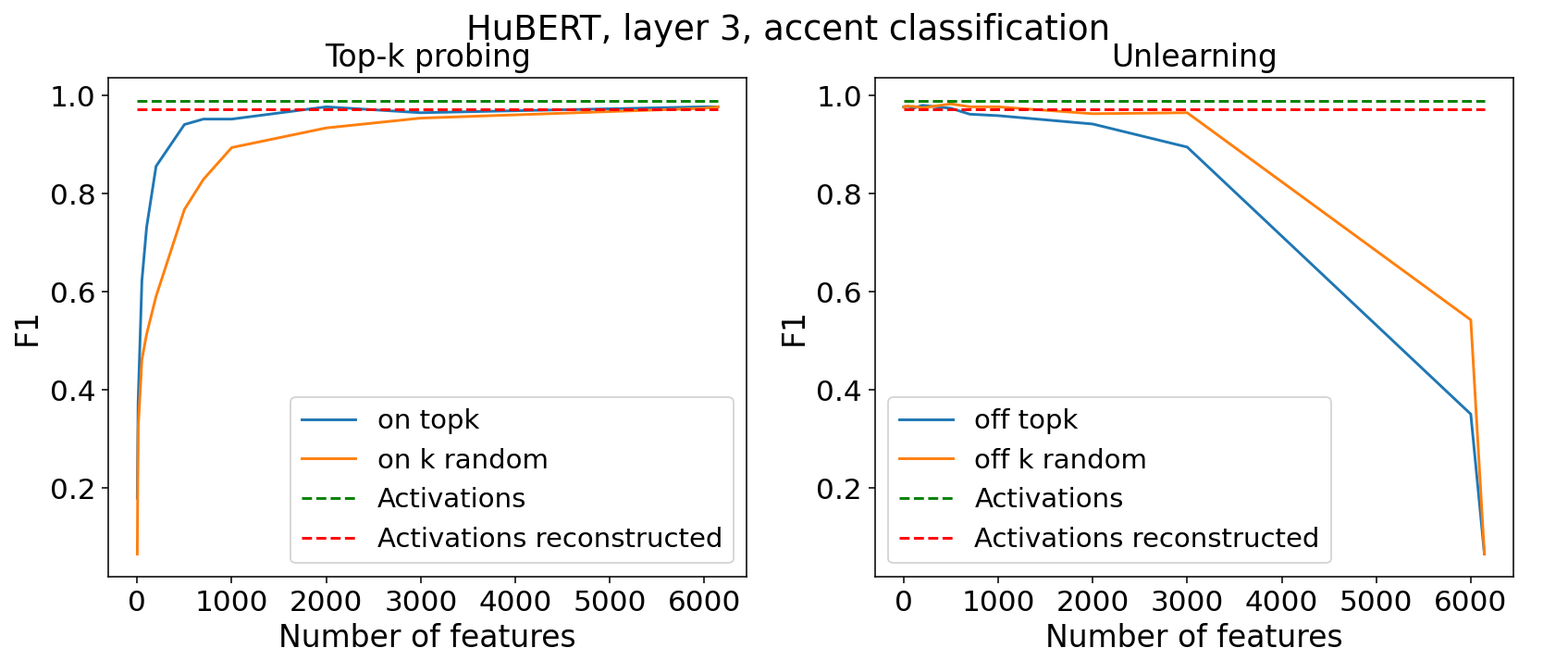}
\includegraphics[width=0.5\linewidth]{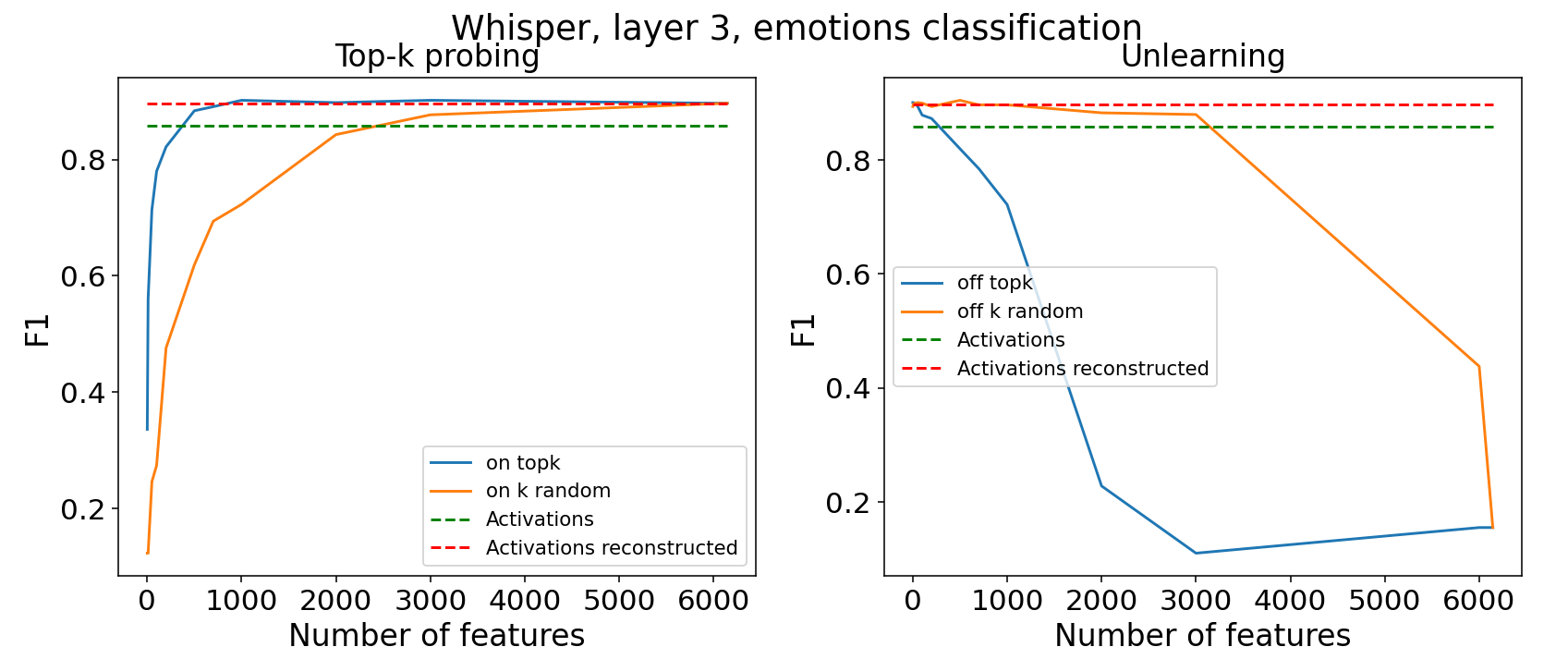}%
\includegraphics[width=0.5\linewidth]{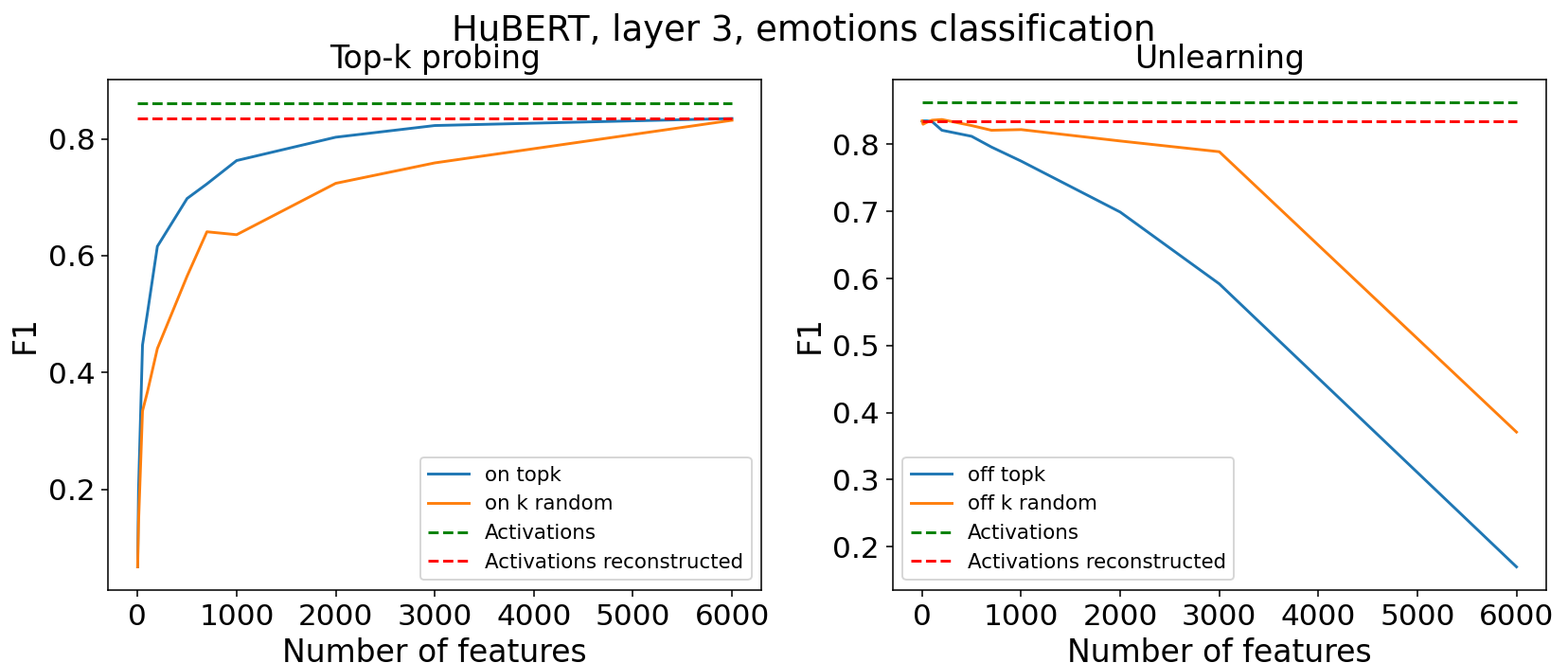}
\caption{Top-k probing and unlearning for four classification tasks}
\label{fig:classification_full}
\end{figure*}

All results are presented in Fig. \ref{fig:classification_full}

\section{Vowel unlearning details}
\label{sec:unlearning_app}

\subsection{Technical details}

In our unlearning experiments, we iteratively removed SAE features in order of their discriminative power (estimated by Fisher score) for a particular spoken letter (vowel) and retrained a LogisticRegression classifier after each removal to measure vowel recognition performance on the remaining features. This allows us to track whether the SAE embeddings still retain information about each vowel class. We employed two distinct regularization approaches to examine their impact:

\textbf{Standard Regularization Setting}: Following established practices \citep{jourdan2024tacotargetedconcepterasure, pmlr-v162-ravfogel22a}, we initially used \texttt{LogisticRegression} from \texttt{scikit-learn} with default hyperparameters. However, during preliminary experiments, we encountered convergence issues with the default 100 iterations, which we resolved by increasing \texttt{max\_iter} to 10000. This \texttt{max\_iter} value was maintained throughout all further experiments.

\textbf{No Regularization Setting}: We conducted our main experiments, featured in the Fig.~\ref{fig:unlearning_vowels_main}, using \texttt{LogisticRegression} with no regularization (\texttt{penalty='none', solver='newton-cg'}) and \texttt{max\_iter=10000}. We find this unregularized approach preferable, as it provides the most rigorous test of information erasure by allowing the classifier to fully exploit any remaining information in the features without the artificial constraints imposed by regularization.

Both experimental settings employed a 5:2 train/test split with stratification across speakers and letters, ensuring balanced representation and preventing bias toward specific speakers or phonemes.

\subsection{Unlearning plots for various letters and regularization setups}

\begin{figure*}[b]
\includegraphics[width=0.99\linewidth]{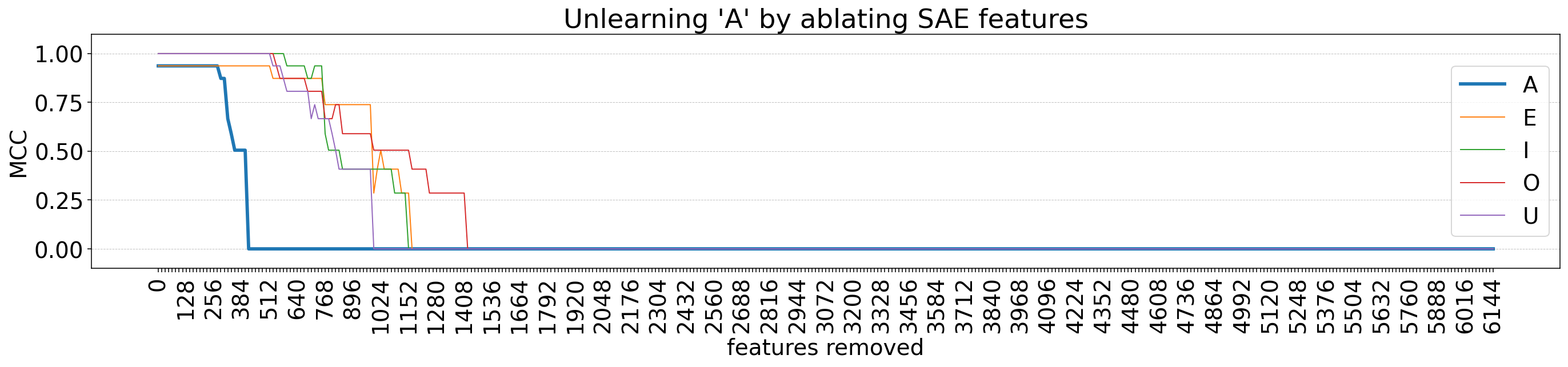}
\includegraphics[width=0.99\linewidth]{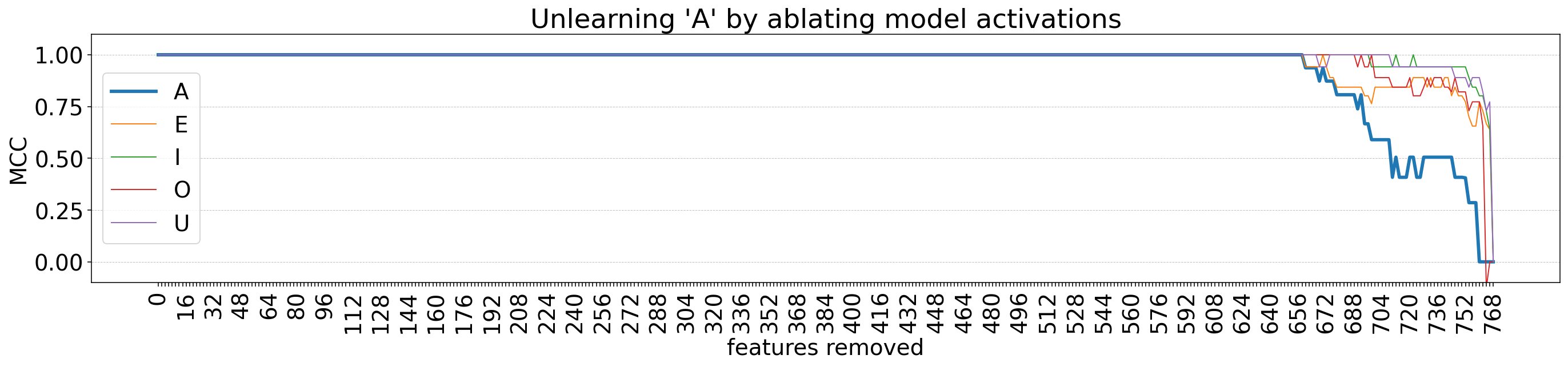}
\includegraphics[width=0.99\linewidth]{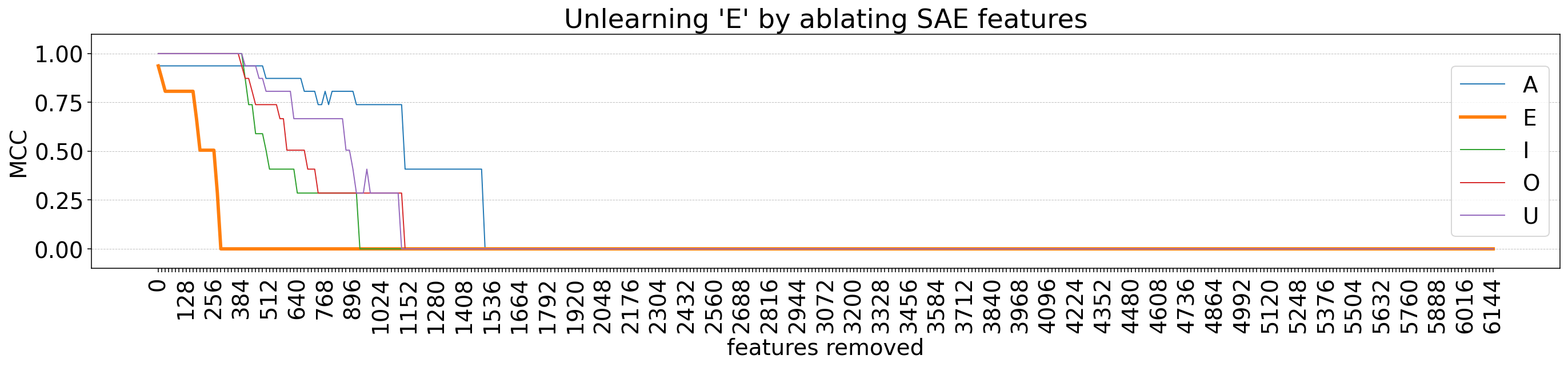}
\includegraphics[width=0.99\linewidth]{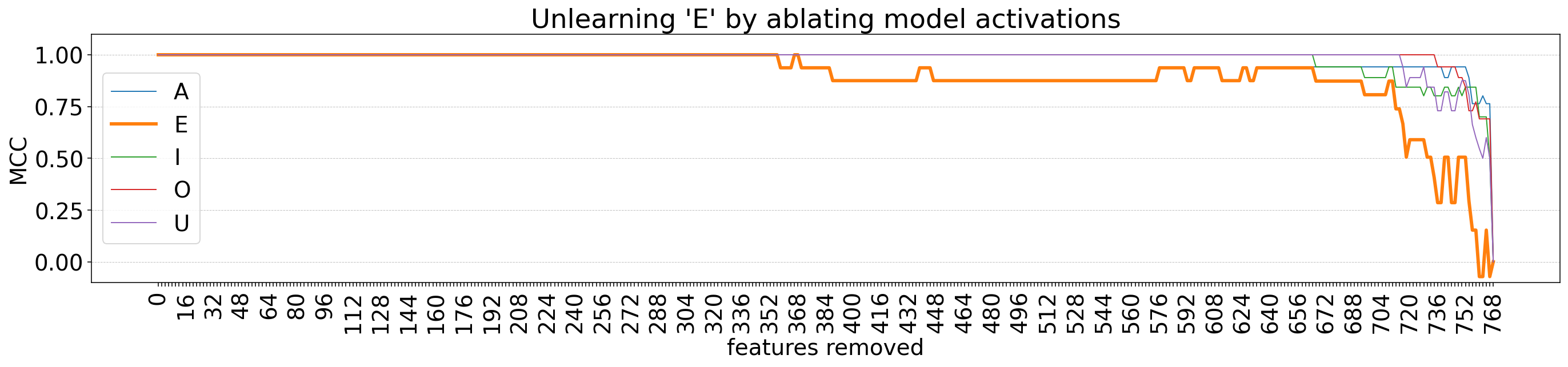}
\includegraphics[width=0.99\linewidth]{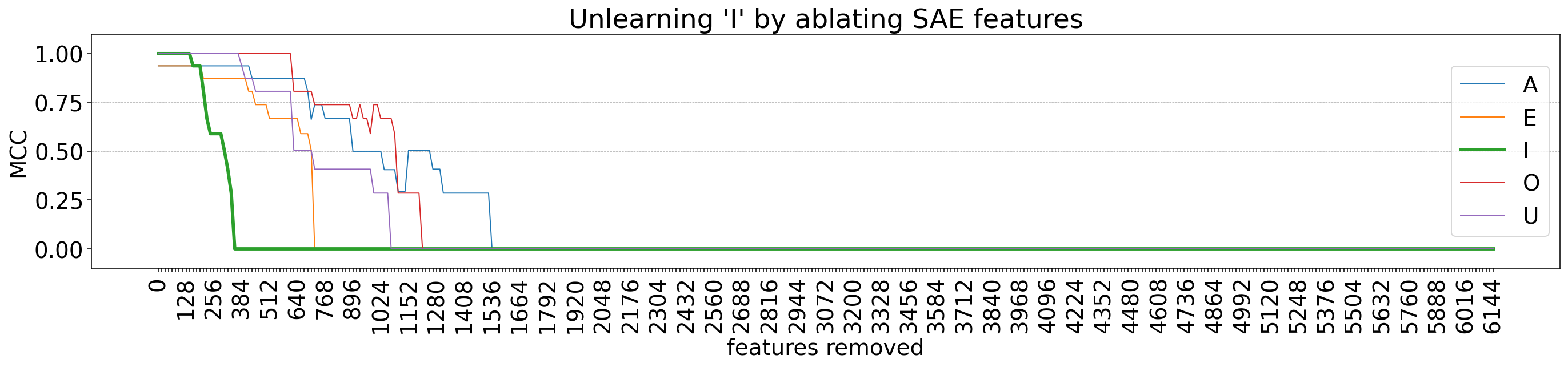}
\includegraphics[width=0.99\linewidth]{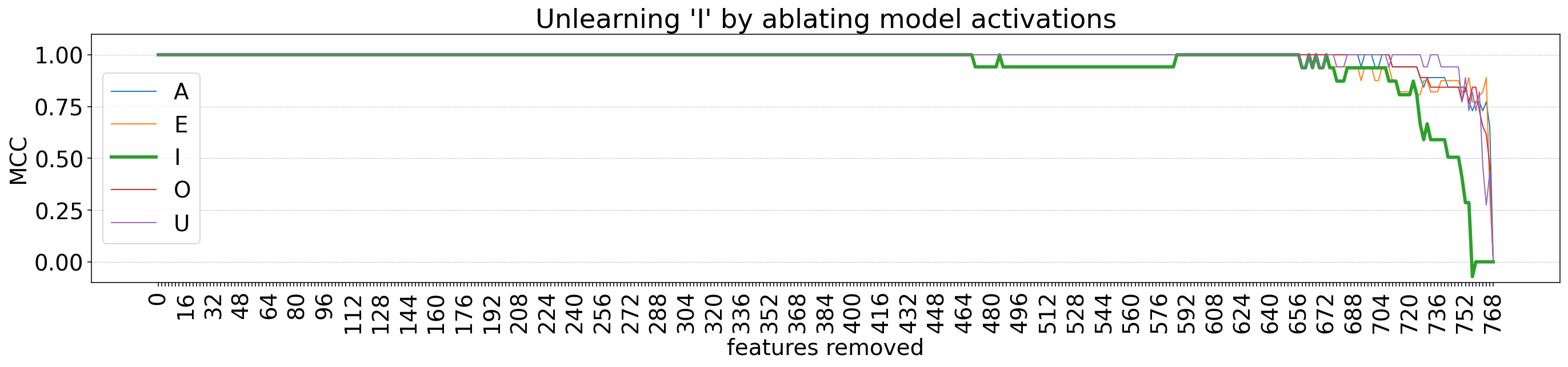}
\caption{Unlearning plots for letters 'A', 'E' and 'I' at the last layer of HuBERT model, using standard \texttt{LogisticRegression} with standard \texttt{penalty='l2'} and \texttt{max\_iter=10000}}
\label{fig:unlearning_app_11_1}
\end{figure*}

\begin{figure*}[b]
\includegraphics[width=0.99\linewidth]{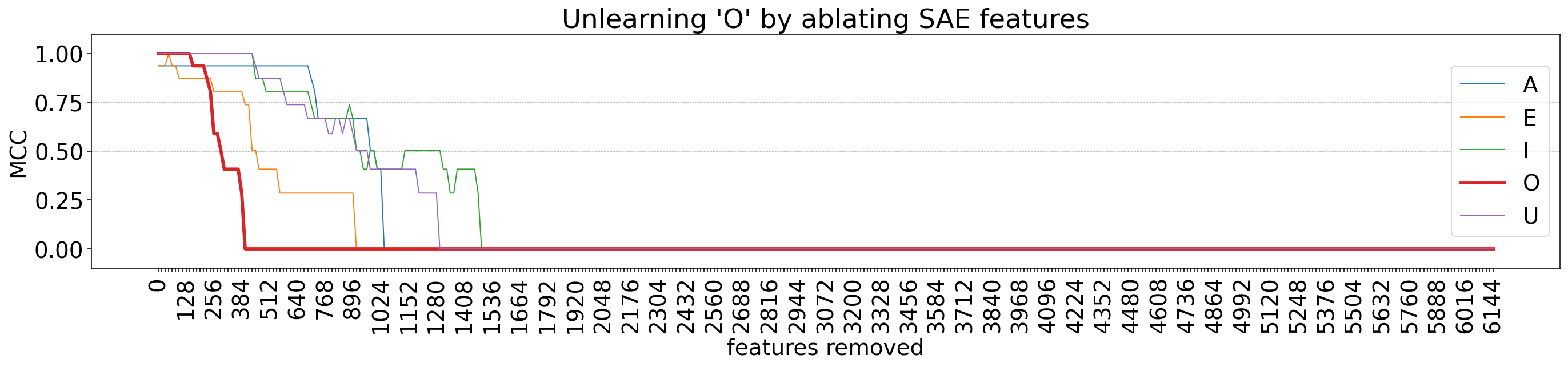}
\includegraphics[width=0.99\linewidth]{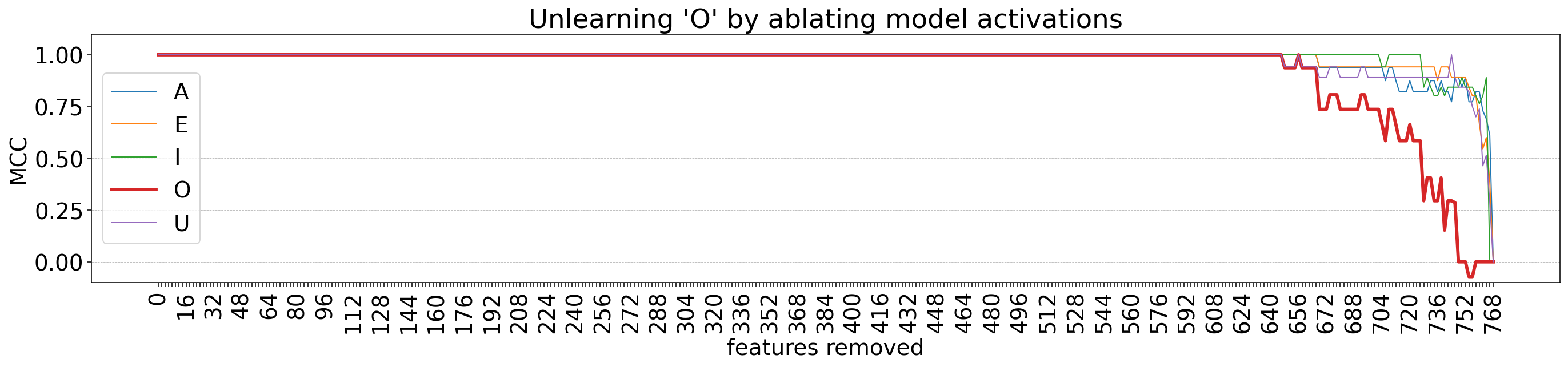}
\includegraphics[width=0.99\linewidth]{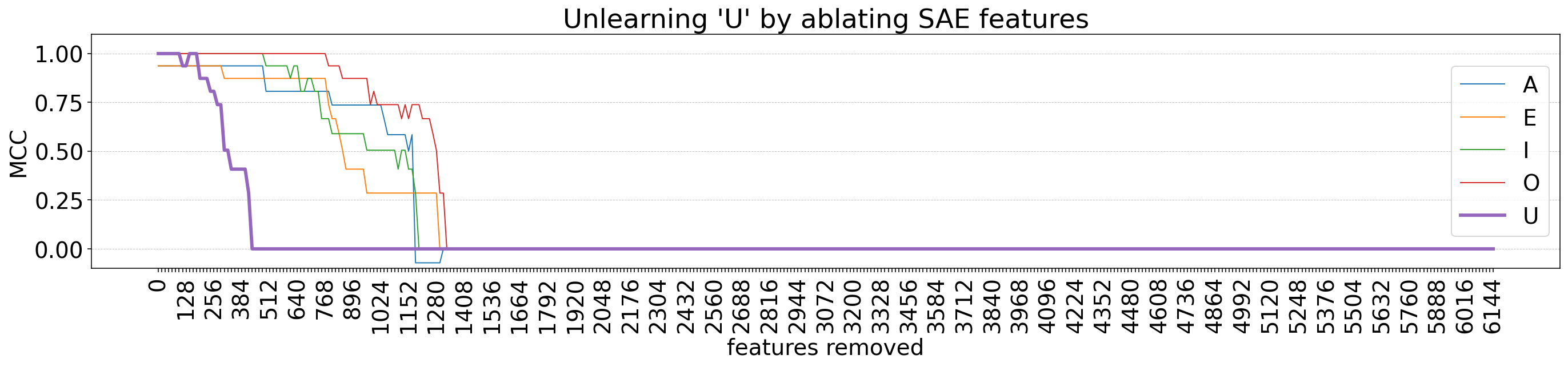}
\includegraphics[width=0.99\linewidth]{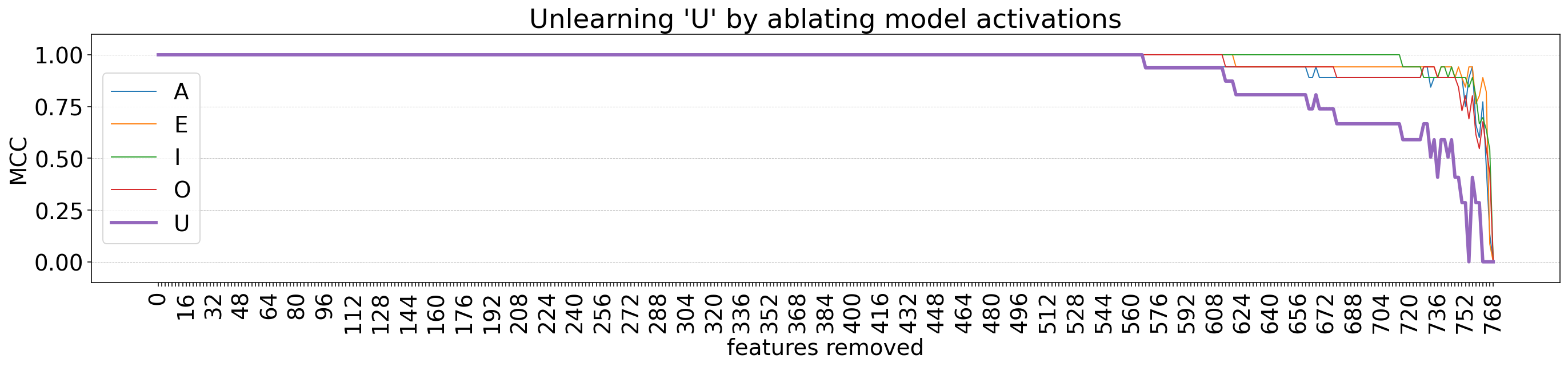}
\caption{Unlearning plots for letters 'O' and 'U' at the last layer of HuBERT model, using standard \texttt{LogisticRegression} with standard \texttt{penalty='l2'} and \texttt{max\_iter=10000}}
\label{fig:unlearning_app_11_2}
\end{figure*}

\begin{figure*}[b]
\includegraphics[width=0.99\linewidth]{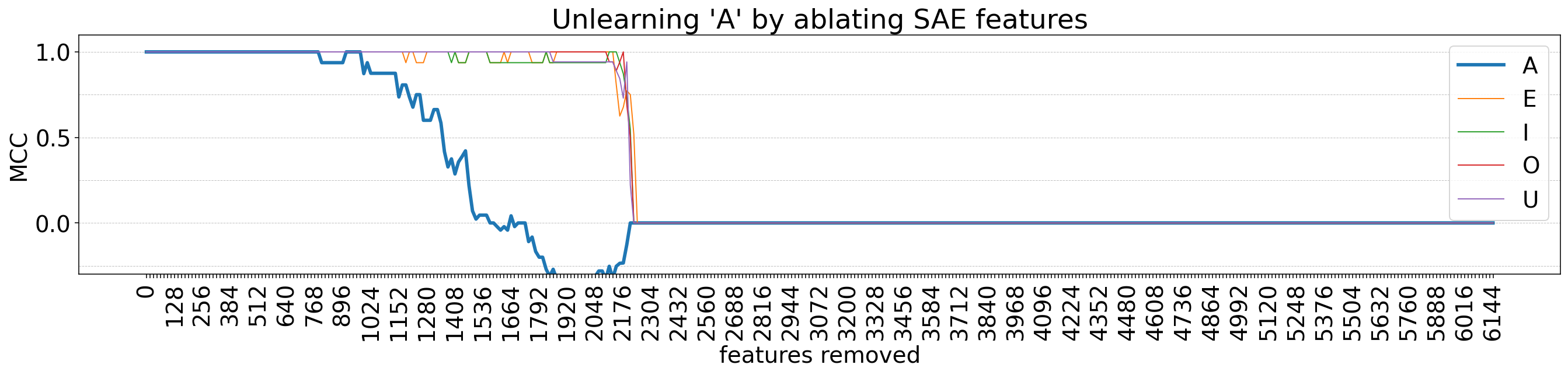}
\includegraphics[width=0.99\linewidth]{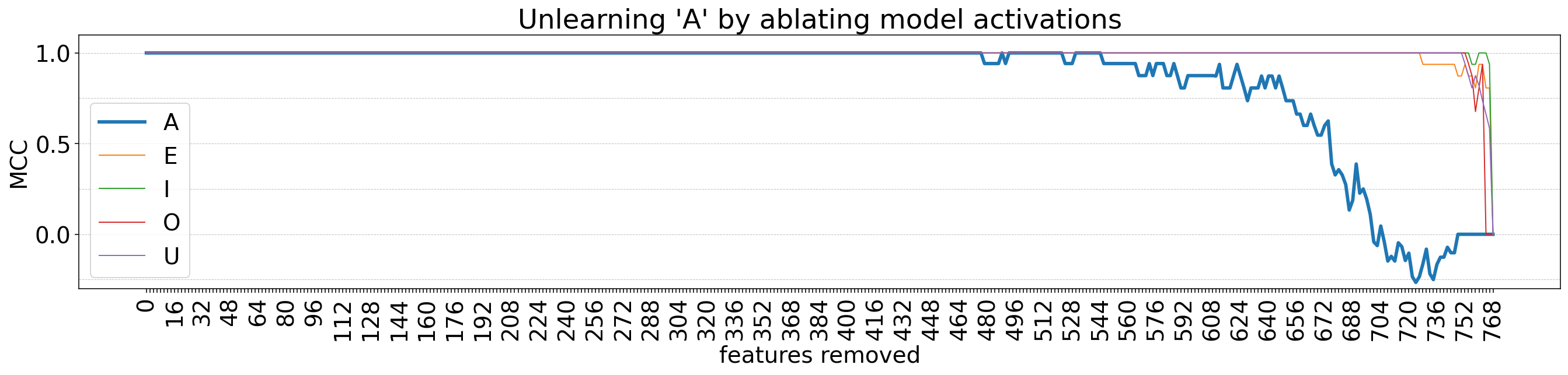}
\includegraphics[width=0.99\linewidth]{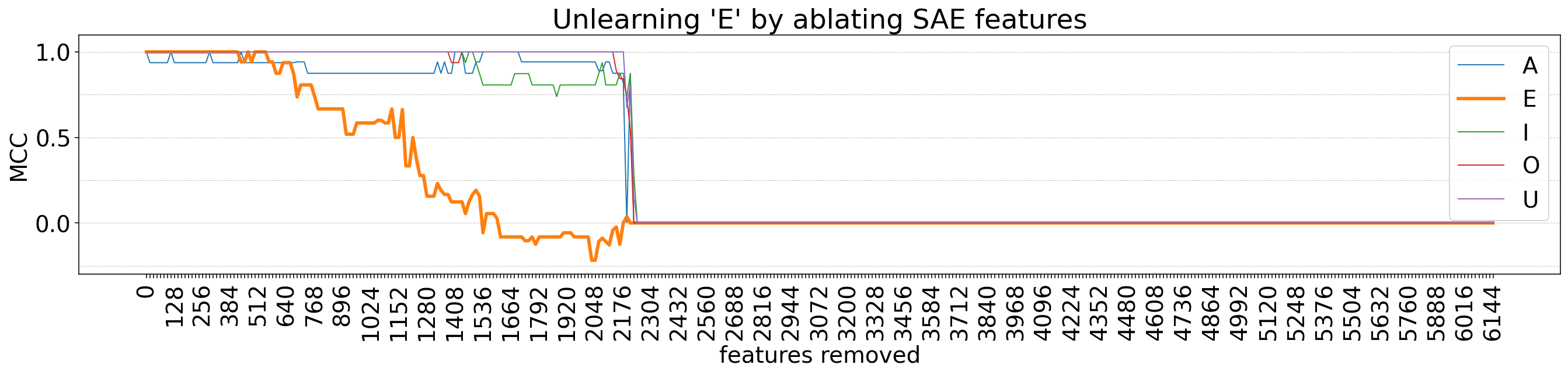}
\includegraphics[width=0.99\linewidth]{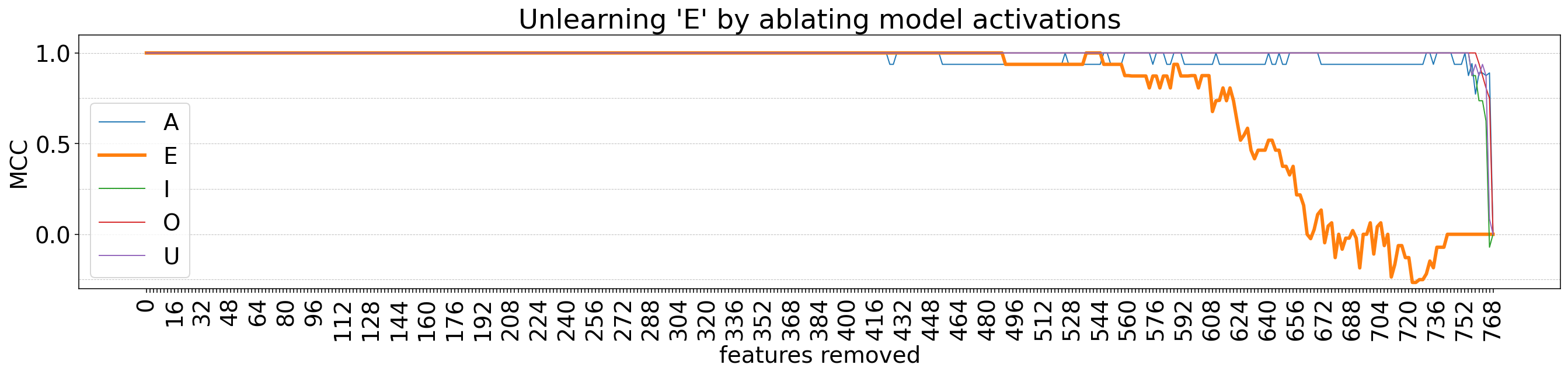}
\includegraphics[width=0.99\linewidth]{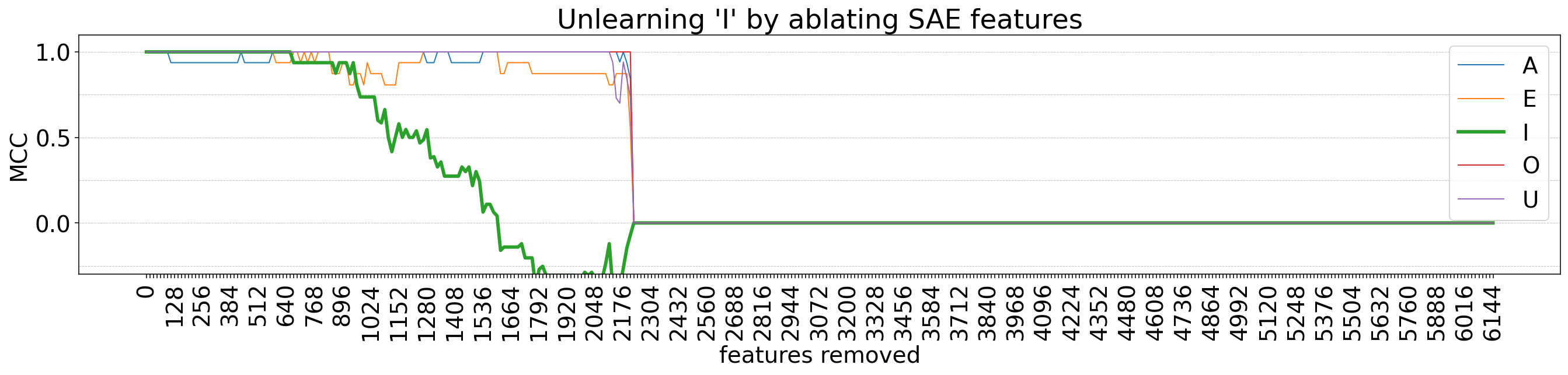}
\includegraphics[width=0.99\linewidth]{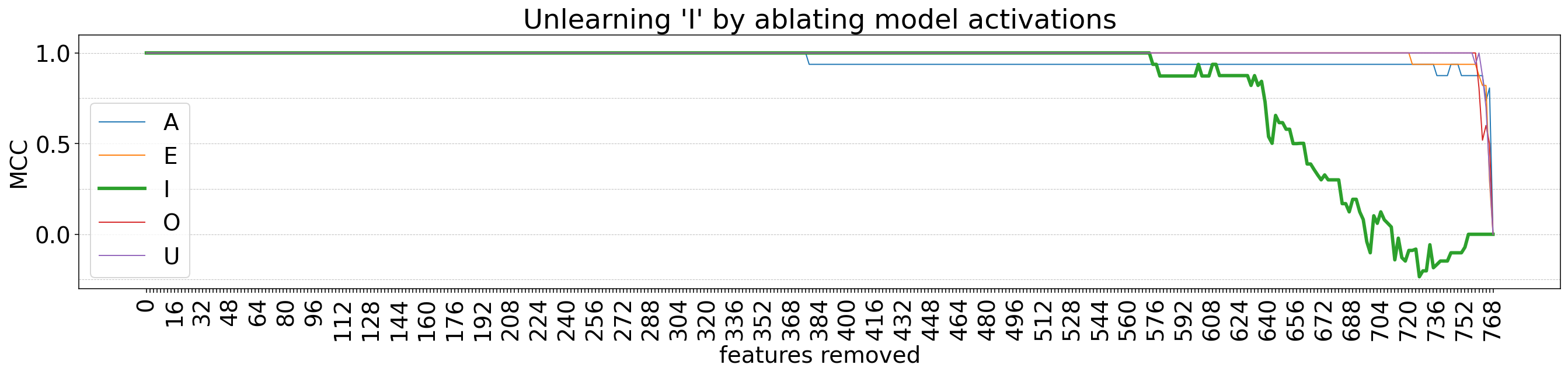}
\caption{Unlearning plots for letters 'A', 'E' and 'I' at the last layer of HuBERT model, using \texttt{LogisticRegression} without regularization and \texttt{max\_iter=10000}}
\label{fig:unlearning_app_11_1_no_reg}
\end{figure*}

\begin{figure*}[b]
\includegraphics[width=0.99\linewidth]{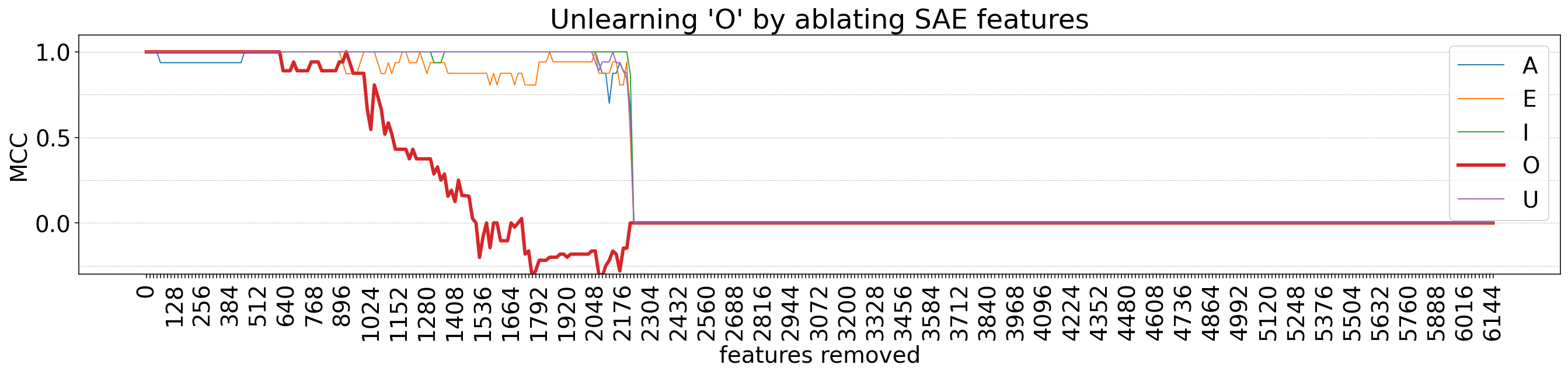}
\includegraphics[width=0.99\linewidth]{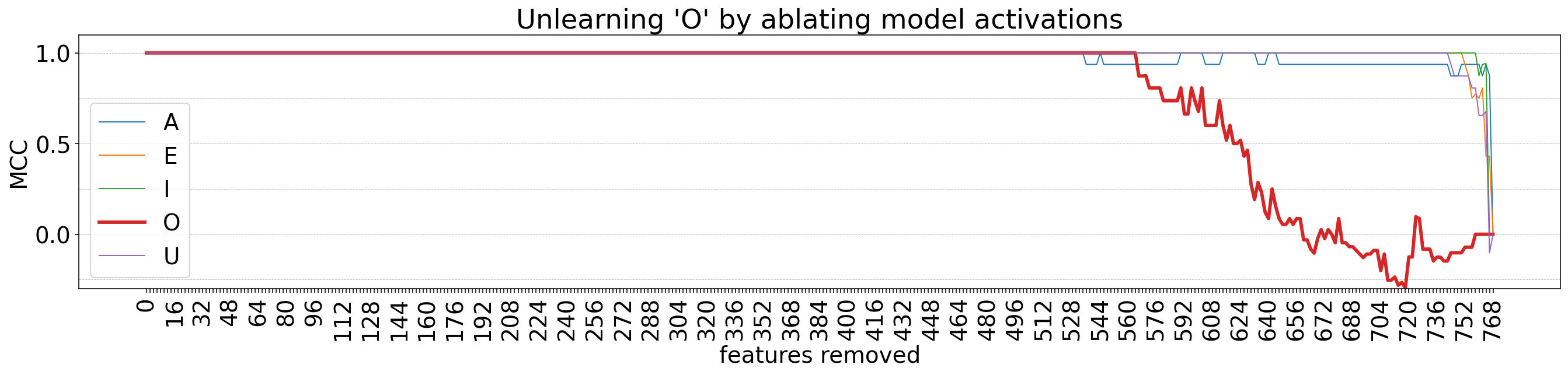}
\includegraphics[width=0.99\linewidth]{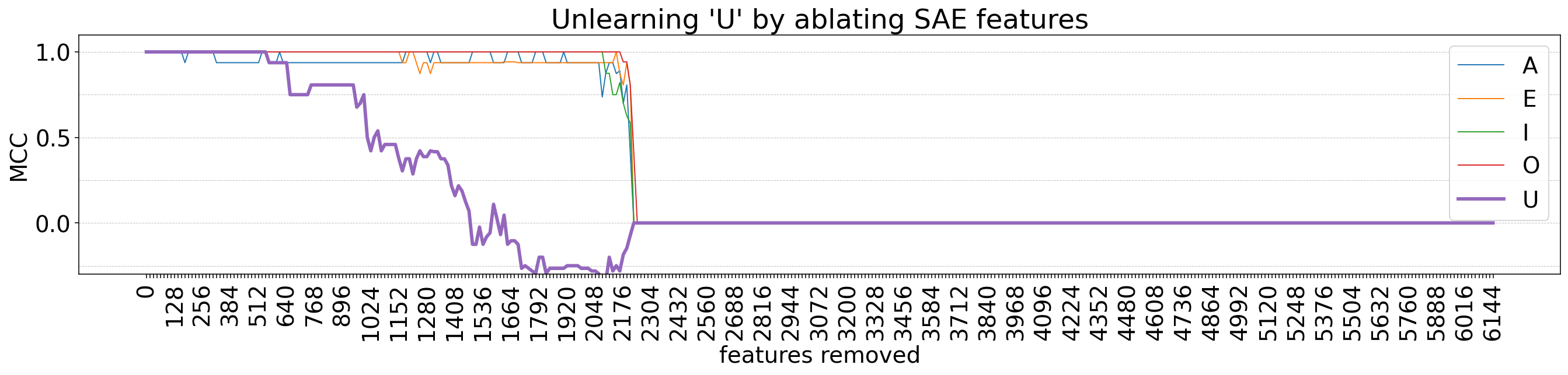}
\includegraphics[width=0.99\linewidth]{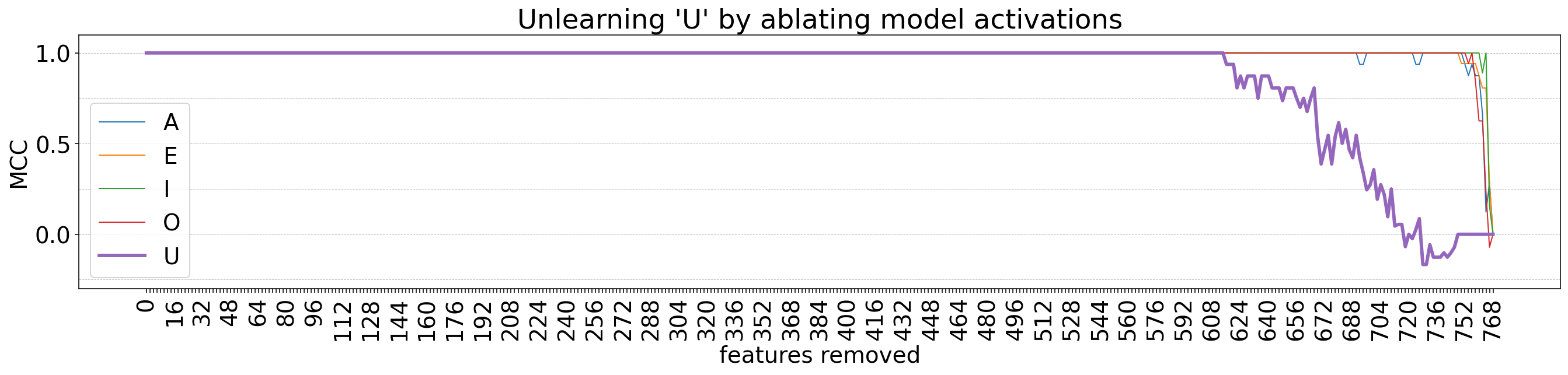}
\caption{Unlearning plots for letters 'O' and 'U' at the last layer of HuBERT model, using \texttt{LogisticRegression} without regularization and \texttt{max\_iter=10000}}
\label{fig:unlearning_app_11_2_no_reg}
\end{figure*}

Fig.~\ref{fig:unlearning_app_11_1} and \ref{fig:unlearning_app_11_2} present vowel unlearning experiments for HuBERT's 12th layer using standard \texttt{LogisticRegression} with default \texttt{l2} regularization, default value \texttt{C=1}, and increased \texttt{max\_iter=10000}.
For experiments without regularization (\texttt{penalty='none'}) see  Fig.~\ref{fig:unlearning_app_11_1_no_reg} and \ref{fig:unlearning_app_11_2_no_reg}.

These experiments %
reveal a significant difference: unregularized logistic regression requires removal of over 1000 features for successful unlearning, while logistic regression with standard L2 regularization achieve comparable results with only 160–400 features (3–6\% of the total). This is even fewer than the features required when working with original HuBERT activations. 

However, we caution that these L2 regularization results may be overly optimistic: regularization artificially constrains the classifier's capacity to extract information, potentially masking the presence of recoverable information rather than confirming its absence. Unregularized classifiers, which can fully exploit all available patterns, provide a more realistic assessment of true information removal, suggesting that genuine unlearning requires the more extensive feature removal observed in our unregularized experiments. In the same time, unregularized classifiers have their own limitations in our experimental setting: since we have more features than training samples, they exhibit poor convergence and numerical instability, making them challenging to work with despite providing more rigorous tests of information erasure.

\subsection{k-probing vowels}

\begin{figure*}[b]
\includegraphics[width=0.49\linewidth]{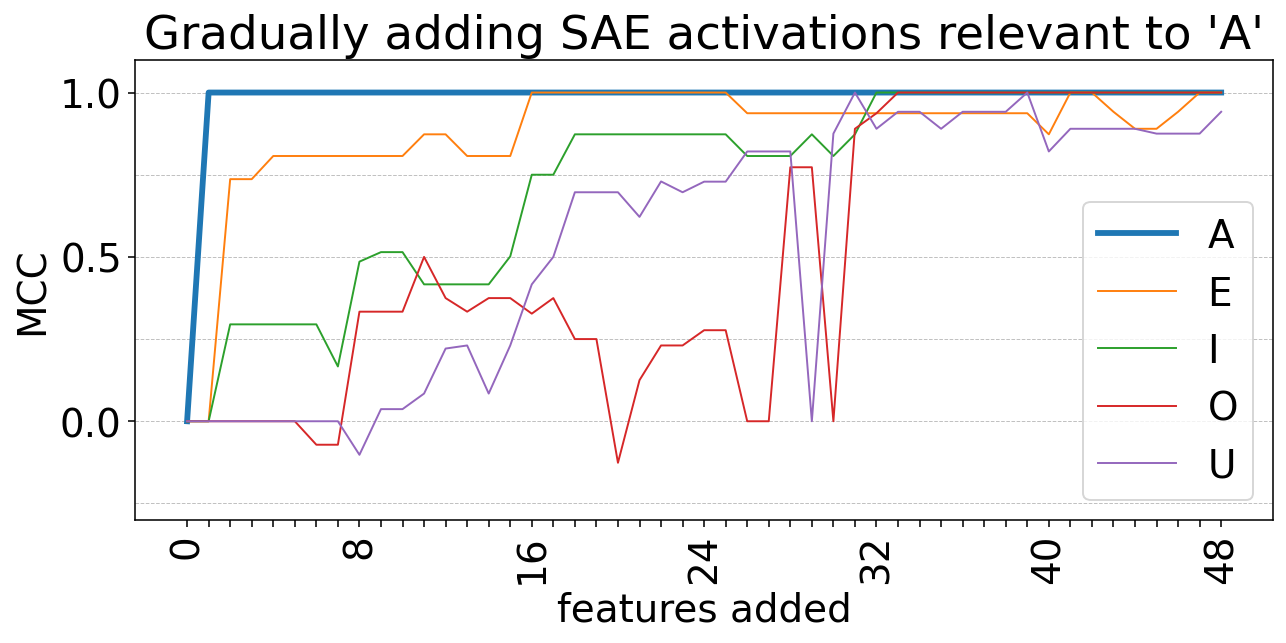}
\includegraphics[width=0.49\linewidth]{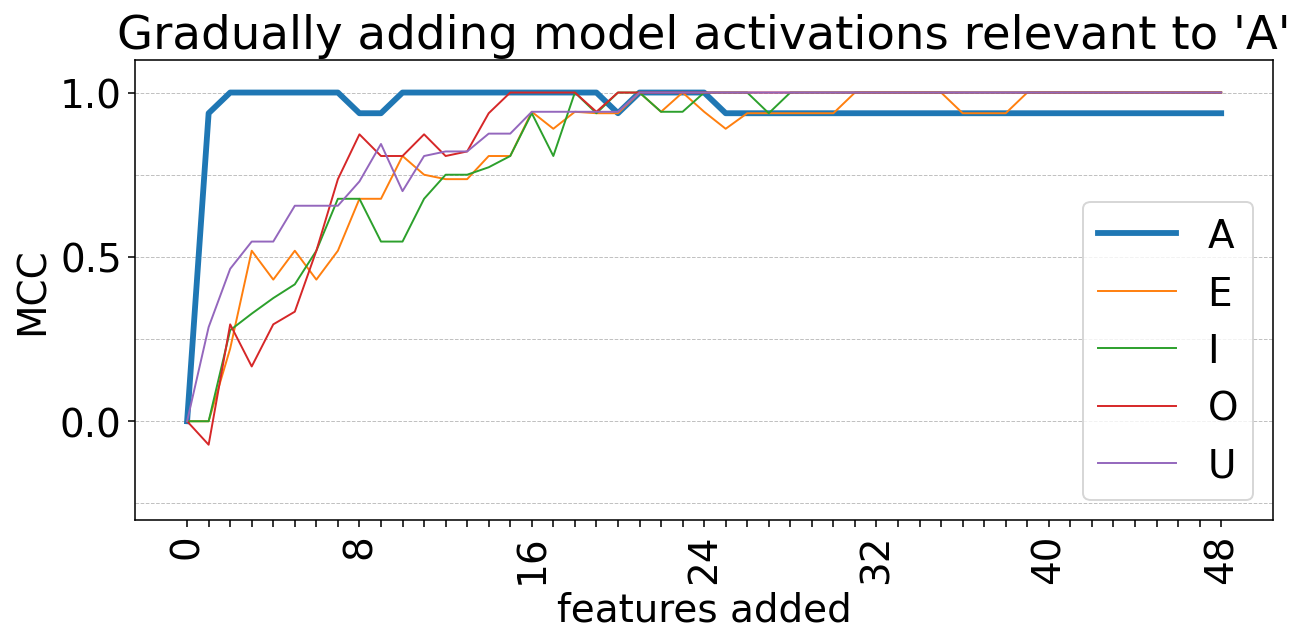}
\includegraphics[width=0.49\linewidth]{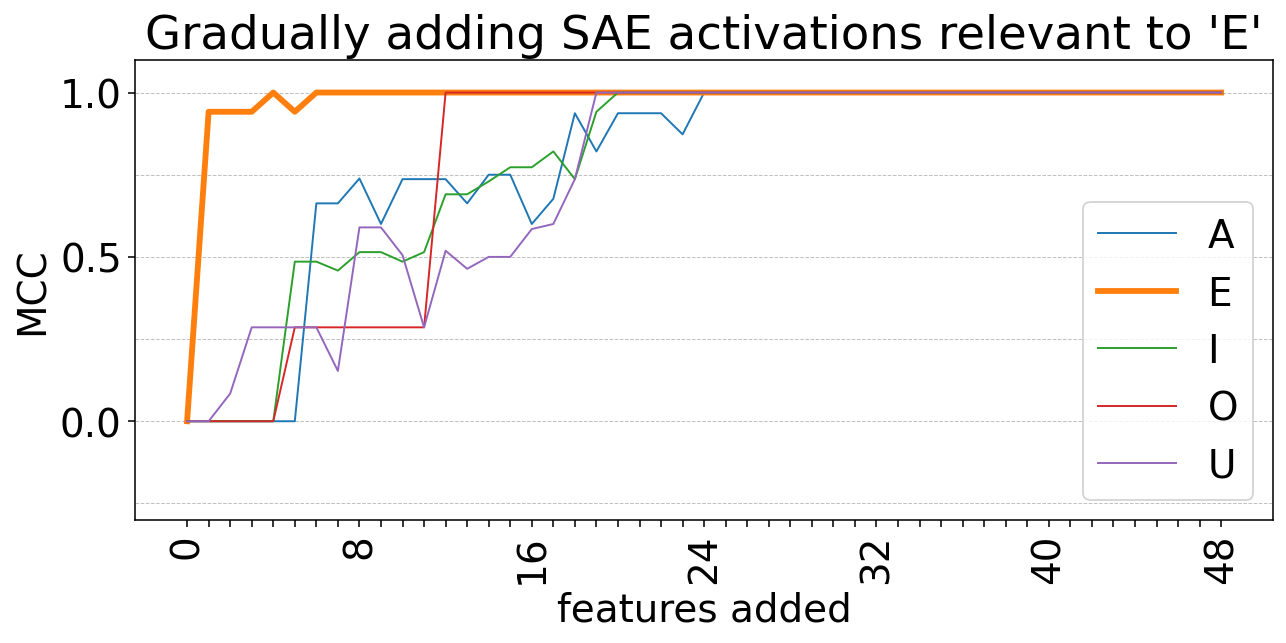}
\includegraphics[width=0.49\linewidth]{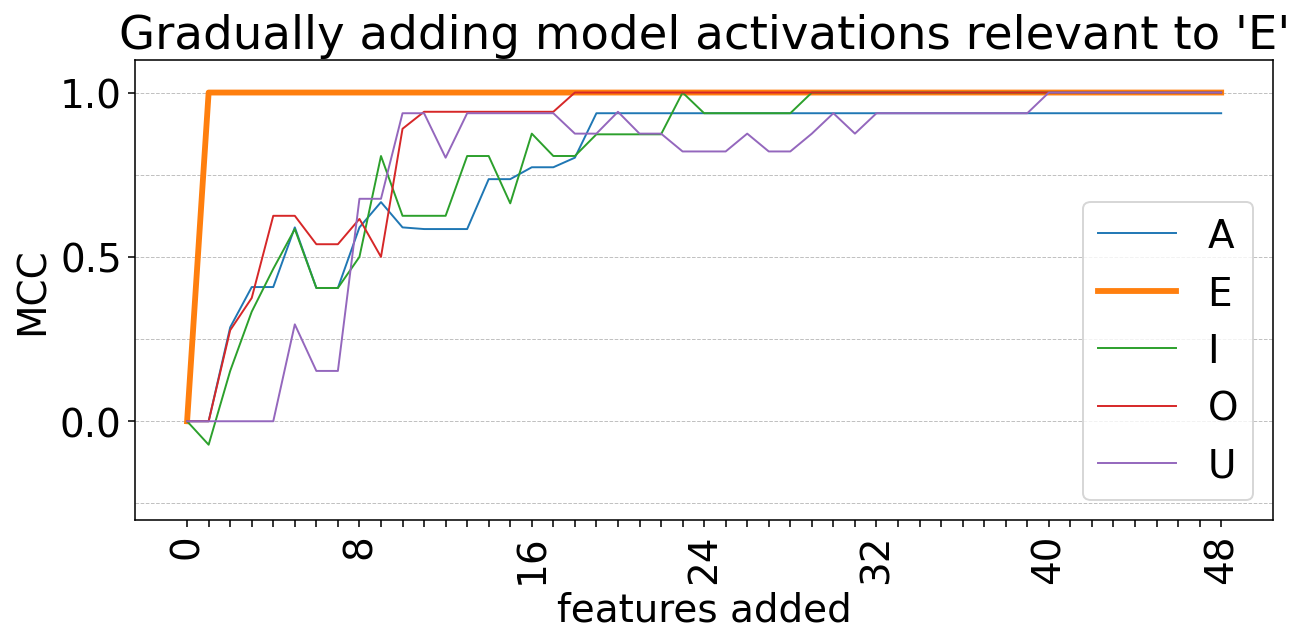}
\includegraphics[width=0.49\linewidth]{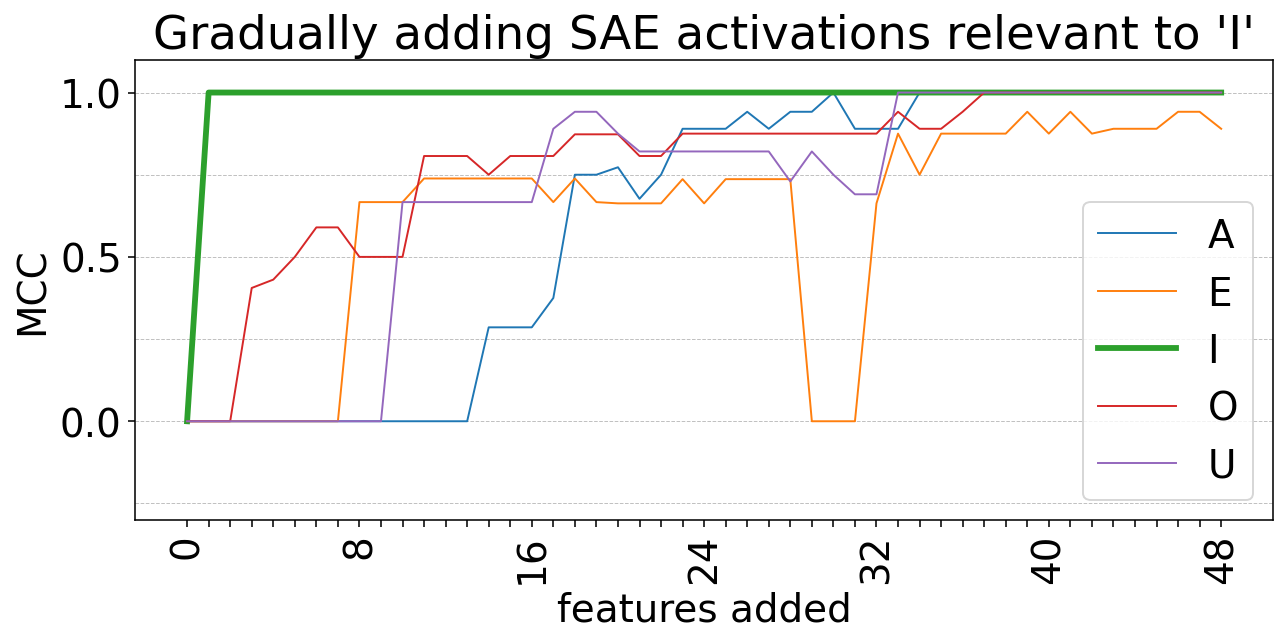}
\includegraphics[width=0.49\linewidth]{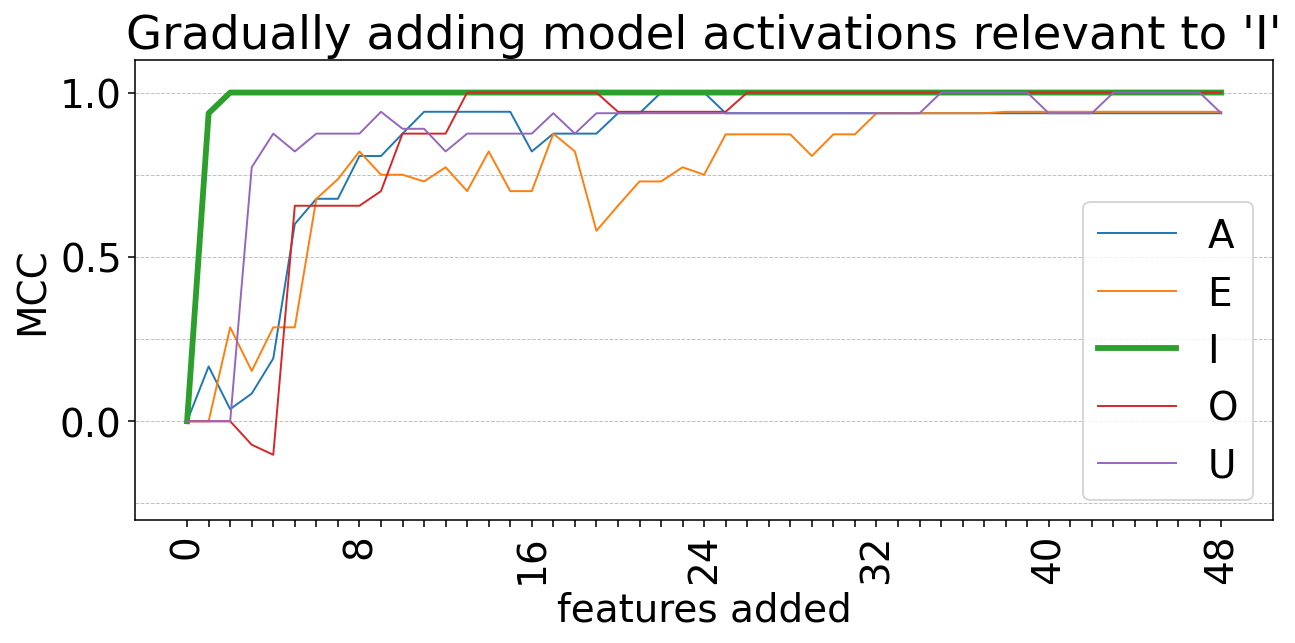}
\includegraphics[width=0.49\linewidth]{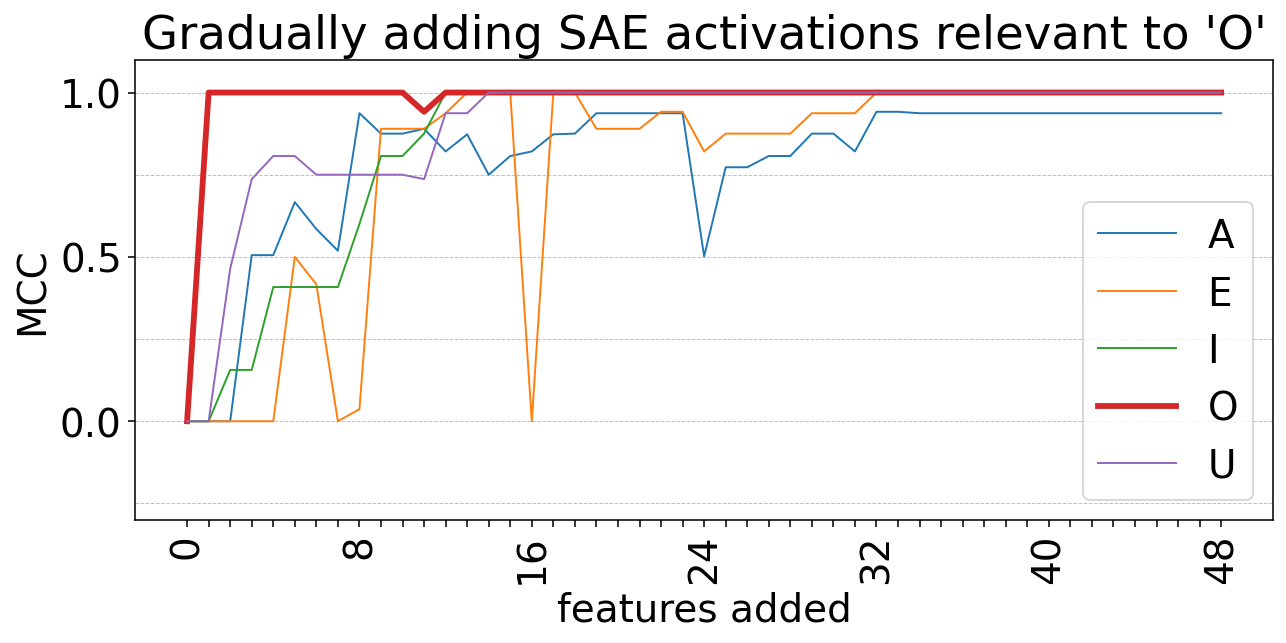}
\includegraphics[width=0.49\linewidth]{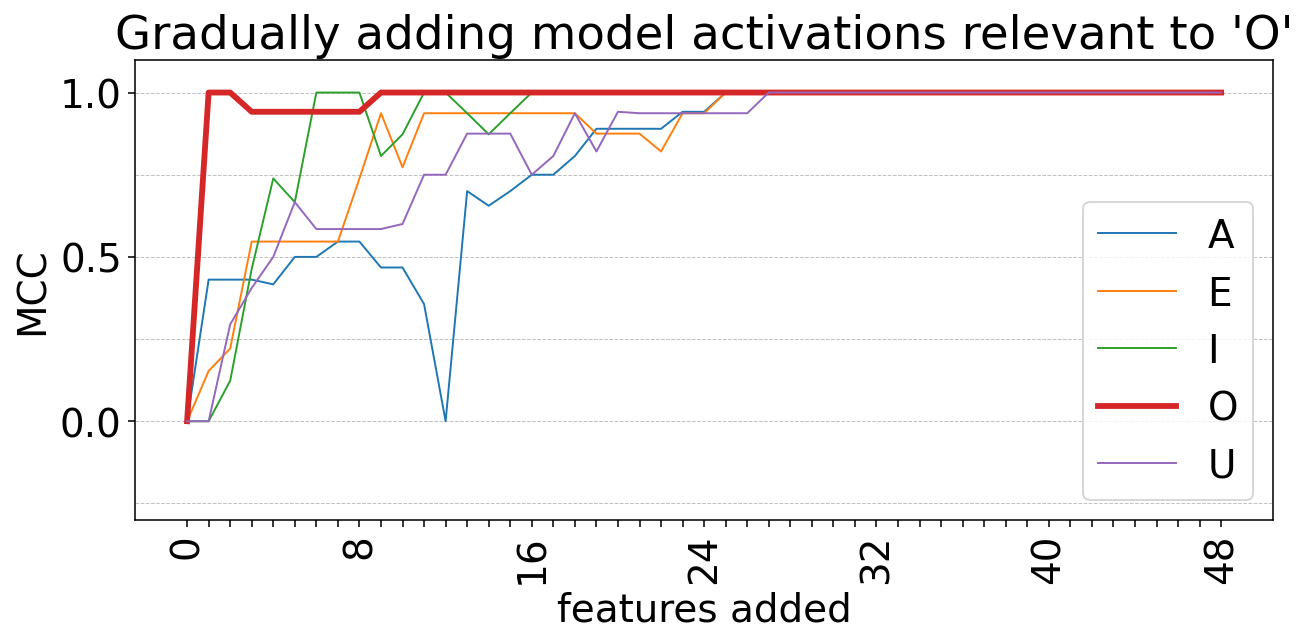}
\includegraphics[width=0.49\linewidth]{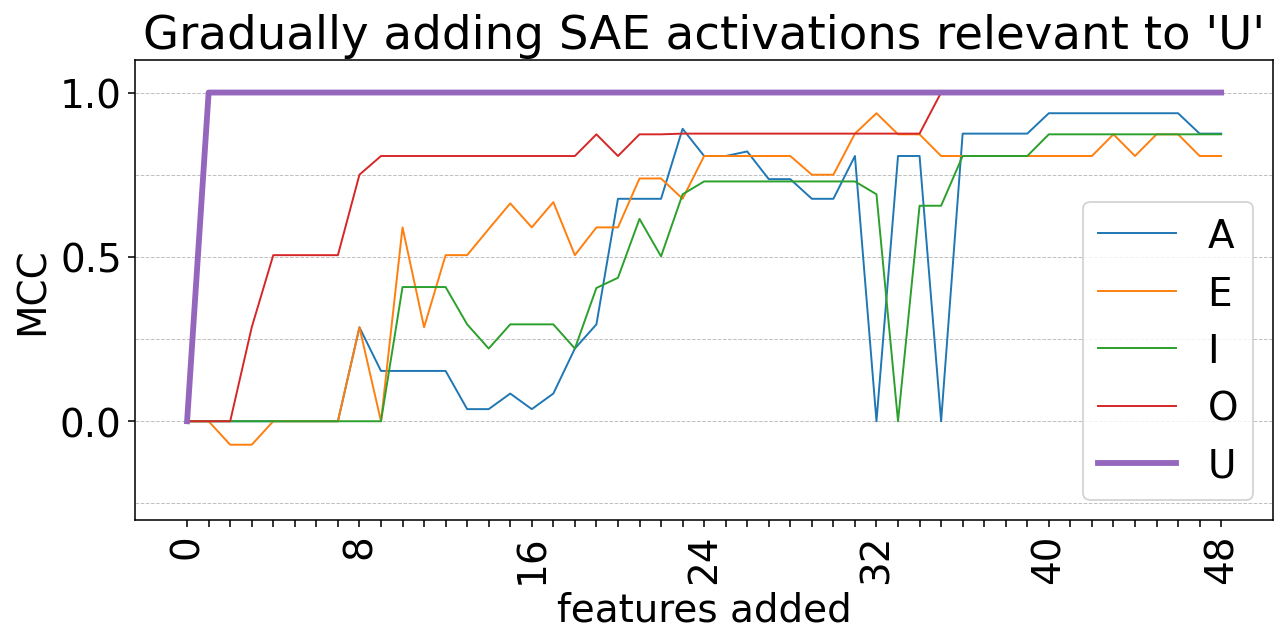}
\includegraphics[width=0.49\linewidth]{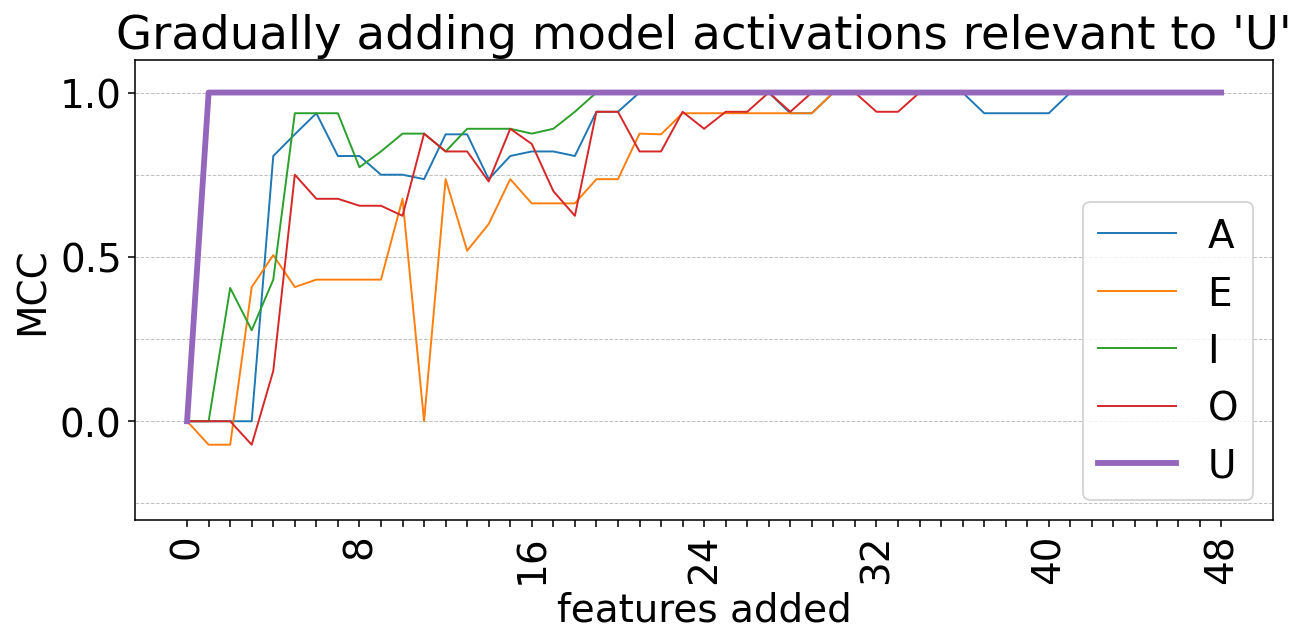}
\caption{K‑probe vowel classification at layer 9. The curves show classification accuracy as the most informative features are added sequentially. Only the first 49 features are displayed; beyond this point, accuracy approaches perfect accuracy for all vowels.}
\label{fig:k_probing_vowels_app}
\end{figure*}

In addition to the experiments where we progressively removed features, we ran a ``reverse'' series of tests in which features were added one by one—starting with the most informative according to the Fisher score and then adding less important ones. We discovered that to regain high accuracy in recognizing a single vowel against the others, it was enough to activate just one or two of the top‑ranked features in both SAE activations and HuBERT embeddings (see Fig.~\ref{fig:k_probing_vowels_app}. This indicates that the highest Fisher-ranked features carry enough of the phonetic information for a reliable classification. For this experiment we also used Logistic Regression without regularization.

\section{Interpretation by labels}
\label{appendix:clf_by_label}

The feature search procedure is as follows: ($1$) identify all latents that are activated on samples with the target labels; ($2$) for each latent, evaluate the F1 score at different thresholds in steps of $0.1$ across the interval from its minimum to maximum activation value; ($3$) if the F1 score at any threshold exceeds $0.5$, consider that feature correlates with the target label.

\begin{figure*}[htb]

\subfloat[HuBERT, layer 4, "laughter"]{\includegraphics[width=0.5\linewidth]{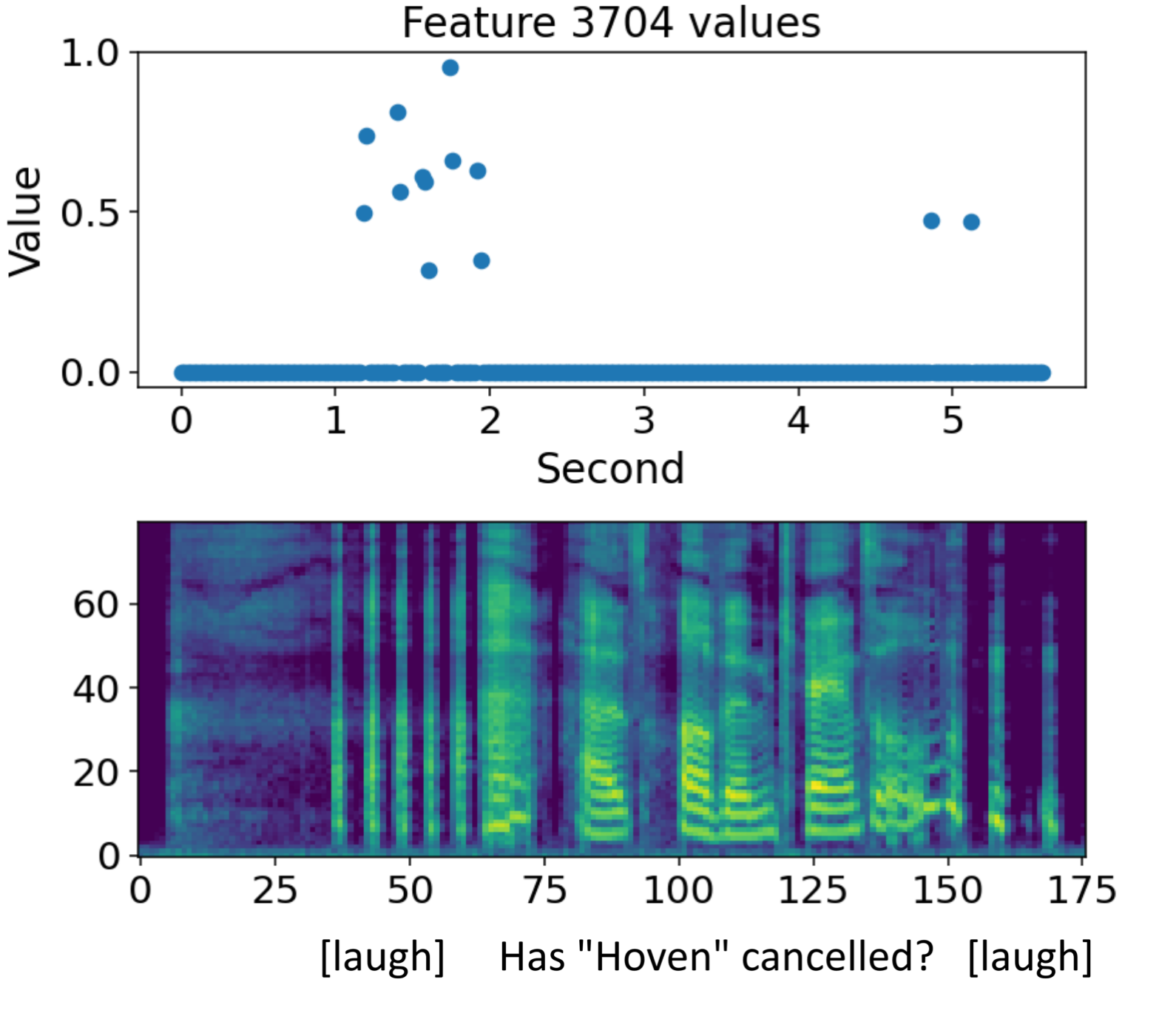}}
\subfloat[HuBERT, layer 4, "repetitive sounds (e.g. siren, ticking)"]{\includegraphics[width=0.5\linewidth]{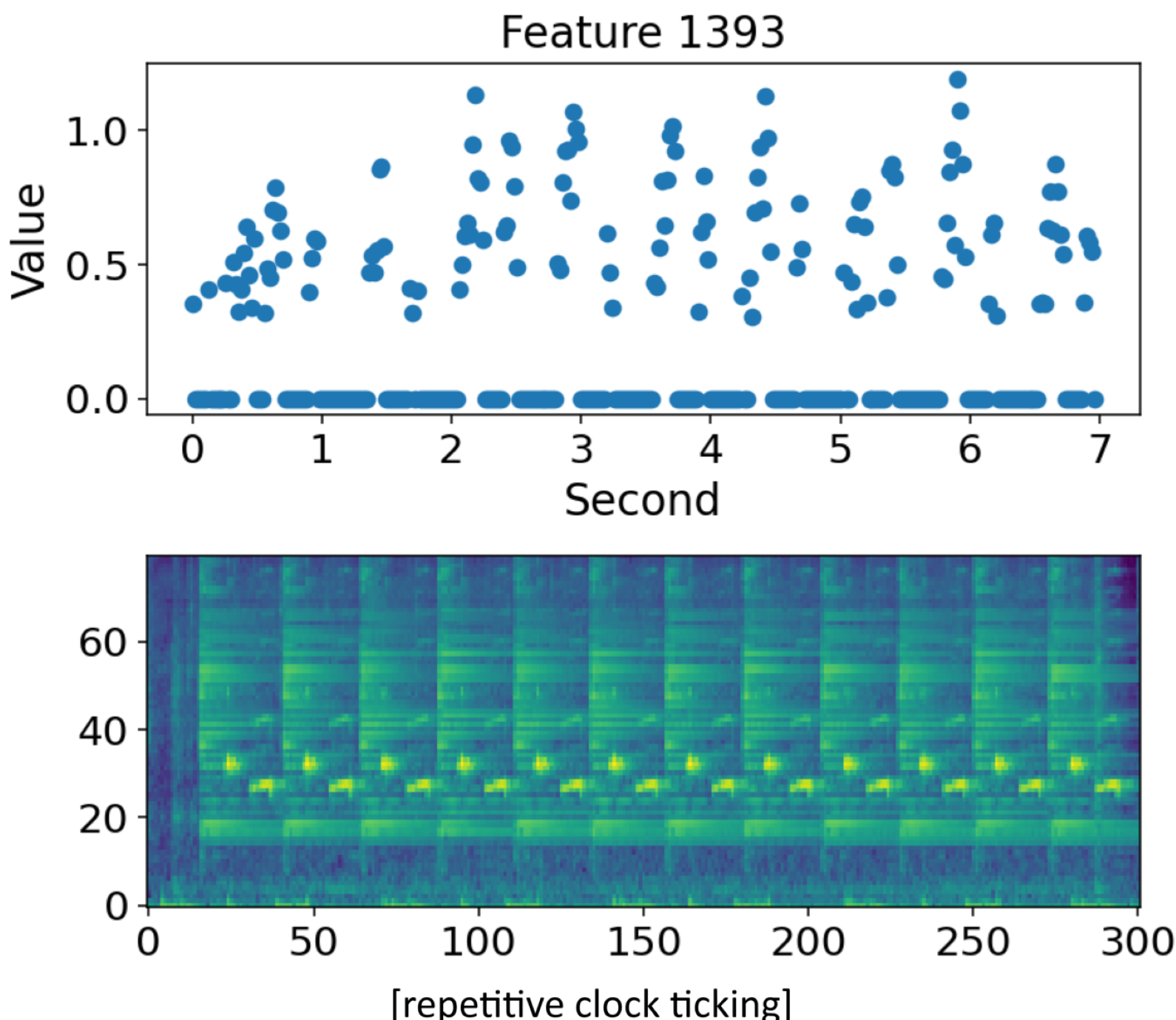 }}
\caption{Additional features found in classification by label experiment.}
\label{fig:clf_by_label}
\end{figure*}

Fig.~\ref{fig:clf_by_label} presents some interesting features found during the experiment and may be illustrated by mel-spectrogram representation and the feature activated frames. The corresponding audio fragments are available on the demo page. 

\section{Auto-interpretation details}
\label{appendix:auto-interpretation}

\begin{figure*}
    \centering \includegraphics[width=1.0\linewidth]{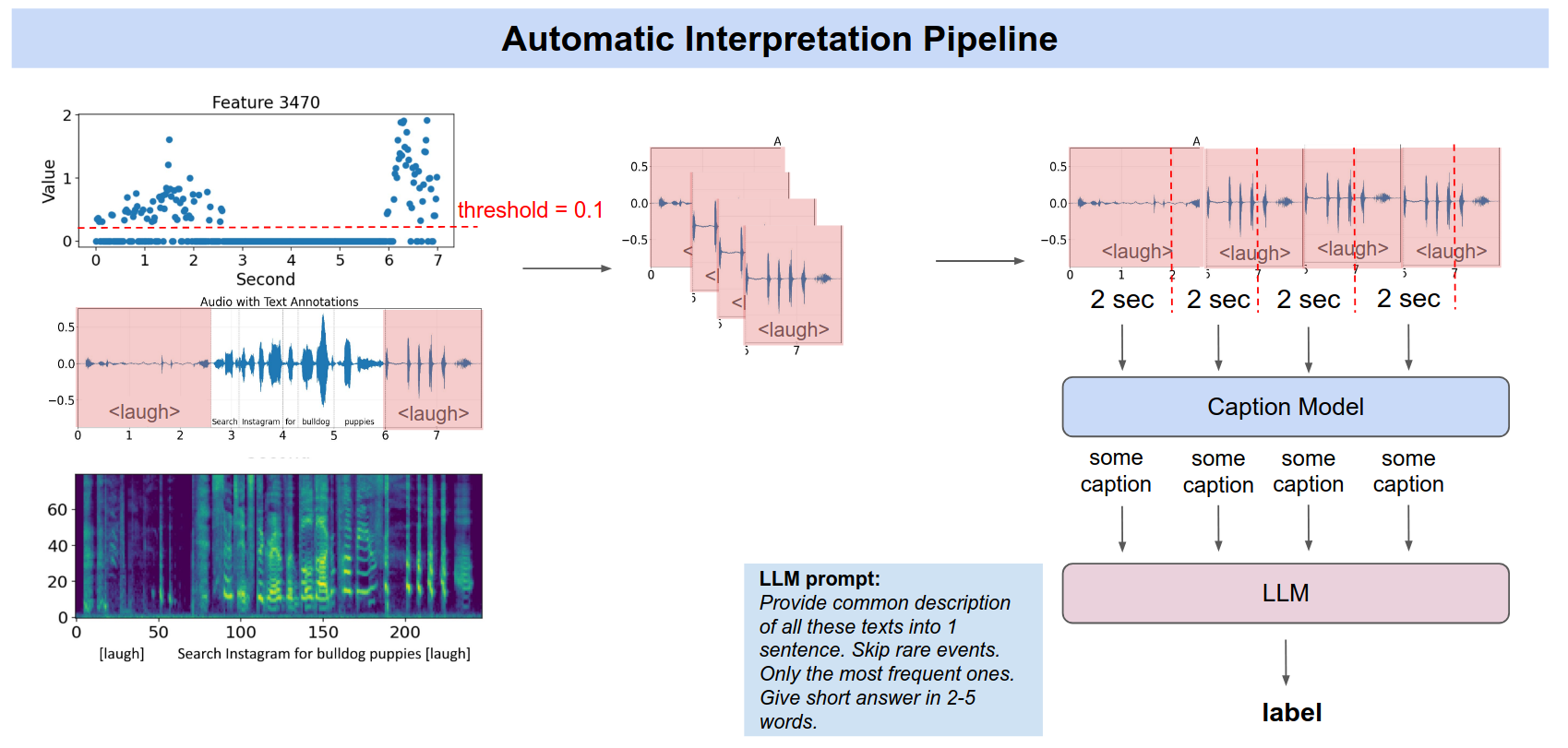}
    \caption{Automatic Interpretation Pipeline}
\label{fig:autointerpretation_pipeline}
\end{figure*}

\begin{table}[t]
    \centering
    \small
    \begin{tabular}{ll|cccc}
        \toprule
        top-$k$ & $\alpha$ & \textbf{Musan} & \textbf{FSD50k} & \textbf{WHAM} \\
        \midrule
        \multicolumn{5}{l}{Musan vec.} \\
        \midrule
        1   & 0.5 & 0.31 & 0.29 & 0.30 \\
        1   & 1.0 & 0.31 & 0.29 & 0.30 \\
        1   & 3.0 & 0.31 & 0.24 & 0.30 \\
        5   & 0.5 & 0.29 & 0.29 & 0.31 \\
        5   & 1.0 & 0.27 & 0.29 & 0.31 \\
        5   & 3.0 & 0.20 & 0.24 & 0.29 \\
        10  & 0.5 & 0.29 & 0.29 & 0.31 \\
        10  & 1.0 & 0.27 & 0.27 & 0.30 \\
        10  & 3.0 & 0.20 & 0.19 & 0.25 \\
        100 & 0.5 & 0.27 & 0.21 & 0.30 \\
        100 & 1.0 & 0.20 & 0.12 & 0.26 \\
        100 & 3.0 & \textbf{0.01} & \textbf{0.00} & \textbf{0.00} \\
        \midrule
        \multicolumn{5}{l}{FSD50k vec.} \\
        \midrule
        1   & 0.5 & 0.24 & 0.23 & 0.24 \\
        1   & 1.0 & 0.24 & 0.23 & 0.24 \\
        1   & 3.0 & 0.24 & 0.20 & 0.23 \\
        5   & 0.5 & 0.23 & 0.23 & 0.24 \\
        5   & 1.0 & 0.21 & 0.23 & 0.24 \\
        5   & 3.0 & 0.16 & 0.20 & 0.23 \\
        10  & 0.5 & 0.23 & 0.23 & 0.24 \\
        10  & 1.0 & 0.21 & 0.21 & 0.24 \\
        10  & 3.0 & 0.16 & 0.14 & 0.19 \\
        100 & 0.5 & 0.20 & 0.15 & 0.23 \\
        100 & 1.0 & \color{ForestGreen}{0.16} & \color{ForestGreen}{0.09} & \color{ForestGreen}{0.20} \\
        100 & 3.0 & \textbf{0.01} & \textbf{0.00} & 0.01 \\
        \midrule
        \multicolumn{5}{l}{WHAM vec.} \\
        \midrule
        1   & 0.5 & 0.36 & 0.36 & 0.36 \\
        1   & 1.0 & 0.36 & 0.35 & 0.36 \\
        1   & 3.0 & 0.36 & 0.32 & 0.35 \\
        5   & 0.5 & 0.36 & 0.36 & 0.36 \\
        5   & 1.0 & 0.35 & 0.35 & 0.36 \\
        5   & 3.0 & 0.32 & 0.32 & 0.33 \\
        10  & 0.5 & 0.36 & 0.34 & 0.35 \\
        10  & 1.0 & 0.35 & 0.32 & 0.35 \\
        10  & 3.0 & 0.32 & 0.24 & 0.30 \\
        100 & 0.5 & 0.34 & 0.25 & 0.34 \\
        100 & 1.0 & 0.32 & 0.13 & 0.29 \\
        100 & 3.0 & 0.04 & \textbf{0.00} & \textbf{0.00} \\
        \bottomrule
    \end{tabular}
    \caption{False Positive Rate (FPR) for SAE steering with different configurations, full version of the Table~\ref{tab:steering-sae-topk-alpha-tuning}. Tuning the scaling factor $\alpha$ and the number of top-$k$ most informative SAE features selected from the hallucination classifier. Dataset specific columns represents the dataset on which FPR was calculated. Table is divided into 3 sections, each section refers to a specific dataset on which the SAE steering vector was formed.}
    \label{tab:steering-sae-topk-alpha-tuning-full}
\end{table}

\begin{figure} %
    \centering
    \begin{subfigure}{0.5\textwidth}
        \centering
        \includegraphics[width=\linewidth]{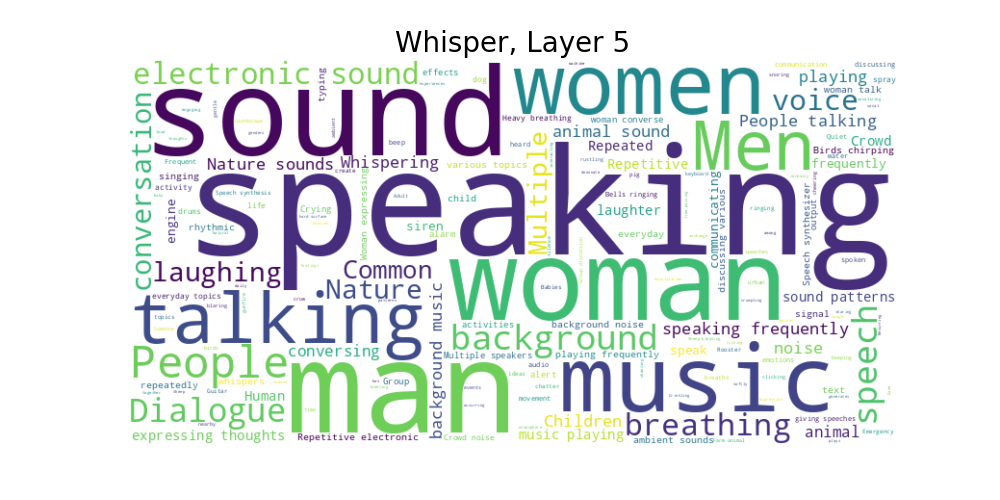}
        \label{fig:wordcloud}
    \end{subfigure}
    \begin{subfigure}{0.5\textwidth}
        \centering
        \includegraphics[width=\linewidth]{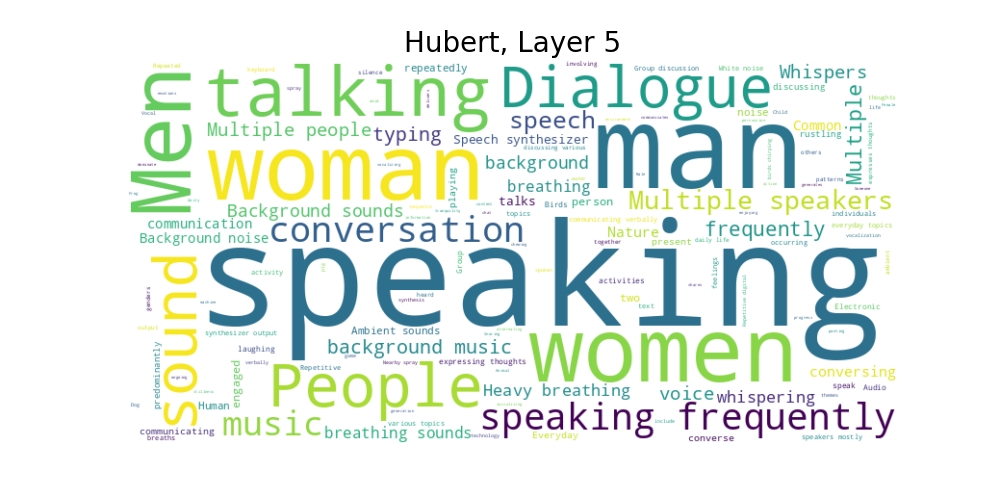}
        \label{fig:frequencies}
    \end{subfigure}
    \caption{Features label frequencies for Whisper and Hubert models.}
    \label{fig:feature_wordmap}
\end{figure}

Fig.\ref{fig:feature_wordmap} presents a word map of characteristic labels, where for both Hubert and Whisper models, the dominant interpretation is related to speech. However, the following limitations should be considered. First, the threshold value of 0.1 was empirically selected. Second, the test data set consists largely of speech data. This bias toward speech may cause small but frequent activations of features, pushing the resulting label toward a speech interpretation and obscuring rarer music or sound events. Additional limitation of this method is its dependence on the capabilities of the underlying audio captioning. This is particularly evident when interpreting phoneme-level features. For instance, a feature responsible for the vowel sound "A" will produce audio chunks consisting of various people producing that isolated sound. This lack of broader acoustic context often confuses the captioning model, which may then misclassify the sound as generic multi-speaker dialogue rather than identifying the specific phoneme.

\section{Steering details}\label{appendix:steering}

As a baseline we propose to mitigate hallucinations in the Whisper speech recognition model by applying steering vectors to its internal activations.
The steering vector is derived by contrasting latent representations of hallucinatory and non-hallucinatory samples, with labels automatically assigned based on Whisper's internal no-speech probability score. 

The core hypothesis is that a direction in the activation space can be identified that, when amplified, suppresses the model's tendency to generate spurious transcriptions for non-speech audio inputs.

We compute the normalized difference between mean activations:
\[
\vec{s} = \frac{\overline{\text{act}_H} - \overline{\text{act}_N}}{\lVert \overline{\text{act}_H} - \overline{\text{act}_N} \rVert},
\]
where \(H\) and \(N\) are Hallucinations and Non-hallucinations clusters respectively, where \(H\) is represented by non-speech samples with $\text{no\_speech\_prob} < \tau$ and \(N\) by non-speech samples with $\text{no\_speech\_prob} \geq \tau$.

\subsection{Experiment setup}
\textbf{Datasets}: Experiments are conducted on three non-speech datasets: FSD50k (sound events), Musan (general noise), WHAM (noisy speech without intelligible speech). For FSD50k samples with speech related lables are filtered. To evaluate the impact on genuine speech recognition performance, we use the LibriSpeech test-clean dataset.

\textbf{Model}: All experiments are based on the Whisper small model on activations after transformer block of AudioEncoder after 8th layer.

\textbf{Metrics}: Our primary metric for evaluating hallucination reduction is the False Positive Rate (FPR), defined as the proportion of non-speech audio clips for which the model generates any transcription with a no\_speech\_prob below a set threshold equals to 0.5. We also report the standard Word Error Rate (WER) on LibriSpeech to ensure that steering does not degrade performance on legitimate speech tasks. Due to the fact that after steering, the no\_speech\_prob parameter distribution on LibriSpeech dataset practically did not shift to the right, WER is a much better estimate of the preservation of the model's ability to recognize speech than True Positive Rate (TPR) or AUC score.

\subsection{Results and visualization}

Identifying the SAE features responsible for hallucinations was accomplished through a classification task using logistic regression. Calculated F1 metric depends on the hyperparameter $k$ (the number of SAE features used in the classification). This allows us to find a tradeoff between classification accuracy and the number of SAE features used. Our intuition was that although hallucinations are a complex concept, we want to find the minimum $k$ with quality at the level of the entire SAE vector classification. The results are presented in Table~\ref{tab:steering-classification}.

For the baseline configuration, we pursued a dual optimization objective: tuning the hyperparameter $\alpha$ while simultaneously identifying the dataset that yields the most effective steering vectors. Steering vectors were calculated independently for each dataset using its corresponding activation distributions. Each dataset's steering vector was then evaluated across a range of $\alpha$ values, applied to all datasets to assess both the hyperparameter sensitivity and the generalization performance of vectors originating from different source datasets. The results clearly show that the best steering vector is obtained on the Musan dataset with $\alpha=3$, as presented in Table~\ref{tab:steering-baseline-alpha-tune}.

To verify that the proposed steering vectors do not degrade standard ASR performance, we evaluate them on the LibriSpeech test-clean set for several values of the steering strength $\alpha$. For each steering configuration we measure the Word Error Rate (WER) of the ASR model. The results are summarized in Table~\ref{tab:steering-vector-asr-validation}. For Musan, FSD50k and WHAM steering vectors, WER remains essentially unchanged with respect to the unsteered model (around $0.05$) across all tested values of $\alpha$, indicating that these steering directions do not harm recognition quality on clean speech. 

Whisper inference with and without SAE and the effect of SAE on FPR are examined separately. Table~\ref{tab:steering-w/wo-sae-fpr-validation} shows that the addition of SAE does not significantly shift the distribution of the parameter no\_speech\_prob in the Musan and FSD50k datasets, but on the WHAM dataset, the FPR decreases from 0.51 to 0.36. This phenomenon requires further study. Furthermore, inference with SAE does not significantly change WER, as shown in Table~\ref{tab:steering-w/wo-sae-wer-validation}.

Unlike baseline hyperparameter optimization experiments, SAE-based steering introduces an additional hyperparameter $k$, the number of SAE features whose activations are steered. These features are selected according to their importance in the hallucination classifier. Thus, SAE steering requires jointly choosing both the scaling factor $\alpha$ and the sparsity level $k$. Table~\ref{tab:steering-sae-topk-alpha-tuning} reports the resulting FPR for each steering vector on each evaluation dataset. Experiments shows that a steering vector constructed on the FSD50k dataset with $k=100$ and $\alpha=3$ drives the FPR close to zero on all evaluation datasets. However, steering should not only suppress hallucinations but also preserve recognition quality. Therefore, we highlight two configurations: an \emph{extreme} setting, which achieves the strongest hallucination suppression with $k=100$ and $\alpha=3$, and an \emph{optimal} setting, which uses the same number of features but a milder scaling, $k=100$ and $\alpha=1$, to better balance hallucination reduction and ASR accuracy.

An analysis of the impact of SAE steering on speech recognition quality is also presented in Table~\ref{tab:steering-sae-wer-validation}, which shows that \emph{extreme} steering significantly degrades the model's ability to perform its original task. Meanwhile, \emph{optimal} steering degrades WER by only 0.3\%, while reducing FPR by 70\% (0.37 -> 0.11).

\begin{table}[t]
    \centering
    \small
    \begin{minipage}{0.48\textwidth}
        \centering
        \begin{tabular}{lcccccc}
            \toprule
                    & all  & 1000 & 100 & 10 & 5 & 1 \\
            \midrule
            Musan   & 0.72 & 0.74 & 0.67 & 0.55 & 0.42 & 0.33 \\
            FSD50k  & 0.72 & 0.74 & 0.73 & 0.55 & 0.51 & 0.35 \\
            WHAM    & 0.82 & 0.83 & 0.82 & 0.80 & 0.73 & 0.70 \\
            \bottomrule
        \end{tabular}
        \caption{F1 scores of the logistic–regression hallucination classifier as a function of the number $k$ of SAE features used. Rows correspond to evaluation datasets (Musan, FSD50k, WHAM). Column \textit{all} uses the full SAE vector, while \textit{top-$k$} columns use only the $k$ most informative SAE features, illustrating the tradeoff between classification quality and feature sparsity.}
        \label{tab:steering-classification}
    \end{minipage}
\end{table}

\begin{figure}
    \centering
    \includegraphics[width=1.0\linewidth]{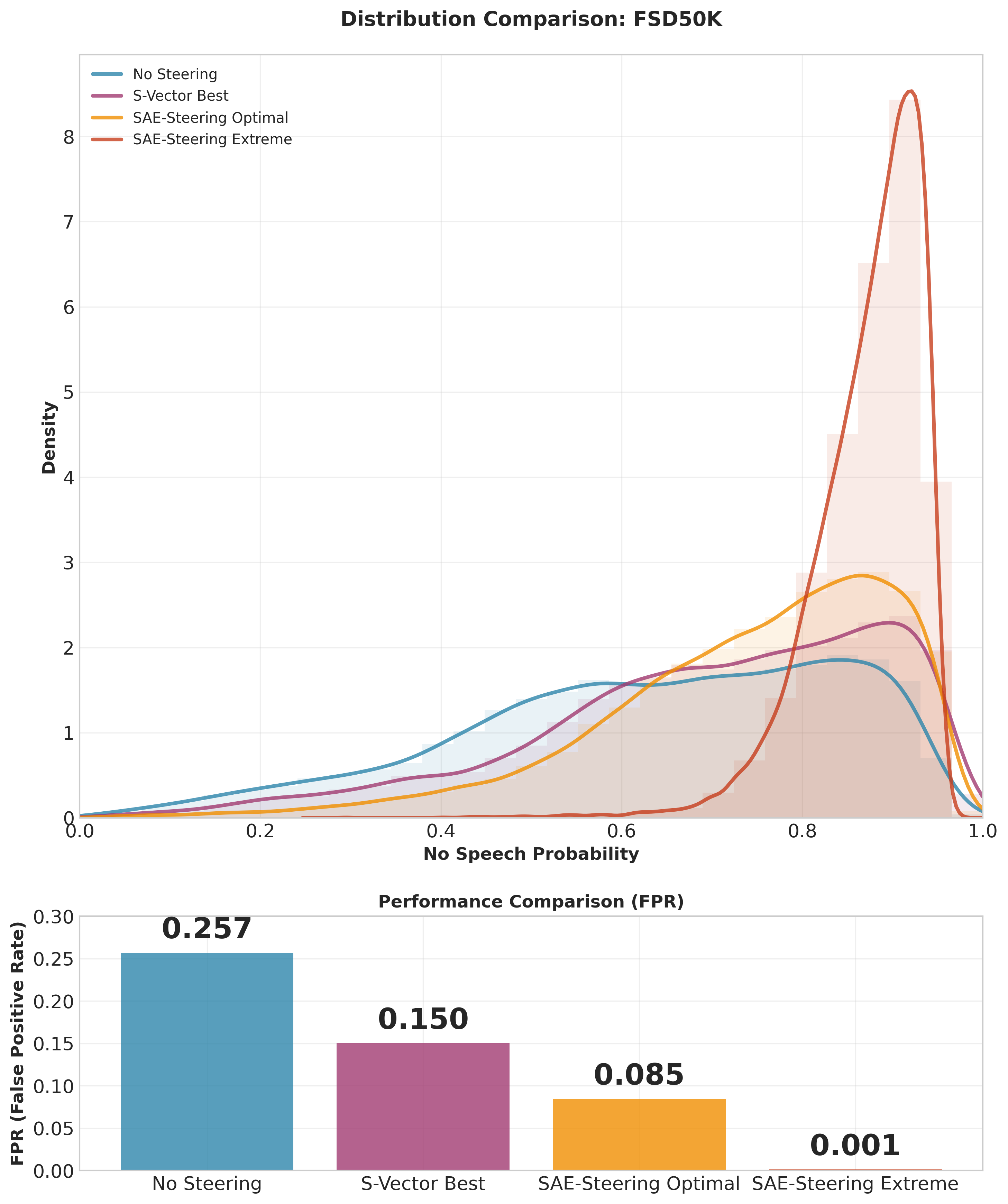}
    \caption{Distribution of no\_speech\_prob on the FSD50k dataset before and after applying steering vectors. The post-steering distribution is skewed towards 1.0.}
    \label{fig:steering-fsd50k}
\end{figure}

\begin{figure}
    \centering
    \includegraphics[width=1.0\linewidth]{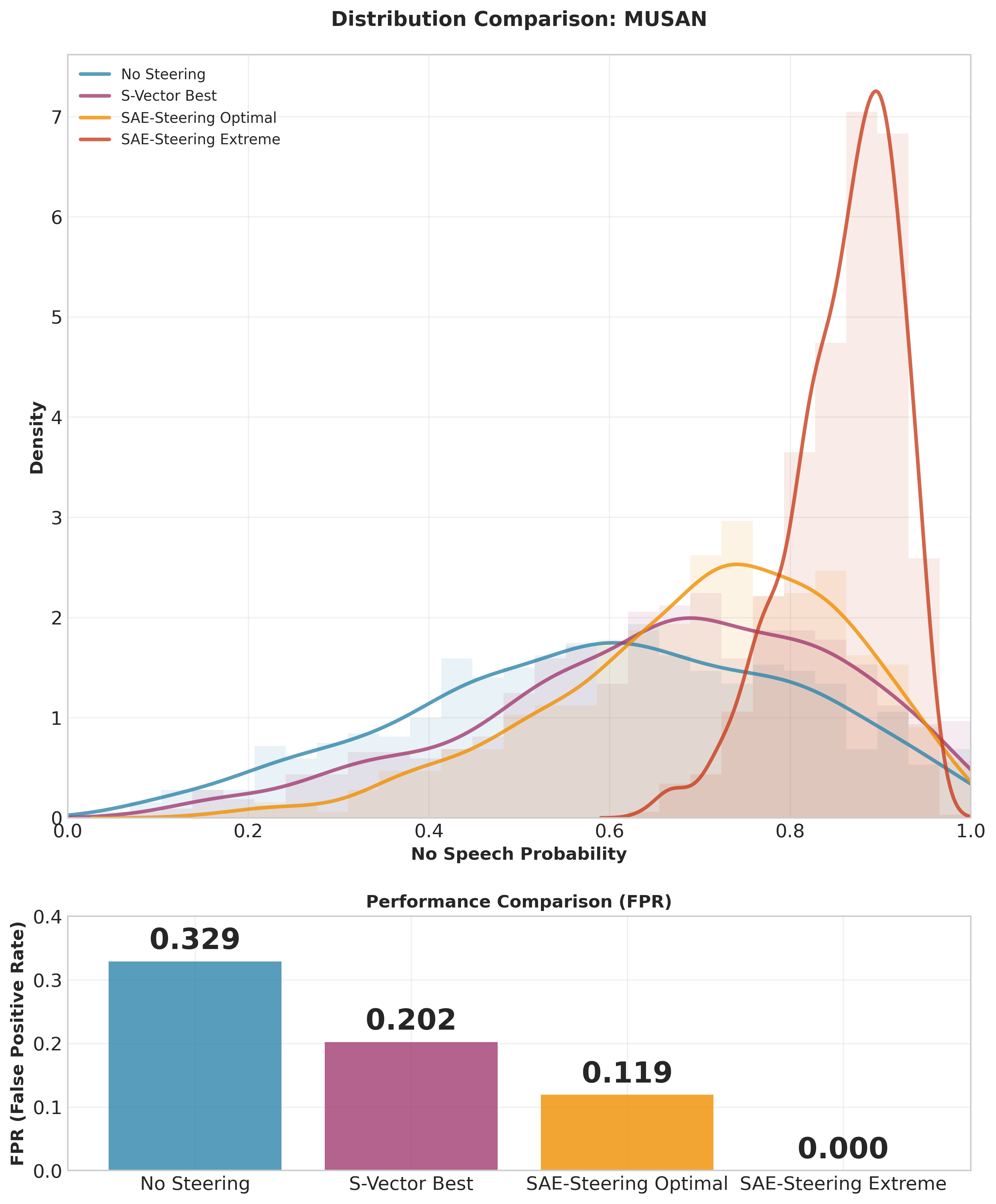}
    \caption{Distribution of no\_speech\_prob on the Musan dataset before and after applying steering vectors. The post-steering distribution is skewed towards 1.0.}
    \label{fig:steering-musan}
\end{figure}

\begin{figure}
    \centering
    \includegraphics[width=1.0\linewidth]{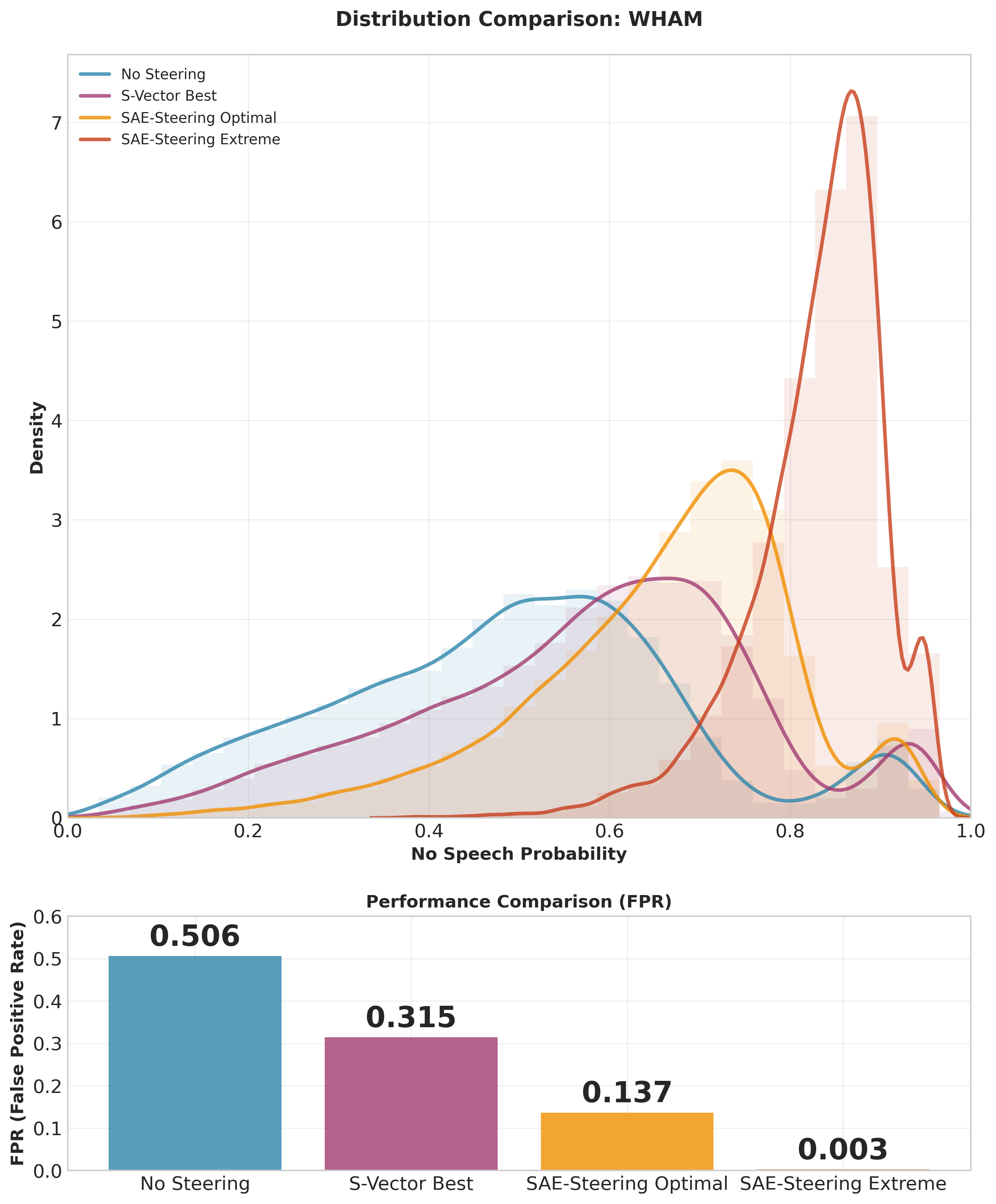}
    \caption{Distribution of no\_speech\_prob on the WHAM dataset before and after applying steering vectors.}
    \label{fig:steering-wham}
\end{figure}

\begin{figure}
    \centering
    \includegraphics[width=1.0\linewidth]{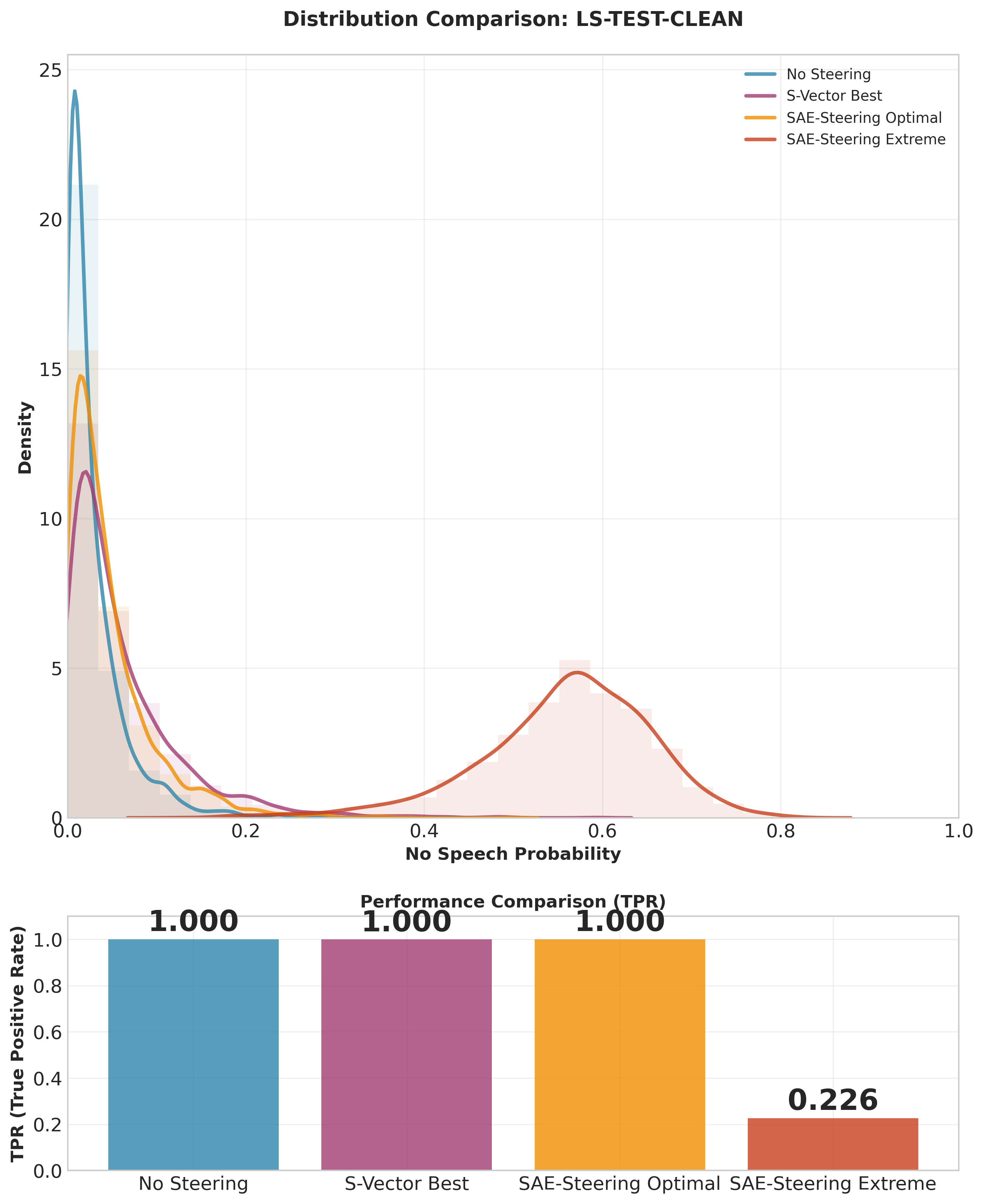}
    \caption{Distribution of no\_speech\_prob on the LibriSpeech test-clean dataset before and after applying steering vectors. The post-steering distribution is skewed towards 1.0.}
    \label{fig:steering-ls-test-clean}
\end{figure}

\begin{table}[t]
    \centering
    \small
    \begin{tabular}{lcccc}
        \multicolumn{5}{c}{$\alpha = 0.5$} \\
        \toprule
            & No Steering & Musan & FSD50k & WHAM \\
        \midrule
        Musan    & 0.33 & \textbf{0.31} & 0.32 & 0.34 \\
        FSD50k   & 0.27 & \textbf{0.24} & 0.25 & 0.27 \\
        WHAM     & 0.51 & \textbf{0.47} & 0.49 & 0.52 \\
        \bottomrule
    \end{tabular}

    \vspace{0.8em}

    \begin{tabular}{lcccc}
        \multicolumn{5}{c}{$\alpha = 1$} \\
        \toprule
            & No Steering & Musan & FSD50k & WHAM \\
        \midrule
        Musan    & 0.33 & \textbf{0.28} & 0.31 & 0.34 \\
        FSD50k   & 0.27 & \textbf{0.22} & 0.24 & 0.28 \\
        WHAM     & 0.51 & \textbf{0.44} & 0.47 & 0.54 \\
        \bottomrule
    \end{tabular}

    \vspace{0.8em}

    \begin{tabular}{lcccc}
        \multicolumn{5}{c}{$\alpha = 3$} \\
        \toprule
            & No Steering & Musan & FSD50k & WHAM \\
        \midrule
        Musan    & 0.33 & \textbf{0.20} & 0.28 & 0.36 \\
        FSD50k   & 0.27 & \textbf{0.15} & 0.22 & 0.30 \\
        WHAM     & 0.51 & \textbf{0.32} & 0.41 & 0.57 \\
        \bottomrule
    \end{tabular}
    \caption{False Positive Rate for baseline steering-vector steering at different $\alpha$ values.
    Columns show error when steering with the corresponding steering vector.}
    \label{tab:steering-baseline-alpha-tune}
\end{table}
The no\_speech\_prob parameter distribution shift after steering plots for the selected configurations (baseline Musan steering vector with $alpha=3$ and SAE steering vectors with top-$k=100$ and $\alpha \in \{1, 3\}$) are also presented. For the Musan (Fig.~\ref{fig:steering-musan}) dataset, FSD50k (Fig.~\ref{fig:steering-fsd50k}) and WHAM (Fig.~\ref{fig:steering-wham}) FPR was calculated, for LibriSpeech test-clean (Fig.~\ref{fig:steering-ls-test-clean}) TPR (higher is better) was calculated.
\begin{table}[t]
    \centering
    \small
    \begin{tabular}{lcccc}
        \toprule
            & \textbf{No Steering} & \textbf{Musan} & \textbf{FSD50k} & \textbf{WHAM} \\
        \midrule
        $\alpha = 0.5$ & 0.051 & 0.051 & 0.051 & 0.051 \\
        $\alpha = 1.0$ & 0.051 & 0.051 & 0.051 & 0.051 \\
        $\alpha = 3.0$ & 0.051 & 0.053 & 0.055 & 0.053 \\
        \bottomrule
    \end{tabular}
    \caption{Steering vector validation on ASR task. Word error rate (WER; lower is better) of the Whisper-small model on LibriSpeech test-clean when steered with different steering vectors and strengths. Rows correspond to steering strength $\alpha \in \{0.5, 1.0, 3.0\}$, while columns indicate which steering vector is used.}
    \label{tab:steering-vector-asr-validation}
\end{table}

\begin{table}[t]
    \centering
    \small
    \begin{tabular}{@{}llcccc@{}}
    \toprule
    \textbf{Test} & \textbf{Steering} & \multicolumn{2}{c}{\textbf{Top-k=100}} & \textbf{k=10} & \textbf{k=1} \\
    \cmidrule(lr){3-4} \cmidrule(l){5-5} \cmidrule(l){6-6}
    \textbf{Dataset} & \textbf{Vector} & $\alpha$=1.0 & $\alpha$=3.0 & $\alpha$=3.0 & $\alpha$=3.0 \\
    \midrule
    Musan & Musan & 0.20 & \textbf{0.01} & 0.20 & 0.28 \\
    & FSD50K & \color{ForestGreen}{0.12} & \textbf{0.00} & 0.19 & 0.31 \\
    & WHAM & 0.26 & \textbf{0.00} & 0.25 & 0.34 \\
    \addlinespace
    FSD50K & Musan & 0.16 & \textbf{0.01} & 0.16 & 0.22 \\
    & FSD50K & \color{ForestGreen}{0.09} & \textbf{0.00} & 0.14 & 0.24 \\
    & WHAM & 0.20 & 0.01 & 0.19 & 0.28 \\
    \addlinespace
    WHAM & Musan & 0.32 & 0.04 & 0.32 & 0.44 \\
    & FSD50K & \color{ForestGreen}{0.13} & \textbf{0.00} & 0.24 & 0.47 \\
    & WHAM & 0.29 & \textbf{0.00} & 0.30 & 0.54 \\
    \bottomrule
    \end{tabular}

    \caption{False Positive Rate (FPR) for SAE steering with different configurations. Tuning the scaling factor $\alpha$ and the number of top-$k$ most informative SAE features selected from the hallucination classifier. Test Dataset column represents the dataset on which FPR was calculated. Steering Vector column according to the dataset from which the SAE steering vector was formed. Short version, all experiments presented in the Table~\ref{tab:steering-sae-topk-alpha-tuning-full}.}
    \label{tab:steering-sae-topk-alpha-tuning}
\end{table}

\begin{table}[t]
    \centering
    \small
    \begin{tabular}{@{}lcc@{}}
    \toprule
    \textbf{Method} & \textbf{No Steering} & \textbf{SAE + No Steering} \\
    \midrule
    Baseline & 0.051 & 0.052 \\
    \bottomrule
    \end{tabular}
    \caption{Whisper small WER calculation on LibriSpeech test-clean without/with SAE.}
    \label{tab:steering-w/wo-sae-wer-validation}
\end{table}

\begin{table}[t]
    \centering
    \small
    \begin{tabular}{@{}lcc@{}}
    \toprule
    \textbf{Dataset} & \textbf{No Steering} & \textbf{SAE + No Steering} \\
    \midrule
    Musan & 0.33 & 0.31 \\
    FSD50K & 0.27 & 0.24 \\
    WHAM & 0.51 & 0.36 \\
    \bottomrule
    \end{tabular}
    \caption{FPR Reduction in inference configurations without/with SAE.}
    \label{tab:steering-w/wo-sae-fpr-validation}
\end{table}

\begin{table}[t]
    \centering
    \small
    \begin{tabular}{@{}lccccc@{}}
    \toprule
    & & \multicolumn{4}{c}{\textbf{Word Error Rate (WER)}} \\
    \cmidrule(l){3-6}
    \textbf{S-Vector} & $\alpha$ & $k=100$ & $k=10$ & $k=5$ & $k=1$ \\
    \midrule
    \multirow{3}{*}{Musan} & 0.5 & 0.053 & 0.052 & 0.052 & 0.052 \\
    & 1.0 & 0.054 & 0.052 & 0.052 & 0.052 \\
    & 3.0 & 0.139 & 0.054 & 0.054 & 0.052 \\
    \addlinespace
    \multirow{3}{*}{FSD50K} & 0.5 & 0.053 & 0.052 & 0.052 & 0.052 \\
    & 1.0 & 0.055 & 0.053 & 0.052 & 0.052 \\
    & 3.0 & 0.984 & 0.054 & 0.052 & 0.052 \\
    \addlinespace
    \multirow{3}{*}{WHAM} & 0.5 & 0.054 & 0.052 & 0.052 & 0.052 \\
    & 1.0 & 0.064 & 0.052 & 0.052 & 0.052 \\
    & 3.0 & 0.975 & 0.055 & 0.053 & 0.052 \\
    \bottomrule
    \end{tabular}
    \caption{WER on LibriSpeech test-clean when steering with SAE-based vectors. The table reports WER while jointly tuning the steering strength parameter $\alpha$ and the number of top-$k$ most informative SAE features. Column S-Vector indicate the dataset used to calculate the SAE steering vector.}
    \label{tab:steering-sae-wer-validation}
\end{table}

\section{Mel-interpretation details}\label{appendix:mel-int}

\begin{figure}[h!]
    \centering
    \includegraphics[width=0.4\linewidth]{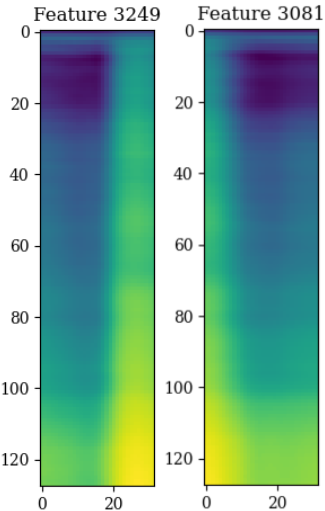}
    \caption{Features 3249 and 3081 of Hubert's SAE from layer 11.}
    \label{fig:mel12}
\end{figure}

The experiment was designed to find features which activates at the beginning or end of a word. It was assumed that the energy in the averaged feature maps from the mel spectrograms would be shifted to the right and left, respectively; that is, for a word-beginning feature, the energy would be to the right of center, and for a word-ending feature, it would be to the left. These maps were found among the top 5 features with the highest activation frequency difference between the speech domain and the music and sound domains, based on frequencies from Domain Specialization experiments. Therefore, it can be concluded that detected features are strongly speech-specific. Moreover, they were present in several layers of HuBERT and Whisper, both at the beginning and closer to the end of the network. However, only a variant for HuBERT layer 11 is presented in the article in Fig.~\ref{fig:mel12}.

\section{Details of EEG experiments}
\label{appendix:eeg}
We chose data from midline parietal electrode Pz collected from $19$ subjects listening to $5$ excerpts from audiobook each $3$ minutes long resulting in $15$ minutes for each participant. SAE features (stimuli $s$) were extracted from these excerpts with models trained on the last $12$-th layer of HuBERT-base and Whisper-base models before normalization. SAE features were normalized to have unit maximum whereas EEG signals (responses $r$) were first processed by band-pass filter keeping frequencies between $1$Hz and $8$Hz and then normalized to have zero median and unit interquartile range. Both EEG signals and SAE features were resampled to $128$Hz. Temporal-response functions (TRFs) were built with mTRFpy Python package.

We randomly chose $1000$ HuBERT and $1000$ Whisper features activating at least once per second on average. For each feature $f$, we found time lag values $\tau_{min}^{(f)}$ and $\tau_{max}^{(f)}$ minimizing and maximizing TRFs respectively on the development set with the total duration of $6$ minutes. Then, for each feature $f$ for TRFs built on the test set with the overall duration of $9$ minutes we performed two one-sided t-tests to check whether corresponding TRFs have statistically significant negative correlation at $\tau_{min}^{(f)}$ and positive correlation at $\tau_{max}^{(f)}$. After that, we applied Holm-Bonferroni correction to process the results of multiple statistical tests at significance level $0.05$. As a result of this procedure, we found around $1\%$ of Whisper and $1.5\%$ of HuBERT features having significant correlation with Pz electrode response at certain time lags.

As one can see from Fig.~\ref{fig:trf}, correlation between SAE and EEG features can occur with almost zero time lag. It may seem counterintuitive since it must take some time for brain to process audio information, but it has to be noted that HuBERT and Whisper feature extractors have access to both left and right audio context and can activate, for example, in the end of a particular word or sound, which can explain this seeming contradiction.

\end{document}